\documentclass[twocolumn,prd,aps,longbibliography,superscriptaddress,preprintnumbers,tightenlines,showpacs,nofootinbib,
amsfonts,amsmath]{revtex4-2}

\pdfoutput=1

\usepackage{epsfig}
\usepackage{graphics}
\usepackage{graphicx}
\usepackage{amsmath,amssymb,mathrsfs}
\usepackage{amsfonts}
\usepackage[usenames,dvipsnames]{xcolor}
\usepackage{xcolor}
\usepackage{wasysym}
\usepackage{times}
\usepackage{mathptmx}
\usepackage{gensymb}
\usepackage{appendix}
\usepackage{listings}
\usepackage{url}
\usepackage[normalem]{ulem}
\usepackage{alltt}
\usepackage[colorlinks]{hyperref}
\usepackage{cleveref}
\usepackage{longtable}
\usepackage{enumitem}
\setlist{nosep}
\usepackage{color}
\usepackage{calc}
\usepackage{tensor}
\usepackage{bm}
\usepackage{times}
\usepackage{multirow}
\usepackage[varg]{txfonts}
\usepackage{float}
\usepackage{dcolumn}
\usepackage[nolist,nohyperlinks]{acronym}
\usepackage{xspace}
\usepackage[english]{babel}
\usepackage[abs]{overpic}
\usepackage{pict2e}
\usepackage[caption=false]{subfig} 
\allowdisplaybreaks[1]
\usepackage[utf8]{inputenc}
\usepackage{gensymb}
\usepackage{bm}
\usepackage{stackengine}
\usepackage{boldline,multirow}
\usepackage{braket}
\usepackage{longtable}
\usepackage{tabularx}
\usepackage{rotating}

\Crefname{figure}{Fig.}{Figs.}

\DeclareMathAlphabet{\mathcalstd}{OMS}{cmsy}{m}{n}
\DeclareMathAlphabet{\mathpzc}{OT1}{pzc}{m}{it}

\definecolor{dodgerblue}{HTML}{1E90FF}
\definecolor{viennared}{HTML}{DA0A14}

\hypersetup{citecolor=dodgerblue}

\newcommand{\gecc}[0]{{e_g}}                 
\newcommand{\egw}[0]{e_{\omega_{22}}}                

\newcommand{\wgw}{\omega_{22}} 
\newcommand{\wgwp}{\omega_{22}^{\rm p}} 
\newcommand{\wgwa}{\omega_{22}^{\rm a}} 
\newcommand{\wgwavg}{\langle\omega_{22}\rangle} 
\newcommand{\wgwavgi}{\langle\omega_{22}\rangle_{i}} 
\newcommand{\wgwradi}{\langle\Omega_{22}^{\rm r}\rangle_{i}} 
\newcommand{\wgwrad}{\langle\Omega_{22}^{\rm r}\rangle} 
\newcommand{\initNRw}{\Omega_0} 
\newcommand{\wref}{\omega_{22}^{\rm ref}} 
\newcommand{\worb}{\Omega_{\rm orb}} 
\newcommand{\worba}{\Omega^a_{\rm orb}} 
\newcommand{\worbp}{\Omega^p_{\rm orb}} 
\newcommand{\worbavg}{\langle\Omega_{\rm orb}\rangle} 
\newcommand{\worbrad}{\langle\Omega^{\rm r}_{\rm orb}\rangle} 

\newcommand{\wphigeo}{\Omega^{}_{\rm geo}} 
\newcommand{\wradgeo}{\Omega^{\rm r}_{\rm geo}} 
\newcommand{\wmnmode}{\varpi_{mn}} 

\newcommand{\wphiPA}{\langle\omega_{\rm 0PA}\rangle} 
\newcommand{\wradPA}{\Omega^{r}_{\rm 0PA}} 

\newcommand{\smr}[0]{\nu} 

\begin{document}


\title{
\hspace*{-0.4cm}Eccentric binary black holes: Comparing numerical relativity and small mass-ratio perturbation theory\hspace*{-0.4cm}
}
\newcommand{\aei}{\affiliation{Max Planck Institute for Gravitational Physics
(Albert Einstein Institute), Am M{\"u}hlenberg 1, 14476 Potsdam, Germany}}
\newcommand{\caltech}{\affiliation{Theoretical Astrophysics, Walter Burke
Institute for Theoretical Physics, California Institute of Technology, Pasadena,
California 91125, USA}}
\newcommand{\cornell}{\affiliation{Cornell Center for Astrophysics and Planetary
    Science, Cornell University, Ithaca, New York 14853, USA}}

\newcommand{\coimbra}{
\affiliation{CFisUC, Department of Physics, University of Coimbra, 3004-516 Coimbra, Portugal}}
\newcommand{\NBIA}{\affiliation{Niels Bohr International Academy, Niels Bohr Institute, Blegdamsvej 17, 2100 Copenhagen, Denmark}}

\author{Antoni Ramos-Buades}\aei 
\author{Maarten van de Meent} \aei\NBIA
\author{Harald P. Pfeiffer} \aei
\author{Hannes R. R{\"u}ter} \coimbra
\author{Mark A. Scheel} \caltech
\author{Michael Boyle} \cornell
\author{Lawrence E. Kidder} \cornell

\date{\today}

\begin{abstract}
	The modelling of unequal mass binary black hole systems is of high importance to detect and estimate parameters from these systems.
	Numerical relativity (NR) is well suited to study systems with comparable component masses, $m_1\sim m_2$, whereas small mass ratio (SMR) perturbation theory applies to binaries where $q=m_2/m_1<< 1$.
	This work investigates the applicability for NR and SMR as a function of mass ratio for eccentric  non-spinning  binary black holes.
	We produce $52$ NR simulations with mass ratios between $1:10$ and $1:1$ and initial eccentricities up to~$0.7$.
From these we extract quantities like gravitational wave energy and angular momentum fluxes and periastron advance, and assess their accuracy.
To facilitate comparison, we develop tools to map between NR and SMR inspiral evolutions of eccentric binary black holes.
	We derive post-Newtonian accurate relations between different definitions of eccentricity.
	Based on these analyses, we introduce a new definition of eccentricity based on the (2,2)-mode of the gravitational radiation, which reduces to the Newtonian definition of eccentricity in the Newtonian limit.
	From the comparison between NR simulations and SMR  results, we quantify the unknown next-to-leading order SMR contributions to the gravitational energy and angular momentum fluxes, and periastron advance.
	We show that in the comparable mass regime these contributions are subdominant and higher order SMR contributions are negligible.
\end{abstract}


\maketitle

\acrodef{PN}{post-Newtonian}
\acrodef{EOB}{effective-one-body}
\acrodef{NR}{numerical relativity}
\acrodef{GW}{gravitational wave}
\acrodef{BBH}{binary black hole}
\acrodef{BH}{black hole}
\acrodef{BNS}{binary neutron star}
\acrodef{NSBH}{neutron star-black hole}
\acrodef{SNR}{signal-to-noise ratio}
\acrodef{aLIGO}{Advanced LIGO}
\acrodef{AdV}{Advanced Virgo}

\newcommand{\PN}[0]{\ac{PN}\xspace}
\newcommand{\EOB}[0]{\ac{EOB}\xspace}
\newcommand{\NR}[0]{\ac{NR}\xspace}
\newcommand{\BBH}[0]{\ac{BBH}\xspace}
\newcommand{\BH}[0]{\ac{BH}\xspace}
\newcommand{\BNS}[0]{\ac{BNS}\xspace}
\newcommand{\NSBH}[0]{\ac{NSBH}\xspace}
\newcommand{\GW}[0]{\ac{GW}\xspace}
\newcommand{\SNR}[0]{\ac{SNR}\xspace}
\newcommand{\aLIGO}[0]{\ac{aLIGO}\xspace}
\newcommand{\AdV}[0]{\ac{AdV}\xspace}

\newcommand{\citeme}[0]{{\color{purple}{Citation!}}}

\newcommand*{\vecL}[0]{${{\mathbf{L}}}$ }

\newcommand{\n}{\newline}

\section{Introduction} \label{sec:Intro}

Binary black hole (BBH) mergers have dominated the gravitational wave (GW)
observations of the LIGO and Virgo detectors \cite{TheLIGOScientific:2014jea, TheVirgo:2014hva} in the first, second and the third observing runs \cite{ TheLIGOScientific:2016pea, LIGOScientific:2018mvr, LIGOScientific:2020ibl,LIGOScientific:2021djp}. One key  parameter of these astrophysical systems is the mass ratio $q=m_2/m_1\leq 1$ of the binaries' components. Current GW observations \cite{LIGOScientific:2018jsj, Venumadhav:2019lyq, LIGOScientific:2020ibl,Nitz:2021zwj} predominantly find mass ratios close to unity with a few observations showing support for low mass ratios \cite{LIGOScientific:2020stg,LIGOScientific:2020zkf}.

With the increasing number of GW detections in the upcoming observing runs by
ground-based detectors \cite{LIGOScientific:2018jsj,
  LIGOScientific:2020kqk}, and space-borne detectors, like the LISA
observatory \cite{Katz:2019qlu,Gair:2010bx}, it is likely that
more binaries with mass asymmetries are found.   In particular, LISA will be sensitive to binaries with mass ratios ranging from $q\sim 1$, over intermediate
  mass-ratio systems ($q\sim 10^{-3}$) to extreme mass ratio inspirals
  at $q\sim 10^{-5}$.  Furthermore, third-generation ground-based
  detectors with improved low frequency sensitivity relative to
  today's ground-based detectors will be able to detect the capture of
  stellar mass black holes (BHs) by intermediate mass BHs with mass-ratios down to
  $q\sim 10^{-3}$~\cite{Jani:2019ffg}. Thus, the modelling of GWs from BBHs at all mass ratios is of preeminent relevance for a correct detection and analysis of these sources.
  
This modelling problem may be tackled by different approaches:  using weak field perturbation theory, like post-Newtonian (PN) theory \cite{Blanchet:2013haa} and post-Minkowskian expansions \cite{Cristofoli:2019neg}, effective methods (like the effective-one-body formalism \cite{Buonanno:1998gg, Damour:2008yg} or phenomenological models \cite{Ajith:2007qp}), small mass ratio (SMR) perturbation theory \cite{Pound:2021qin} and numerical relativity (NR), i.e., solving numerically the full Einstein equations \cite{Lehner:2014asa}.  

Orbital eccentricity is another important parameter describing binary systems.  While emission of gravitational waves generally does reduce eccentricity~\cite{Peters:1963ux,PhysRev.131.435}, orbital eccentricity is a key parameter to constrain the formation scenario of these binaries and thus the astrophysical origin of GW sources~\cite{PortegiesZwart:1999nm, Mandel:2009nx,  Samsing:2013kua, Rodriguez:2018rmd, Fragione:2018vty, Zevin:2018kzq,Zevin:2021rtf}.  For current ground-based detectors there are ongoing efforts to search for signatures of orbital eccentricity in the detected GW signals \cite{Romero-Shaw:2019itr,Gayathri:2020coq,Romero-Shaw:2020thy,Bustillo:2020syj,Gamba:2021gap,Romero-Shaw:2021ual,Romero-Shaw:2022xko,Iglesias:2022xfc,Clarke:2022fma,Knee:2022hth,Bonino:2022hkj}.
For future GW detectors, especially for high-mass ratio binaries in LISA, it is expected that emission of GWs has not circularized most binaries yet. Therefore, the correct modelling of orbital eccentricity effects is fundamental to accurately describe such systems in future detectors, in particular for extreme mass-ratio inspirals, which are described by SMR perturbation theory.

Recently, Ref.~\cite{vandeMeent:2020xgc} demonstrated that NR
  simulations at modest mass-ratios ($q\gtrsim 0.1$) can be used to
  gain insight into the accuracy of the SMR expansion, confirming the
  known leading-order term, and predicting next-to-leading order
  contributions.  Ref.~\cite{vandeMeent:2020xgc} considered
  non-eccentric (quasi-circular) binaries only, with both BHs non-spinning.  
  Here, we begin to extend the analysis in
\cite{vandeMeent:2020xgc} to eccentric 
BBHs, while still keeping both BHs non-spinning.  The non-circularity of the binary's orbit introduces a new
  timescale to the two-body problem, the timescale of the periastron
  precession, which induces oscillations in the dynamical and GW
  quantities of the binary system complicating substantially the
  analysis relative to the quasi-circular case described in
  \cite{vandeMeent:2020xgc}.

An additional difficulty arises from the fact that eccentricity is a gauge dependent parameter in general relativity, thus complicating the comparison between SMR evolutions and NR simulations. In order to overcome this problem, we develop tools to extract gauge invariant quantities from both SMR and NR waveforms. Using PN theory, we derive relations among the eccentricity defined from the orbital and (2,2)-mode gravitational wave frequency, and the PN temporal eccentricity. We show that a commonly used definition of eccentricity based on the (2,2)-mode frequency ---Eq.~(\ref{eq:eq7}) below--- does \textit{not} reduce to the Newtonian definition of eccentricity. We therefore adopt a new definition of eccentricity, $e_{\rm gw}$ in Eq.~(\ref{eq:eqEccDef}) below.  This new definition continues to be based on the frequency of the (2,2) GW-mode, but also satisfies the correct Newtonian limit.

NR simulations of BBHs  have been  routinely performed since more than a decade ago. Motivated by expectation of small eccentricities for most of the GW signals in the frequency band of ground-based detectors, most of the NR groups have focused on the production of simulations of quasi-circular BBH mergers \cite{Hannam:2007wf,Hannam:2010ec,Purrer:2012wy,Mroue:2013xna,Jani:2016wkt,Healy:2017psd,Healy:2017xwx, Ramos-Buades:2018azo, Boyle:2019kee,Healy:2019jyf, Healy:2020vre, Healy:2022wdn}, with the exception of a limited number of studies exploring BBH coalescences in eccentric orbits \cite{Hinder:2008kv, Mroue:2010re,Lewis:2016lgx, Hinder:2017sxy,Islam:2021mha, Huerta:2019oxn,Habib:2019cui, Ramos-Buades:2019uvh,Healy:2022wdn}.
The Spectral Einstein Code (\texttt{SpEC})~\cite{SpECwebsite}
  is an accurate and efficient NR code that has been used to study
  quasi-circular inspirals in great depth~\cite{Mroue:2013xna,
    Boyle:2019kee}.  Eccentric inspirals at low eccentricity were
  studied to some
  extent~\cite{Mroue:2010re,Lewis:2016lgx,Hinder:2017sxy,Islam:2021mha}.
  We expand \texttt{SpEC}'s capabilities for accurate simulations of
  binaries with larger eccentricities, $0.2\lesssim e\lesssim 0.8$,
  which are characterized by large variations of GW frequency and
  amplitude between apastron and periastron passages.  We have
  produced a set of $52$ non-spinning eccentric simulations between mass
  ratios $1:10$ and $1:1$, and with initial eccentricities of up to
  0.7.  The number of orbits is typically $\gtrsim 20$, yielding a
  dataset with the longest evolutions and highest initial
  eccentricities up to date,
  which sets it also apart from the simulations of other groups
  \cite{Huerta:2019oxn,Ramos-Buades:2019uvh,Healy:2022wdn}.

The main purpose of this article is to compare NR and SMR
  calculations.  This requires a map from the instantaneous state of
a NR simulation to the geodesic on which the point-particle
instantaneously moves in its motion around the central black hole.
We characterize the instantaneous state of SMR and NR simulations by symmetric mass ratio, $\nu= m_1 m_2 /(m_1+m_2)^2$,  eccentricity, $e_{\rm gw}$,  and   orbit-averaged frequency of the 22-mode, $\wgwavg$. These quantities can be uniquely determined in SMR and NR configurations and they generate an  unambiguous map between SMR and NR configurations, as described in Sec.~\ref{sec:IndependentQuantities}.

We find that the leading order prediction in the SMR expansion for the energy and angular momentum fluxes agree with the NR results to within $10\%$.
The next-to-leading order SMR contributions to the fluxes can be estimated by rescaling the difference of the NR and leading order SMR contribution by a factor of the symmetric mass-ratio. The result has a very small dispersion in symmetric mass ratio, which implies that the next-to-next-to leading order SMR contribution is small, even for comparable masses. This is compatible with the findings of~\cite{Warburton:2021kwk} in the quasi-circular case. Comparing the zero-eccentricity limit of our next-to-leading order estimate to the exact results of~\cite{Warburton:2021kwk} we find the results to be comparable with an overall small shift likely due to the orbit-averaging procedure applied to extract quantities from the eccentric NR simulations.
A similar analysis is done for the periastron advance. In this case the NR results are within $8 \%$ of the leading-order (geodesic) SMR result, the next-to-leading SMR contribution is compatible with previous exact calculations in the quasi-circular limit~\cite{Barack:2011ed,vandeMeent:2016hel}, and the next-to-next-to-leading SMR contribution appears small in the comparable mass regime.

This article is organized as follows. In Sec.~\ref{sec:NRsimulations} we present a detailed description of the new dataset of eccentric  non-spinning NR simulations produced for this work.  We investigate the relations between different eccentricity definitions in Sec.~\ref{sec:Eccdef}, and we provide a definition of eccentricity based on the (2,2)-mode frequency, which reduces to the Newtonian definition of eccentricity in the Newtonian limit.
Section~\ref{sec:SMRevolutions} describes the SMR evolutions performed in this work, and in Sec.~\ref{sec:IndependentQuantities} discusses the mapping between SMR and NR configurations. In Sec.~\ref{sec:SMRNRcomparison} we compare the quantities extracted from the NR simulations to the SMR  perturbation theory results,  and provide constraints on the values of the next order terms in the SMR expansion for the GW energy and angular momentum fluxes, as well as the periastron advance. In Sec.~\ref{sec:Conclusions} we summarize our main conclusions and discuss future work.
The appendices contain additional technical details: Appendix \ref{sec:AppendixIC} describes our method to set the initial parameters in the NR simulations, in Appendix \ref{sec:AppendixNRconvergence} we assess the quality of the NR waveforms, and in Appendix \ref{sec:AppendixeOm22} we provide details of the derivation of the relations between different definitions of eccentricity using PN theory.

\section{Numerical relativity simulations}
\label{sec:NRsimulations}

The NR simulations produced in
this work are performed with the \texttt{SpEC} code
\cite{SpECwebsite}, utilizing  numerical techniques summarized in \cite{Mroue:2013xna,Boyle:2019kee}.  In particular,
\texttt{SpEC} evolves a first-order representation of the generalized
harmonic evolution system~\cite{Lindblom:2005qh} using a multi-domain
spectral method~\cite{Kidder:1999fv, Scheel:2008rj, Szilagyi:2009qz,
  Hemberger:2012jz}.  At the outer boundary constraint-preserving
boundary conditions~\cite{Lindblom:2005qh, Rinne:2006vv, Rinne:2007ui}
are employed, whereas black hole excision is used inside the apparent
horizons~\cite{Scheel:2008rj, Szilagyi:2009qz, Hemberger:2012jz,
  Ossokine:2013zga}.  The transition to ringdown is accomplished with
the techniques described in Refs.~\cite{Scheel:2008rj,
  Hemberger:2012jz}.
Initial data are constructed with the eXtended Conformal-Thin Sandwich (XCTS)
approach \cite{York:1998hy,Pfeiffer:2002iy,Cook:2004kt}, and
we describe in Sec.~\ref{subsec:NRdataset} and Appendix~\ref{sec:AppendixIC} how
we achieve binaries with a desired value of orbital eccentricity.

For improved performance for eccentric systems we adopt part of the
modifications developed in \cite{Rueter:2021a} to produce accurate
simulations of hyperbolic encounters.  Most notably, adaptive
  mesh refinement and GW output is triggered more
  frequently to adjust to periastron passages which happen on fast
  timescales, and which cause pulses of higher-frequency
  GWs that travel through the computational grid.

\subsection{Numerical relativity dataset}
\label{subsec:NRdataset}

\begin{table*}[!]
\begin{tabular}{ l | D{.}{.}{3.2} D{.}{.}{3.2} D{.}{.}{3.6} D{.}{.}{2.4} | D{.}{.}{4.2} D{.}{.}{5.0} D{.}{.}{5.2} D{.}{.}{1.4} D{.}{.}{1.3} }
 & \multicolumn{4}{c|}{Initial data} & \multicolumn{5}{c}{Physical properties}   \\
\hline  
\hline
&&&&&&&&&\\[-1.1em]   
 \multicolumn{1}{c}{$\text{SXS ID}$}
& \multicolumn{1}{c}{$1/q$}
& \multicolumn{1}{c}{$D_0/M_0$}
& \multicolumn{1}{c}{$M_0 \Omega_0$}
& \multicolumn{1}{c|}{$a_0 \times 10^6$}
& \multicolumn{1}{c}{$N_{\text{orbits}}$}
& \multicolumn{1}{c}{$\;T_{\text{merger}}/M$}
& \multicolumn{1}{c}{$\;(T_{\text{ref}}-T_{\text{merger}})/M\;$}
& \multicolumn{1}{c}{$e_{\rm gw}^{\text{ref}}$}
& \multicolumn{1}{c}{$l_{\text{ref}}/(2 \pi)$}  \\
&&&&&&&&&\\[-1.1em]
\hline 
\hline 
SXS:BBH:2517 & 1.0 & 16.03 & 0.0142 & -2.898 & 19.0 & 5199 & -3177.8 & 0.024 & 0.428 \\
SXS:BBH:2518 & 1.0 & 22.02 & 0.0089 & -0.808 & 40.3 & 16510 & -14489.9 & 0.027 & 0.251   \\
SXS:BBH:2519 & 1.0 & 20.03 & 0.0102 & -1.183 & 32.0 & 11593 & -9573.4 & 0.028 & 0.234  \\
SXS:BBH:2520 & 1.0 & 18.03 & 0.0114 & -1.804 & 18.2 & 4963 & -3011.5 & 0.105 & 0.028  \\
SXS:BBH:2521 & 1.0 & 26.02 & 0.0061 & -0.416 & 25.2 & 8799 & -6985.0 & 0.183 & 0.273 \\
SXS:BBH:2522 & 1.0 & 28.02 & 0.0053 & -0.310 & 25.0 & 8930 & -7170.0 & 0.207 & 0.249  \\
SXS:BBH:2523 & 1.0 & 34.02 & 0.0038 & -0.144 & 27.3 & 10993 & -9314.3 & 0.239 & 0.463  \\
SXS:BBH:2524 & 1.0 & 60.01 & 0.0013 & -0.015 & 30.0 & 17074 & -15625.3 & 0.313 & 0.567  \\
SXS:BBH:2525 & 1.0 & 45.01 & 0.0022 & -0.047 & 22.5 & 9820 & -8415.9 & 0.330 & 0.402  \\
SXS:BBH:2526 & 1.0 & 70.01 & 0.0010 & -0.008 & 26.2 & 15735 & -14412.3 & 0.352 & 0.845   \\
SXS:BBH:2527 (*) & 1.0 & 130.00 & 0.0003 & -0.001 & 19.9 & 16380 & -15345.6 & 0.437 & 0.579  \\
SXS:BBH:2528 & 1.0 & 65.01 & 0.0010 & -0.011 & 15.0 & 7137 & -6132.4 & 0.445 & 0.621  \\
SXS:BBH:2529 & 2.0 & 18.03 & 0.0120 & -1.279 & 28.1 & 9025 & -6803.6 & 0.024 & 0.080 \\
SXS:BBH:2530 & 2.0 & 20.02 & 0.0098 & -0.838 & 25.7 & 8060 & -5905.2 & 0.097 & 0.369  \\
SXS:BBH:2531 & 2.0 & 26.02 & 0.0061 & -0.294 & 28.2 & 9924 & -7930.6 & 0.180 & 0.944  \\
SXS:BBH:2532 & 2.0 & 28.02 & 0.0053 & -0.219 & 27.8 & 10050 & -8117.4 & 0.206 & 0.880  \\
SXS:BBH:2533 & 2.0 & 34.01 & 0.0038 & -0.102 & 30.3 & 12315 & -10478.3 & 0.238 & 0.285   \\
SXS:BBH:2534 & 2.0 & 60.01 & 0.0013 & -0.011 & 33.3 & 18990 & -17396.7 & 0.312 & 0.595  \\
SXS:BBH:2535 & 2.0 & 65.01 & 0.0010 & -0.008 & 16.3 & 7798 & -6707.7 & 0.449 & 0.263  \\
SXS:BBH:2536 & 3.0 & 22.02 & 0.0089 & -0.343 & 53.4 & 22385 & -19827.7 & 0.024 & 0.992   \\
SXS:BBH:2537 & 3.0 & 22.02 & 0.0085 & -0.344 & 38.0 & 13645 & -11146.6 & 0.087 & 0.201   \\
SXS:BBH:2538 & 3.0 & 28.01 & 0.0056 & -0.132 & 48.6 & 20622 & -18189.4 & 0.125 & 0.321  \\
SXS:BBH:2539 & 3.0 & 22.02 & 0.0083 & -0.344 & 31.6 & 10465 & -8036.2 & 0.126 & 0.036   \\
SXS:BBH:2540 & 3.0 & 17.30 & 0.0120 & -37.880 & 19.2 & 4898 & -2477.2 & 0.130 & 0.071   \\
SXS:BBH:2541 & 3.0 & 28.01 & 0.0055 & -0.132 & 39.9 & 15662 & -13320.9 & 0.163 & 0.231  \\
SXS:BBH:2542 & 3.0 & 26.02 & 0.0061 & -0.177 & 32.7 & 11606 & -9310.5 & 0.180 & 0.419  \\
SXS:BBH:2543 & 3.0 & 28.01 & 0.0053 & -0.132 & 32.2 & 11710 & -9499.1 & 0.206 & 0.293 \\
SXS:BBH:2544 & 3.0 & 55.01 & 0.0015 & -0.009 & 29.4 & 14782 & -13129.2 & 0.350 & 0.874  \\
SXS:BBH:2545 & 4.0 & 18.02 & 0.0120 & -0.476 & 37.9 & 12437 & -9510.8 & 0.021 & 0.347   \\
SXS:BBH:2546 & 4.0 & 20.02 & 0.0098 & -0.312 & 34.5 & 10987 & -8148.1 & 0.096 & 0.776   \\
SXS:BBH:2547 & 4.0 & 26.01 & 0.0061 & -0.109 & 37.5 & 13397 & -10781.8 & 0.180 & 0.054   \\
SXS:BBH:2548 & 4.0 & 28.01 & 0.0053 & -0.082 & 37.1 & 13523 & -10988.7 & 0.206 & 0.892   \\
SXS:BBH:2549 & 4.0 & 34.01 & 0.0038 & -0.038 & 40.3 & 16521 & -14126.4 & 0.239 & 0.005  \\
SXS:BBH:2550 & 4.0 & 55.01 & 0.0015 & -0.006 & 33.5 & 16935 & -15062.7 & 0.351 & 0.259   \\
SXS:BBH:2551 & 4.0 & 65.01 & 0.0010 & -0.003 & 20.9 & 10100 & -8730.3 & 0.451 & 0.568   \\
SXS:BBH:2552 & 6.0 & 18.02 & 0.0119 & -0.212 & 48.4 & 16030 & -12334.2 & 0.021 & 0.866  \\
SXS:BBH:2553 & 6.0 & 20.01 & 0.0098 & -0.139 & 43.7 & 14068 & -10489.5 & 0.097 & 0.401  \\
SXS:BBH:2554 & 6.0 & 26.01 & 0.0061 & -0.049 & 47.6 & 17088 & -13796.0 & 0.181 & 0.481  \\
SXS:BBH:2555 & 6.0 & 28.01 & 0.0053 & -0.036 & 46.8 & 17218 & -14034.1 & 0.207 & 0.224   \\
SXS:BBH:2556 & 6.0 & 34.01 & 0.0038 & -0.017 & 50.8 & 21006 & -18008.4 & 0.241 & 0.205  \\
SXS:BBH:2557 & 6.0 & 45.01 & 0.0022 & -0.006 & 40.6 & 18243 & -15790.5 & 0.334 & 0.662  \\
SXS:BBH:2558 & 6.0 & 65.00 & 0.0010 & -0.001 & 25.5 & 12597 & -10937.6 & 0.453 & 0.066   \\
SXS:BBH:2559 & 8.0 & 14.52 & 0.0164 & -0.267 & 34.6 & 9005 & -4512.5 & 0.009 & 0.718 \\
SXS:BBH:2560 & 8.0 & 20.01 & 0.0097 & -0.073 & 53.1 & 17191 & -12852.9 & 0.097 & 0.076  \\
SXS:BBH:2561 & 8.0 & 26.01 & 0.0061 & -0.026 & 57.8 & 20874 & -16885.6 & 0.183 & 0.033  \\
SXS:BBH:2562 & 8.0 & 28.01 & 0.0053 & -0.019 & 56.6 & 20959 & -17129.4 & 0.209 & 0.654  \\
SXS:BBH:2563 & 8.0 & 28.01 & 0.0052 & -0.019 & 43.9 & 14883 & -11378.5 & 0.261 & 0.160  \\
SXS:BBH:2564 & 10.0 & 14.51 & 0.0164 & -0.156 & 40.3 & 10495 & -5209.6 & 0.012 & 0.312   \\
SXS:BBH:2565 & 10.0 & 15.01 & 0.0156 & -0.136 & 43.9 & 11870 & -6587.3 & 0.015 & 0.147 \\
SXS:BBH:2566 & 10.0 & 30.00 & 0.0045 & -0.008 & 49.2 & 17153 & -13297.8 & 0.289 & 0.938  \\
SXS:BBH:2567 & 10.0 & 28.01 & 0.0050 & -0.011 & 39.8 & 12581 & -8926.6 & 0.315 & 0.047  \\
SXS:BBH:2568 & 10.0 & 45.00 & 0.0022 & -0.002 & 57.8 & 26144 & -22709.4 & 0.335 & 0.299  \\
\hline 
\hline 
\end{tabular}

\caption{Properties of the NR simulations used in this work.
  Columns 2--5 give the initial data parameters needed to
    reproduce each simulation (see main text), whereas columns 6--10
    give some physical properties: the number of orbits,
  $N_\text{orbits}$, the time to merger $T_{\text{merger}}/M$, and the
  reference time $T_{\text{ref}}/M$ corresponding to a frequency of
  the (2,2)-mode $\wref =0.042$, at which the eccentricity
  $e^{\text{ref}}_{\rm gw}$ and mean anomaly $l_{\text{ref}}/(2 \pi)$
  are extracted from the simulation. Here, $M$ is the total mass of
  the binary after initial transients have settled
  down~\cite{Boyle:2019kee}. These parameters and additional properties can be found at numerical precision in the metadata files accompanying each simulation. (*) Simulation performed with SKS initial data, differently from the rest of simulations, which used SHK initial data (see main text for details). }
\label{tab:tabNR}
\end{table*}

We have produced 52 new numerical relativity simulations of binary black holes on eccentric orbits. The simulations are summarized in Table~\ref{tab:tabNR}; for each of the mass-ratios $q=1, 1/2, 1/3, 1/4, 1/6, 1/8$ and $1/10$, simulations with several different eccentricities $e_{\rm gw}$ are computed. Within the XCTS formalism to construct initial data, the simulations in Table \ref{tab:tabNR} were produced using superposed harmonic Kerr (SHK) initial data \cite{Varma:2018sqd}, except for the simulation SXS:BBH:2527, which used superposed Kerr-Schild (SKS) initial data \cite{Lovelace:2010ne} as this is the simulation with largest initial separation and eccentricity, and initial tests with SHK initial data were not successful\footnote{After tuning some settings in the linear solvers of the \texttt{SpEC} initial data code, the SHK initial data was successfully computed, but in order to save computational resources the evolution with SHK initial data was not produced.}.

\begin{figure}[tbp!]
    \centering
\includegraphics[width=\columnwidth]{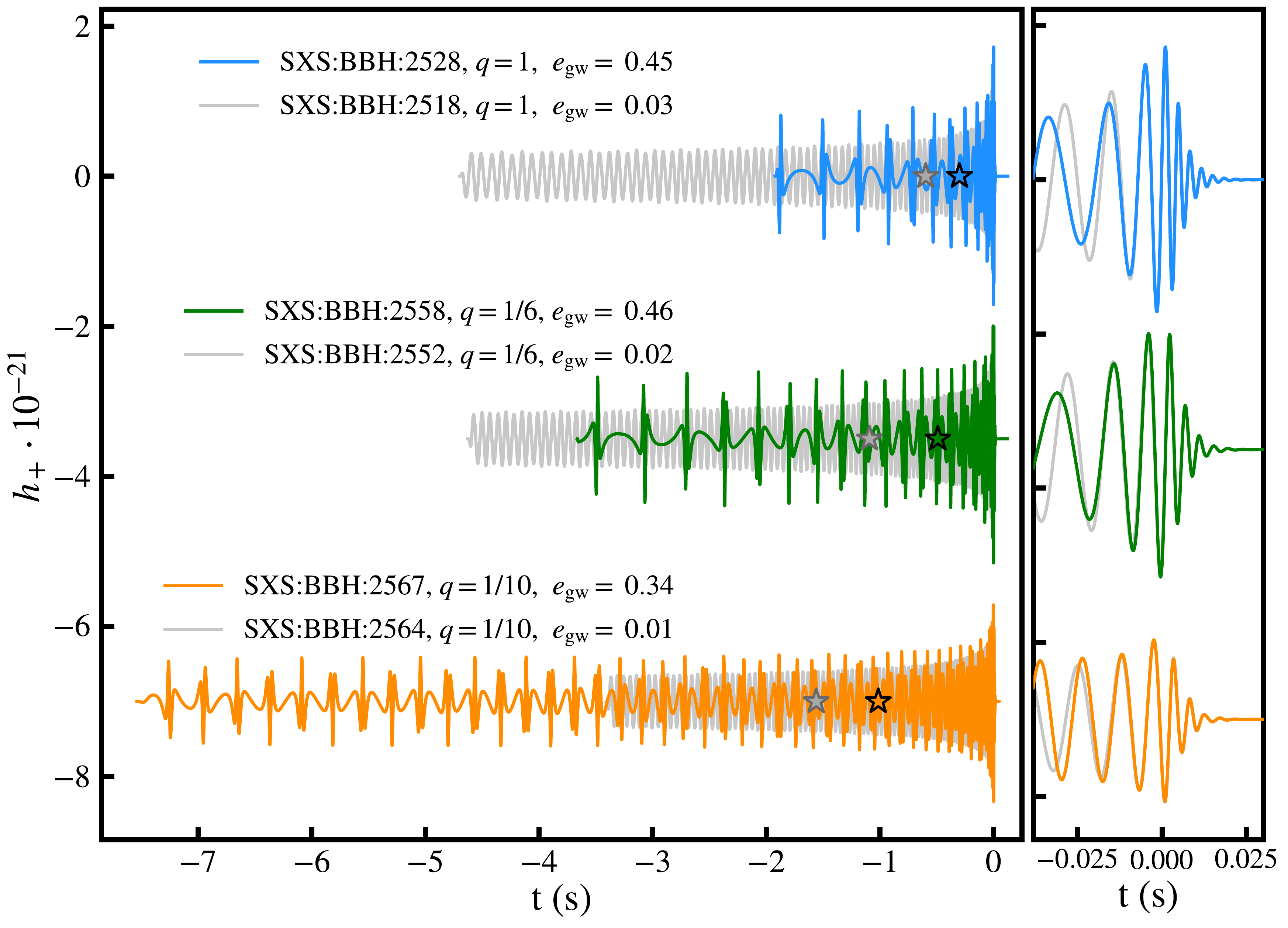}
\caption{Visualization of simulations at two eccentricities each for three different mass-ratios.  Shown is $h_+$ at inclination angle
  $\iota=\pi/3$ and coalescence phase $\phi=0$, for a binary of total
  mass of $60M_\odot$ at a distance of $430$ Mpc.  For ease of
  plotting, the waveforms are offset vertically.  On each waveform,
  the location is marked where the orbit-averaged GW frequency
  $\wgwavg$ equals our reference value $M \wref =0.042$; for
  $M=60M_\odot$ this corresponds to a GW frequency of $22.6$ Hz, near
  the start of the frequency  band of current GW detectors. The
  right panel enlarges the merger part of the signals.
    \label{fig:waveformPlot}}
\end{figure}

For each simulation, Table~\ref{tab:tabNR} reports on the parameters values necessary to reproduce the initial data with the techniques described in \cite{Ossokine:2015yla}:
the inverse mass ratio $1/q=m_2/m_1\geq 1 $, the orbital separation $D_0/M_0$, where $M_0$ is the initial ADM mass, the initial orbital frequency $M_0 \initNRw$, and the initial radial velocity parameter $a_0$ \cite{Pfeiffer:2007yz,Buonanno:2010yk}. The procedure to determine the initial parameters of the simulations is described in Appendix \ref{sec:AppendixIC}. The simulations are started at or very close to apastron due to limitations of the radial map used by the \textit{dual-frame method} \cite{Scheel:2006gg} employed to solve the Einstein equations in \texttt{SpEC} \cite{Boyle:2007ft}. Specifically, the radial mapping of Eq.~(9) in \cite{Boyle:2007ft} connecting the comoving and inertial frames does not allow the orbital separation to increase more than 1.5 times the initial separation.
We note that this limitation has been recently overcome in \texttt{SpEC} by defining a new radial map, however, it is not applied for simulations in this publication, and we leave it to future work to report on this new feature.

To convey a sense of the physical properties of the BBHs studied, Table~\ref{tab:tabNR} also lists  the number of orbits to merger,  $N_\text{orbits}$ and the time to merger $T_{\text{merger}}/M$, where $M$ is the total mass.
We also specify the time (before merger)  $T_{\text{ref}}/M$ where the orbit averaged frequency of the (2,2)-mode reaches the value $\wref = 0.042$, as well as eccentricity $e^{\text{ref}}_{\rm gw}$ and mean anomaly $l_{\text{ref}}/(2 \pi)$ at this reference time.  These quantities are defined with the procedures outlined below in Sec.~\ref{sec:Eccdef}.
 The reference frequency is chosen to be consistent with the length\footnote{We consider the length as measured after the relaxation time, i.e., the time after which is considered that the burst of junk radiation has dissipated.} of the shortest simulation, which corresponds to SXS:BBH:2520 with $4963M$ of evolution and 18 orbits. Apart from this particular case, most of the simulations have typically a time to merger $> 10^4M$. This makes our dataset of eccentric NR waveforms the one with the longest evolutions of eccentric binary black holes to date.

 We extract the gravitational radiation from each simulation using
 the same techniques as in~\cite{Boyle:2019kee}, and decompose
 \begin{equation}\label{eq:hlm}
   h=h_+-i h_\times =  \sum_{lm} h_{lm}\; {_{-2}Y_{lm}}.
\end{equation}
   Each mode $h_{lm}$ is further split into real amplitude and phase as
\begin{equation}
  h_{lm}(t)=A_{lm}(t)e^{-i\phi_{lm}(t)},
\end{equation}
with an associated GW mode frequency of
\begin{equation}
  \omega_{lm}=\dot\phi_{lm}.
\end{equation}

A sample of the computed numerical waveforms are shown in Fig.~\ref{fig:waveformPlot}. One can observe that the highly eccentric configurations develop a very complex structure in the waveform due to the eccentricity of the orbits followed by the BHs. Figure~\ref{fig:waveformPlot} also shows a zoom-in of the merger part on the waveforms, to highlight the similarity of the merger and ringdown parts of the waveform with different eccentricities\footnote{The waveforms in Fig.~\ref{fig:waveformPlot} were time-shifted for the merger to occur at $t=0$.  Furthermore, the low-eccentricity simulations (shown in grey) were phase-shifted to have the same phase at merger as the plotted high-eccentricity simulation.}. Merger and ringdown of the high and low eccentricity inspirals agree well with each other, indicating that the circularization hypothesis is accurately fulfilled for our dataset, consistently with the findings in \cite{Hinder:2008kv, Huerta:2019oxn,Ramos-Buades:2019uvh}. We note that recently some unexpected dependence of the kick velocity on eccentricity was found in  \cite{Radia:2021hjs}. A similar analysis of the kick velocity can be performed on our dataset, and we leave such study of the final velocity as well as other remnant properties for future work.

\subsection{Eccentricity, azimuthal frequency \& mean anomaly}
\label{subsec:NR-eccentricty}

\begin{figure}[tbp!]
\includegraphics[width=\columnwidth,trim=0 15 0 0]{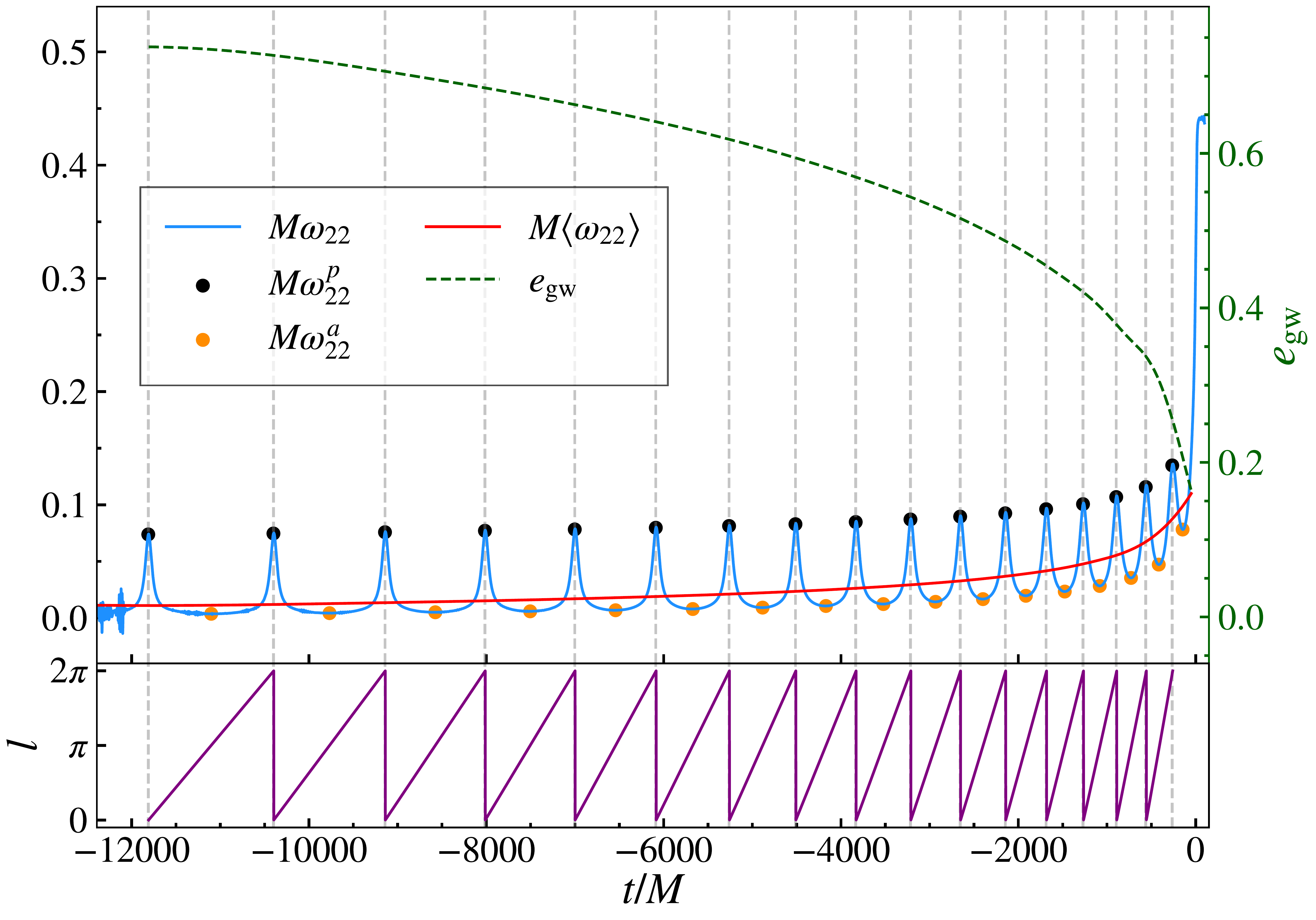}
\caption{\textit{Top panel:} Time evolution of the frequency of the (2,2)-mode (solid blue line) for the simulation SXS:BBH:2558. The values of the (2,2)-mode frequency at periastron and apastron are indicated with orange and black dots, respectively. These are used to compute the orbit-averaged frequency of the (2,2)-mode (solid red curve), and the eccentricity  $e_{\rm gw}$  (dashed green curve) through Eq.~\eqref{eq:eqEccDef}.  \textit{Bottom panel:} Time evolution of the mean anomaly (solid purple line) computed using Eq.~\eqref{eq:eq8} for the same simulation as in the top panel. The vertical dashed gray lines in both panels correspond to the times of the periastron passages.}
\label{fig:fig1a}
\end{figure}

We start with the eccentricity definition proposed by Mora \&
Will~\cite{Mora:2002gf},
\begin{equation}\label{eq:e-MoraWill}
e_{\worb} =  \frac{\sqrt{\worb^p}-\sqrt{\worb^a}}{\sqrt{\worb^p}+\sqrt{\worb^a}},
\end{equation}
where $\worb^p$ and $\worb^a$ are the values of the orbital frequency
at consecutive periastron and apastron passages, i.e. maxima and
minima of $\worb(t)$.  Equation~(\ref{eq:e-MoraWill}) is easy to
compute from orbital trajectories and reduces precisely to the normal
eccentricity in the Newtonian limit~\cite{Mora:2002gf}.  $e_{\worb}$
was for instance used in \cite{Lewis:2016lgx} to analyse generic
precessing \& eccentric BBH inspirals.  To avoid the
coordinate-dependence of $\worb$, recent papers
(e.g.~\cite{Ramos-Buades:2019uvh}) have applied
Eq.~(\ref{eq:e-MoraWill}) to frequencies directly defined from the
gravitational radiation:
\begin{equation}
    e_{\omega_{22}} = \frac{\sqrt{\wgw^p}-\sqrt{\wgw^a}}{\sqrt{\wgw^p}
      + \sqrt{\wgw^a}},
    \label{eq:eq7}
\end{equation}
where $\omega_a, \omega_p$ refer to the (2,2)-mode frequency
$\omega_{22}$ at apastron and periastron, respectively.  This
procedure is illustrated in the top panel of Fig.~\ref{fig:fig1a}: The
time-dependent $\wgw(t)$ has maxima ${\wgw}^p_{,i}$ and minima
${\wgw}^a_{,i}$ indicated with the black and orange dots, where the
integer $i$ labels the extrema.  The maxima and minima correspond to
periastron and apastron passages, respectively, and occur at times
$t^p_i$ and $t^a_i$.

We show below in Sec.~\ref{sec:Eccdef} that $e_{\wgw}$
  disagrees with $e_{\worb}$; most notably, $e_{\wgw}$ does not have
  the correct Newtonian limit. Therefore, we introduce a
new eccentricity definition $e_{\rm gw}$ measured from the frequency
of the (2,2)-mode, which has the correct Newtonian limit:
\begin{subequations}
\label{eq:eqEccDef}
\begin{align}
  e_{\rm gw} &= \cos (\psi/3)- \sqrt{3} \sin (\psi/3)
  \intertext{with}
\psi  &= \arctan\left( \frac{1-e_{\omega_{22}}^2}{2 e_{\omega_{22}}} \right).
\end{align}
\end{subequations}
 This new gravitational-wave frequency $e_{\rm gw}$ is also
  plotted in the top panel of Fig.~\ref{fig:fig1a}.  The
  dashed curve for $e_{\rm gw}$ is obtained by constructing
  interpolating functions through all maxima $\{{\wgw}^p_i\}$ and
through all minima $\{{\wgw}^a_i\}$, and then evaluating
  Eqs.~(\ref{eq:eq7}) and (\ref{eq:eqEccDef}) for these interpolating
  functions.

The average azimuthal frequency from the (2,2)-mode
  for the interval between  the $i$-th and $i+1$-th  periastron passages is defined as
\begin{align}
\label{eq:eq102}
  \wgwavgi   & = \frac{1}{t^p_{i+1}-  t^p_{i} } \int^{t^p_{i+1}}_{t^p_{i}} \wgw(t)\, dt\\
  & = \frac{\phi_{22}(t^p_{i+1})-\phi_{22}(t^p_i)}{t^p_{i+1}-  t^p_{i} }.
\end{align}
We associate this frequency with the temporal midpoint
\begin{equation}
\hat{t}^p_i = \frac{1}{2}\left( t^p_{i+1}+  t^p_{i} \right),
\label{eq:eq101}
\end{equation}
and interpolate the discrete $\left\{(\hat{t}^p_i, \wgwavgi)\right\}$
data to obtain a continuous $\wgwavg(t)$ curve. This curve is also
included in Fig.~\ref{fig:fig1a}.

The mean anomaly of the eccentric binary is defined as~\cite{Schmidt:2017btt}
\begin{equation}
    l = 2\pi \frac{t-t^p_i}{t^p_{i+1}-t^p_{i}},
    \label{eq:eq8}
\end{equation}
where $t^p_i$ and $t^p_{i+1}$ are the times of the periastron-passages
immediately before and after the time $t$ of interest, and is plotted in the lower panel of Fig.~\ref{fig:fig1a}.

\begin{figure}
    \centering
\includegraphics[width=0.98\columnwidth,trim=0 8 0 3]{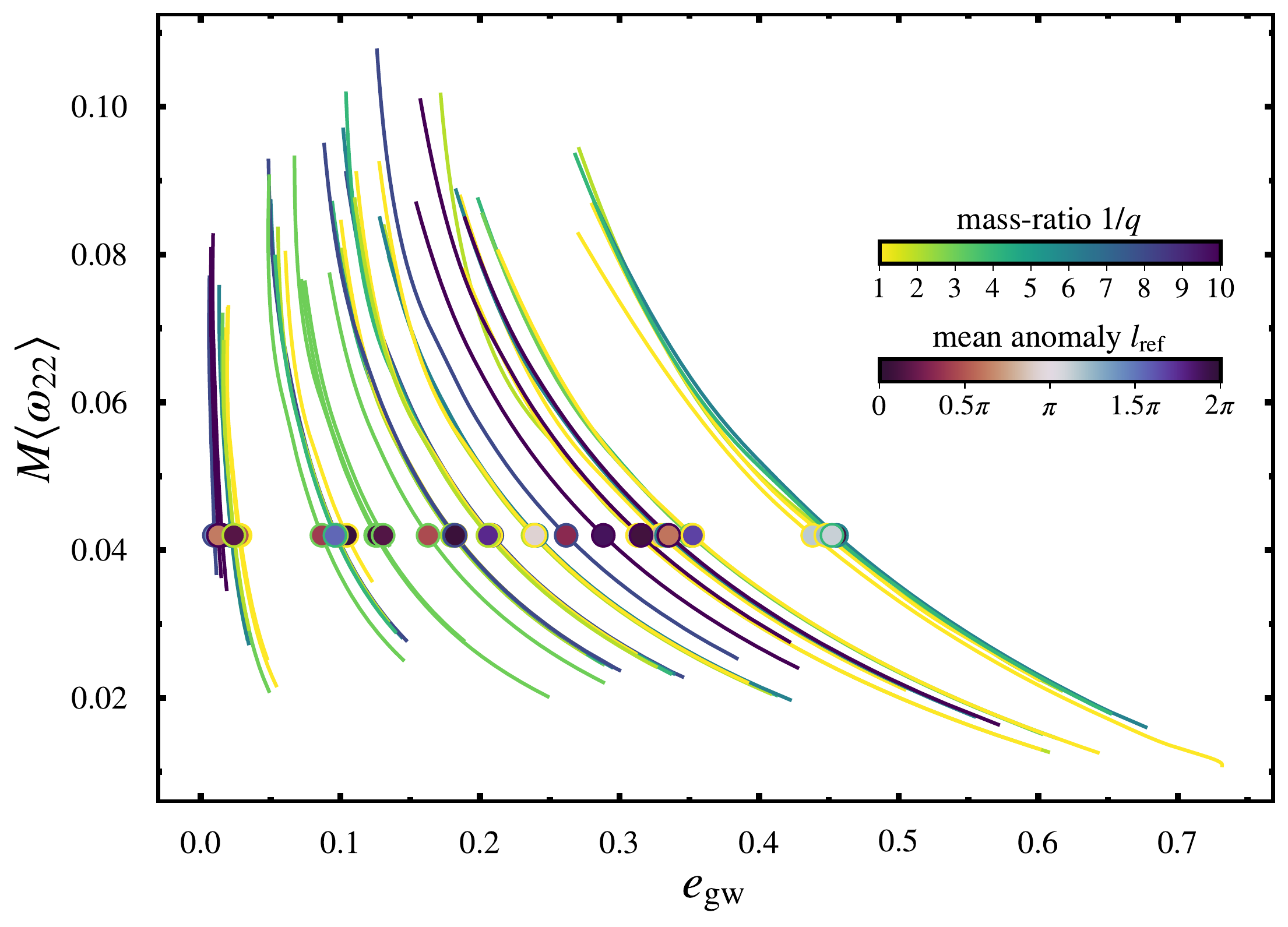}
    \caption{Parameter space coverage of the NR simulations produced in this work. Each curve corresponds to one NR simulation in the orbit-averaged (2,2)-mode frequency, $\wgwavg$, and eccentricity, $e_{\rm gw}$, plane. The simulations start at high eccentricity and low frequencies (bottom right side), and along the evolution the eccentricity decays with increasing orbital frequency (left top part of the panel). The curves are colored according to the inverse mass ratio $1/q$ of the simulation, and we indicate also the values of the mean anomaly at the reference frequency, $l_\text{ref}$ , of $M \wref = 0.042$, at which the comparison to SMR results is performed in Sec.~\ref{sec:SMRNRcomparison}.}
    \label{fig:NRdataset}
\end{figure}

The NR quantities introduced so far are used in Fig.~\ref{fig:NRdataset} to illustrate the entire NR dataset produced in this work.  Figure~\ref{fig:NRdataset} shows the tracks of each simulation in the parameter space spanned by the orbit-averaged (2,2)-mode frequency, $\wgwavg$ and the eccentricity, $e_{\rm gw}$. Each simulation is color-coded by its mass ratio.  We also indicate the value of the mean anomaly at the reference frequency used to perform the analysis. One can observe that the mean anomaly at the reference frequency is randomly distributed. We assess the accuracy of the simulations by computing the unfaithfulness between waveforms at different resolution in Appendix \ref{sec:AppendixNRconvergence}, and we obtain that our dataset of simulations has a median maximum mismatch between different resolutions of $< 10^{-3}$, indicating a convergent behavior of the waveforms with increasing resolution.

\subsection{Quantities for comparisons with small mass-ratio theory}
\label{subsec:NR-Quantities}
In our comparisons with small mass-ratio perturbation theory,
  we will also utilize several more quantities extracted from the NR
  simulations.  We define an orbit-averaged radial frequency based on
the periastron passages as
\begin{equation} \label{eq:eq103}
\wgwradi = \frac{2 \pi }{t^p_{i+1}-  t^p_{i} },
\end{equation}
which is interpolated to a continuous $\wgwrad(t)$ curve.  From this,
we compute periastron advance $K$ as the ratio between the azimuthal
and radial frequencies \cite{LeTiec:2011bk},
\begin{equation}
K   = \frac{\wgwavg/2 }{ \wgwrad}.
\label{eq:eq105}
\end{equation}

The instantaneous energy and angular momentum fluxes are computed from the GW modes, $h_{lm}$, using the expressions~\cite{Ruiz:2007yx}
\begin{align}
\label{eq:eq107}
\dot{E}^{\rm gw} & = \frac{1}{16 \pi}\sum^{\infty}_{l=2} \sum_{m=-l}^{+l} |\dot{h}_{lm}(t)|^2,\\
\dot{J}^{\rm gw}_z & = \frac{1}{16 \pi}\sum^{\infty}_{l=2} \sum_{m=-l}^{+l} (-m) \Im \left[ \dot{h}^*_{lm}(t) h_{lm}(t) \right] ,
\label{eq:eq1071}
\end{align}
where $\dot{h}=d h/dt$, $\Im$ the indicates the imaginary part and $\dot{h}^*_{lm}$ denotes the complex conjugate of $\dot{h}_{lm}$. In the case of non-spinning binaries only the z-component of the angular momentum flux is non-zero.
Analogous to Eq.~\eqref{eq:eq102} we define the orbit average of either of these fluxes as,
\begin{equation}
 \langle X \rangle_i = \frac{1}{t^p_{i+1}-  t^p_{i} } \int^{t^p_{i+1}}_{t^p_{i}} X(t)\, dt.
\label{eq:eq108}
\end{equation}
We associate these discrete averages over each radial oscillation
period with the mid-time $\hat t^p_i$, and interpolate to obtain
continuous functions $\langle\dot{E}^{\rm gw}\rangle(t)$ and
$\langle\dot{J}_z^{\rm gw}\rangle(t)$. A first estimate of the peaks is computed using an envelope subtraction method as in \cite{Lewis:2016lgx}. Each estimate of the peak is used to set a window of $\sim 30 M $ on which a polynomial fit is performed. Finally, this polynomial fit is used to compute the value of the peak.

 \section{Discussion about eccentricity definitions} \label{sec:Eccdef}
There is a large variety of measures of eccentricity in use in general
relativity \cite{Loutrel:2018ydu}. Many of these measures derive from the trajectories of the binaries, and are therefore coordinate dependent. This makes them generally unsuitable for
comparisons between different modelling approaches, which may be computed in different gauges or where there may be no well-defined notion of trajectory at all. However, one gauge invariant observable common to all approaches to modelling gravitational waves from compact binaries is the waveform itself. In this sense, it may seem more reasonable to define eccentricity in terms of gravitational wave quantities rather than quantities dependent on the trajectories of the black holes.

\begin{figure}[tbp!]
\hspace*{0.03cm}	\includegraphics[width=0.97\columnwidth,trim=0 -25 0 0]{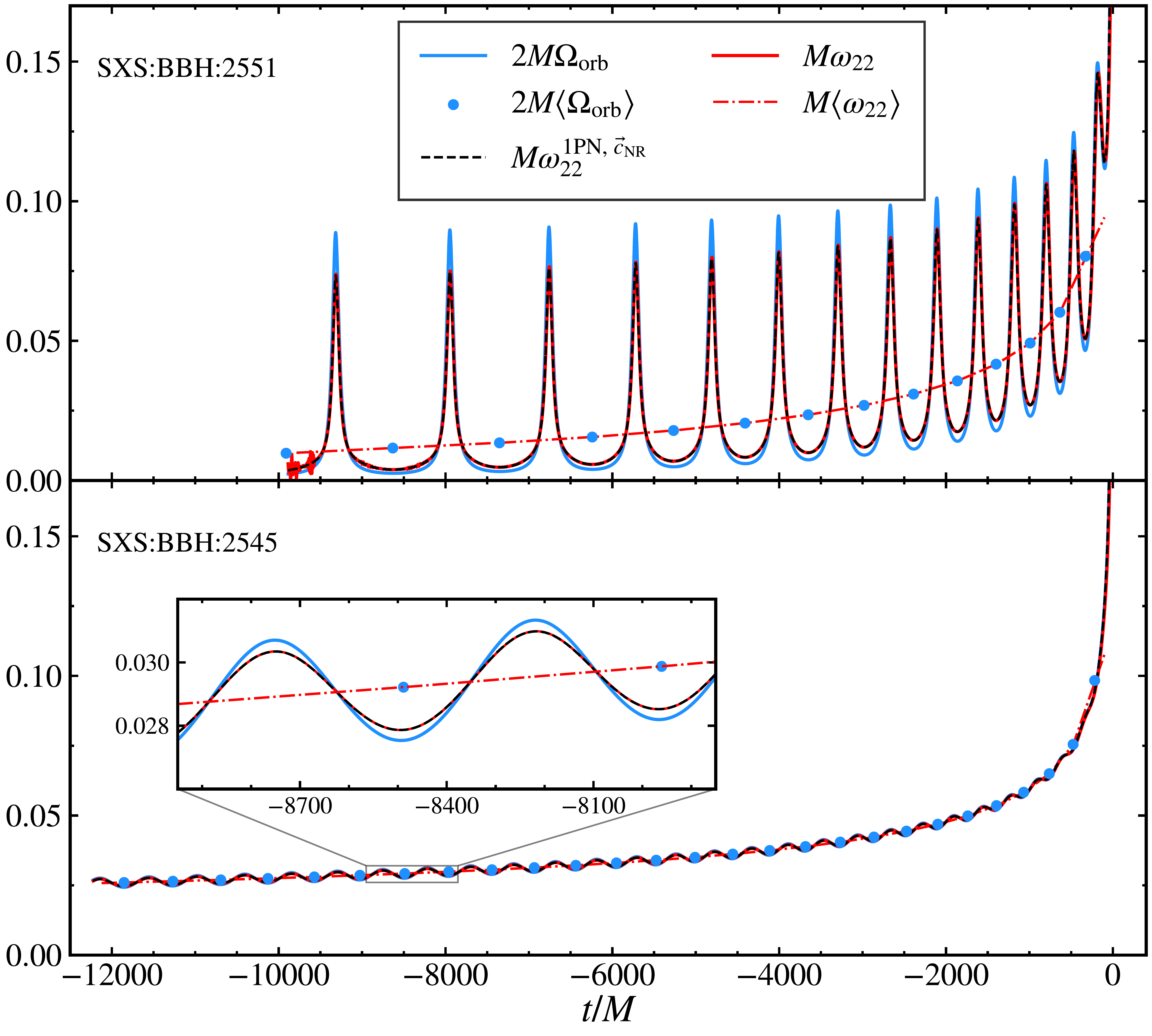}
		\includegraphics[width=0.986\columnwidth,trim=0 25 0 0]{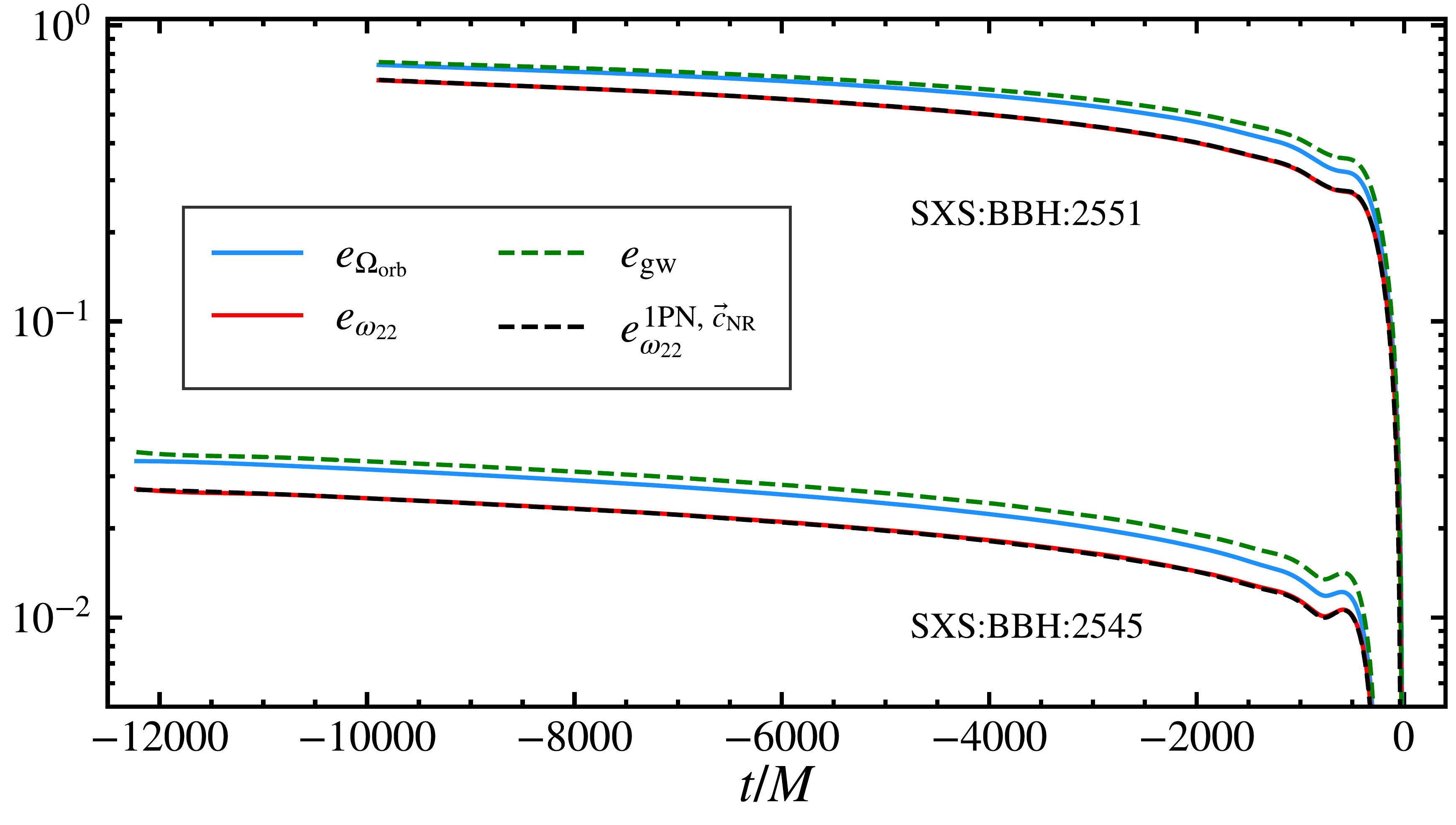}
	\caption{\textit{Top panel:} Time evolution of the (2,2)-mode frequency extracted for two mass-ratio $1/q=6$ NR simulations (SXS:BBH:2545 and SXS:BBH:2551 described in Table \ref{tab:tabNR}) with two different initial eccentricities. For each simulation twice the orbital frequency $2 \worb$ (blue solid lines),  the frequency of the (2,2)-mode (red solid lines), $\wgw$,  and the 1PN expression for the frequency of the (2,2)-mode from Eq.~\eqref{eq:eq110} evaluated using the NR coordinates, $\omega_{22}^{\text{1PN}, \, \vec{c}_{\text{NR}}}$,  (black dashed lines) are shown. Additionally, the orbit-averaged values of the frequency of the (2,2)-mode (red dashdotted lines), $\wgwavg$,  and twice the orbit-averaged orbital frequency (blue dots), $2 \worbavg$, are displayed for each configuration. \textit{Bottom panel:} Eccentricity evolution computed from the orbital and (2,2)-mode frequencies using Eqs.~\eqref{eq:eq7} and \eqref{eq:eqEccDef}, and the 1PN expression for the eccentricity of the (2,2)-mode computed from \eqref{eq:eq111} using NR coordinates,  $e^{\text{1PN}, \, \vec{c}_{\text{NR}}}_{\omega_{22}}$ (black dashed lines).
}
	\label{fig:OmegaPNNRcomparison}
\end{figure}

A gravitational wave mode (see Sec.~\ref{sec:NRsimulations}), has an instantaneous frequency $\omega_{lm}=\dot\phi_{lm}$, which can be related in the inspiral regime to the instantaneous
orbital $\worb  = \dot{\phi}_{\text{orb}}$ by the approximation \cite{Blanchet:2013haa}
 \begin{equation}
 \omega_{lm} \approx m \worb.
\label{eq:eq109}
\end{equation}
However, as eccentricity increases the approximation of Eq.~\eqref{eq:eq109} is no longer valid as can be observed in the top panel of Fig.~\ref{fig:OmegaPNNRcomparison}, where the left and right-hand sides of Eq.~\eqref{eq:eq109} in the case of the $(l,m)=(2,2)$ multipole are displayed. In the top plot of Fig.~\ref{fig:OmegaPNNRcomparison}, the upper and bottom panels correspond to a $q=1/6$ configuration with two different initial eccentricities $e^0_{\wgw }=0.03, 0.63$, respectively. The relation between the orbital and the (2,2)-mode frequency is no longer the simple factor $2$, as in the quasi-circular case.
In order to derive the relation between both frequencies in the more generic eccentric case, we use PN theory. Specifically, we compute $\wgw$ at 1PN order using the instantaneous gravitational modes  from \cite{Mishra:2015bqa}. We obtain a 1PN-accurate expression for $\wgw$ in harmonic coordinates of the form,
\begin{equation}
 \wgw^{\text{1PN}}  =  \mathcal{F}(\nu, r, \dot{r}, \ddot{r}, \dot{\phi}, \ddot{\phi}),
\label{eq:eq110}
\end{equation}
where  $\ddot{r}$ denotes two time derivatives on $r$. The explicit expression for $\mathcal{F}$ is given in Eq.~\eqref{eq:eqD05} in Appendix \ref{sec:AppendixeOm22Orb} together with details of the derivation. Because of $\dot{\phi} = \worb$, Eq.~\eqref{eq:eq110} is a relation between $\wgw$ and $\worb$.

The top panel of Fig.~\ref{fig:OmegaPNNRcomparison} shows that the use of Eq.~\eqref{eq:eq110} with NR coordinates ($\omega_{22}^{\text{1PN}, \, \vec{c}_{\text{NR}}}$ in the figure) agrees notably better with the (2,2)-mode NR frequency than $2\worb$.  The deviations in Eq.~\eqref{eq:eq109} increase with eccentricity.  The relative error can be larger than $10\%$, whereas Eq.~(\ref{eq:eq110}) leads to differences smaller than $1\%$.

It is important to note that the scaling relation in Eq.~\eqref{eq:eq109} between orbital and gravitational wave frequencies is still satisfied in an orbit-averaged sense. This is shown in the upper panel of Fig.~\ref{fig:OmegaPNNRcomparison} for the orbit-averaged frequencies $\wgwavg$ (solid red lines) and $\worbavg$ (blue dots).

Let us now turn to eccentricity defined from the extrema of a frequency.  Equation~(\ref{eq:eq7}) can be
  evaluated from $\omega_{22}$ (as written), or from the orbital frequency $\worb$.  Because $\omega_{22}(t)$ and $\worb(t)$ have modulations of different amplitude (as seen in the top panels of Fig.~\ref{fig:OmegaPNNRcomparison}), the corresponding eccentricities $e_{\omega_{22}}$ and $e_{\worb}$ are also different, as visible in the lower panels of Fig.~\ref{fig:OmegaPNNRcomparison}.\footnote{The eccentricity curves in the lower panels show a spurious bump close to merger arising from the interpolation of the maxima and minima close to the plunge.  Our analysis focuses on the inspiral regime and is not affected by this feature.  We leave to future work the improvement of the eccentricity measurement in the transition from inspiral to plunge.}
Given the remarkable agreement of the PN approximation to $\wgw$ with respect to the NR values, one can insert Eq.~\eqref{eq:eq110} into the right hand side of Eq.~\eqref{eq:eq7}, expand the corresponding expressions up to 1PN, and obtain an approximation for $e_{\wgw}$ in terms of the coordinates as,
\begin{equation}
 e^{\text{1PN}}_{\wgw} =  \mathcal{G}(\nu, r_{a,p}, \worb^{a,p}, \ddot{r}_{a,p}).
\label{eq:eq111}
\end{equation}
where the expression for $\mathcal{G}$ is given by \eqref{eq:eqD08} in Appendix~\ref{sec:AppendixeOm22Orb}, and the subscripts/superscripts $a,p$ refer to the apastron and periastron, respectively. The bottom panel of Fig.~\ref{fig:OmegaPNNRcomparison} shows that Eq.~\eqref{eq:eq111} successfully reproduces $e_{\wgw}$.  Given the overall agreement, we do not pursue to explore higher PN orders, or possible resummations of this PN expression to improve its behavior in the strong field regime, and we leave possible extensions of these expressions, like the inclusion of spin effects, for future work.

The relations in Eqs.~\eqref{eq:eq110} and \eqref{eq:eq111} allow one to obtain an estimate of the eccentricity measured from the (2,2)-mode frequency from the coordinates of the system. This can be useful, for instance, to set an eccentricity reduction or eccentricity control procedure based on the eccentricity measured from the waveforms instead of the trajectories without having to evolve the system such that the gravitational waves reach the extraction radii, and thus, saving computational time.

Equation~(\ref{eq:eq111}), as used in the lower panel of
  Fig.~\ref{fig:OmegaPNNRcomparison}, still utilizes the NR
  trajectory.  If one substitutes in a PN trajectory in
  the quasi-Keplerian parameterization \cite{Memmesheimer:2004cv}, one
  obtains relations between $e_{\worb}$ or $e_{\wgw}$ and the PN
  eccentricity parameters, most notably the widely used $e_t$ \cite{Yunes:2009yz,Huerta:2014eca,Mishra:2015bqa,Loutrel:2017fgu,Klein:2018ybm,Moore:2018kvz,Moore:2019xkm,Tanay:2019knc,Tiwari:2020hsu}.

A detailed derivation of the relation $e_{\worb}-e_t$ up to 3PN order for non-spinning binaries can be found in  Appendix \ref{sec:AppendixeOmOrbet}. We focus here on the relation $e_{\wgw}-e_t$, which is derived up to 1PN order in Appendix \ref{sec:AppendixeOm22et}, providing
\begin{equation}
\begin{split}
e^{\text{1PN}}_{\wgw} = & \frac{\sqrt{2-e_t} \left(1+e_t\right) - \left(1-e_t\right)\sqrt{2+e_t} }{\sqrt{2-e_t} \left(1+e_t\right) + \left(1-e_t\right)\sqrt{2+e_t} } \\
& -\gamma x e_t \frac{ (54 \eta +101) e_t^2+192 \eta -1380 }{84 \left(e_t^4-5 e_t^2+2 \sqrt{4-e_t^2}+4\right)},
\end{split}
\label{eq:eq112}
\end{equation}
where $x=\worb^{2/3}$, and $\gamma = 1/c^2$ is a bookkeeping parameter identifying the 1-PN corrections. At Newtonian order, Eq.~\eqref{eq:eq112} reduces to
\begin{equation}
  e^{\text{0PN}}_{\wgw} = \frac{\sqrt{2-e_t} \left(1+e_t\right) - \left(1-e_t\right)\sqrt{2+e_t} }{\sqrt{2-e_t} \left(1+e_t\right) + \left(1-e_t\right)\sqrt{2+e_t} }.
\label{eq:eq113}
\end{equation}
While Eq.~(\ref{eq:eq113}) achieves the right limits for circular and parabolic orbits ---$e^{\text{0PN}}_{\wgw}(e_t=0) = 0$ and   $e^{\text{0PN}}_{\wgw}(e_t=1) = 1$ --- it disagrees otherwise.  This can be easily seen by expanding  Eq.~\eqref{eq:eq113} for small eccentricities,
\begin{equation}
e^{\text{0PN}}_{\wgw} = \frac{3}{4} e_t + \frac{11}{64}e_t^3 + \mathcal{O}(e_t^5),
\label{eq:eq114}
\end{equation}
which explicitly demonstrates that for small eccentricities  in the Newtonian limit,   $e_{\wgw}$ does not reduce to $e_t$, but rather to $3/4e_t$. in the Newtonian limit, but its relation is  $e_{\wgw} \sim 3 e_t/4 $. An expansion of Eq.~\eqref{eq:eq113} in the large eccentricity limit $1-e_t\ll 1$ yields
\begin{equation}
  1-e^{\text{0PN}}_{\wgw} = \sqrt{3}(1-e_t) + \mathcal{O}\left((1-e_t)^2\right),
  \label{eq:eq115}
\end{equation}
which also exhibits a wrong slope ($\sqrt{3}$) for $e_t$ near 1.
Equations~\eqref{eq:eq114} and \eqref{eq:eq115} show that the definition of eccentricity based on the $(2,2)$-mode frequency will be different from the Newtonian definition of eccentricity in the two limits of the bound case.  Additional PN orders will introduce higher frequency corrections to the Newtonian behaviour, whose impact in the leading Newtonian correction factors between the eccentricity will depend on the region of the parameter space considered.

The relation $e_t(e_{\wgw})$ at Newtonian order can be obtained by inverting Eq.~\eqref{eq:eq113},
\begin{equation}
\begin{split}
\psi & = \arctan\left( \frac{1-e_{\wgw}^2}{2 e_{\wgw}} \right),\\
e^{\text{0PN}}_t &= \cos (\psi/3)- \sqrt{3} \sin (\psi/3).
\end{split}
\label{eq:eq116}
\end{equation}
Applying Eq.~\eqref{eq:eq116} to $e_{\wgw}$ will yield an eccentricity-definition that reduces to the Newtonian definition of eccentricity.

As a consequence of the previous analysis we propose a new definition of eccentricity measured from the frequency of the (2,2)-mode, which corrects the naive result $e_{\wgw}$ obtained from the extrema of $\wgw$ by Eq.~\eqref{eq:eq116},
\begin{equation}
e_{\rm gw}  \equiv e^{\text{0PN}}_t (e_{\wgw}).
\label{eq:eq117}
\end{equation}
By construction, $e_{\rm gw}$ reduces to the Newtonian definition of eccentricity in the Newtonian limit. In the bottom panel of Fig.~\ref{fig:OmegaPNNRcomparison}, $e_{\rm gw}$ is shown to be closer to $e_{\worb}$ than $e_{\wgw}$. Both $e_{\rm gw}$ and $e_{\worb}$ have the correct Newtonian limit, and the differences may be explained due to coordinate effects affecting $e_{\worb}$, and higher PN terms, as $e_{\rm gw}$ is obtained from Eq.~\eqref{eq:eq113}.

 This new definition of eccentricity is adopted throughout the rest of the paper, and its applications to data analysis are further investigated in upcoming work  \cite{gw_eccentricity}.

\section{SMR theory and data}\label{sec:SMRevolutions}

In the small mass-ratio (SMR) limit, the dynamics of a black hole binary can be described through the gravitational self-force formalism. For the inspiral part of the waveform, this formalism leads to a systematic expansion of the waveform in integer powers of the symmetric mass-ratio $\nu$. This expansion is known as the post-adiabatic (PA) expansion. In this section, we introduce the necessary parts of this formalism to produce SMR eccentric inspirals for comparison to our NR data. For a more in depth review of the formalism see e.g.~\cite{Barack:2018yvs, Pound:2021qin}.

\subsection{Equations of motion}
In the SMR limit an eccentric inspiral of non-spinning black holes can be described as a series of evolving (perturbed) eccentric orbits in a Schwarzschild background.  Eccentric orbits in Schwarzschild are often identified by their semi-latus rectum $p$ and geodesic eccentricity $\gecc$, which in turn are defined through the periastron and apastron positions, $r_{  p}$ and $r_{a}$,
\begin{align}
  p       &= \frac{2 r_{ a} r_{  p}}{r_{  a}+r_{  p}},
  \intertext{and}
    \gecc   &= \frac{r_{  a}-r_{ p}}{r_{ a}+r_{ p}}.
\end{align}
The position along the eccentric orbit is tracked by a phase $q_r$ conjugate to the radial action, defined such that $q_r = 0\mod 2\pi$ corresponds to the orbit being at periastron.
The equations of motion for the evolution of the inspiral can be described as an expansion in the symmetric mass ratio $\smr$ (keeping the total mass $M$ fixed),
\begin{subequations}\label{eq:SMREoM}
\begin{align}
\frac{dp}{dt} &= 0 +\smr F_p(p,\gecc,q_r)  +\mathcal{O}(\smr^2),\\
\frac{d\gecc}{dt} &= 0+\smr F_\gecc(p,\gecc,q_r) +\mathcal{O}(\smr^2),\\
\frac{dq_r}{dt} &= \wradgeo(p,\gecc) + \smr f_r(p,\gecc,q_r) +\mathcal{O}(\smr^2),\\
\frac{d\phi}{dt} &= \wphigeo(p,\gecc) + \smr f_r(p,\gecc,q_r) +\mathcal{O}(\smr^2),
\end{align}
\end{subequations}
where $t$ is retarded time at future null infinity, $\wradgeo$ and $\wphigeo$ are the geodesic radial and azimuthal frequencies (w.r.t $t$), and the $F$'s and $f$ are the first order (gravitational self-force) corrections to the equations of motion.

 By applying a  near-identity (averaging) transformation Eqs.~\eqref{eq:SMREoM} can be put in an orbit averaged form (without loss of generality) \cite{VanDeMeent:2018cgn}. The leading terms give rise to the adiabatic (or 0-post-adiabatic, 0PA) approximation to the inspiral equations of motion,

 \begin{subequations}\label{eq:SMREoM0PA}
	\begin{align}
	\frac{dp}{dt} &= \smr \langle F_p\rangle(p,\gecc), \\
	\frac{d\gecc}{dt} &= \smr \langle F_\gecc \rangle(p,\gecc),\\
	\frac{dq_r}{dt} &= \wradgeo(p,\gecc), \\
	\frac{d\phi}{dt} &= \wphigeo(p,\gecc).
	\end{align}
\end{subequations}

The next order in $\smr$ in the approximation -- the 1-post-adiabatic or 1PA order -- requires knowledge of the average parts of the second order $F_p$ and $ F_\gecc$, i.e. the second order gravitational self-force. Despite major progress in calculating the second order self-force and corresponding 1PA corrections for non-spinning quasi-circular inspirals \cite{Pound:2019lzj,Warburton:2021kwk,Wardell:2021fyy}, there are no second-order self-force results yet for eccentric inspirals. Without the input of the second order self-force, any 1PA corrections based purely on the conservative part of the first-order self-force are not gauge invariant \cite{Lynch:2021ogr}, and not suitable for comparison with NR. Consequently, for this work we will focus on comparisons with the adiabatic (0PA) SMR results.

\subsection{Gravitational wave strain}

The gravitational wave strain produced by a test particle orbiting a Schwarzschild black hole can be found by solving the Teukolsky equation for $\psi_4$.  We write $\psi_4$ at future null infinity as

\begin{equation}\label{eq:teuk}
  \lim_{r\to\infty} r \psi_4 = 
  \smr\sum_{lmn} Z_{lmn}\; {_{-2}Y_{lm}}(\theta_{obs},\phi_{obs})e^{-i\wmnmode t},
\end{equation}
where $\wmnmode = m\wphigeo+n\wradgeo$, and $Z_{lmn}$ are the mode amplitudes. The strain-modes at infinity,  Eq.~(\ref{eq:hlm}), are then given as
\begin{align}
    h_{lm} &= -2\smr \sum_{n} \frac{Z_{lmn}}{\wmnmode^2} e^{-i\wmnmode t}\\
    &= \smr \sum_{n} \mathcal{A}_{lmn}(p,\gecc) e^{-i(m \phi+n q_r)} \label{eq:hlmsmr},
\end{align}
where in the last step we have written the strain explicitly in terms of the variable evolved by Eq.~\eqref{eq:SMREoM0PA}. To obtain the strain produced by an adiabatic (0PA) inspiral, one simply elevates the geodesics variables $(p,\gecc,q_r, q_\phi)$ in Eq.~\eqref{eq:hlmsmr} to their inspiral (evolving) counterparts in Eq.~\eqref{eq:SMREoM0PA}.

\subsection{SMR data and interpolation}
To produce SMR 0PA waveforms\footnote{In the language of \cite{Wardell:2021fyy} this would be a 0PAT1 waveform.} we need the various quantities appearing on the right-hand sides of Eqs.~\eqref{eq:SMREoM0PA} and \eqref{eq:hlmsmr}. The 0th order ``frequencies'' $\Omega$ are known analytically~\cite{Schmidt:2002qk}, while the $\langle F_p\rangle$,$\langle F_{\gecc}\rangle$, and $\mathcal{A}_{lmn}$ need to be calculated numerically. All three may be obtained by solving the Teukolsky equation sourced by a test mass following an eccentric geodesic to obtain the $Z_{lmn}$'s in Eq.~\eqref{eq:teuk}, which we do using the arbitrary precision frequency domain code developed in~\cite{vandeMeent:2014raa,vandeMeent:2015lxa,vandeMeent:2016pee}.

Specifically, we calculate $\langle F_p\rangle$, $\langle F_{\gecc}\rangle$, and $\mathcal{Z}_{lmn}$  on a grid of Chebyshev nodes in $x =(M\wphigeo)^{2/3}$ (18 nodes between $0.001$ and $0.130$) and $\gecc$ (12 nodes between $0$ and $0.5$), and interpolate the results using Chebyshev polynomials. The resulting interpolant has a typical relative interpolation error of about $10^{-5}$.

Note that the SMR 0PA inspiral waveforms generated here could in principle have been generated with Fast EMRI Waveforms (FEW) framework~\cite{Chua:2020stf, Hughes:2021exa, Katz:2021yft}.  We chose a different approach because FEW was not yet publicly available when this project started and to retain a better control over numerical errors in the model. In particular, the FEW model was not designed to faithfully reproduce the minima and maxima of the waveform frequency $\wgw$.

\subsection{Frequencies}

 From a (0PA) SMR inspiral we have two distinct ways of obtaining the average orbital and radial frequencies. We can apply the procedure of Secs.~\ref{subsec:NR-eccentricty} and \ref{subsec:NR-Quantities} to extract the average orbital $\wgwavg$ and radial frequencies $\wgwrad$ from the SMR 0PA waveform. We will denote these frequencies $\wphiPA$ and $\wradPA$. Alternatively, we have the instantaneous geodesic frequencies $\wphigeo$ and $\wradgeo$ as they appeared in Eq.~\eqref{eq:SMREoM0PA}. In the $\smr\to 0$ limit, i.e. when there is  no inspiral, Eq.~\eqref{eq:hlmsmr} gives the following expression of the waveform frequency $\omega_{lm}$, 
 \begin{align}\label{eq:omega22smr}
\omega_{lm} &= -\Im\frac{d}{dt}\log\left(  \sum_{n} \mathcal{A}_{lmn}(p,\gecc) e^{-i(m \phi+n q_r)}\right) \\
&= \Re \frac{\sum_{n} (m\frac{d\phi}{dt}+n\frac{dq_r}{dt}) \mathcal{A}_{lmn}(p,\gecc) e^{-i(m \phi+n q_r)}}{\sum_{n} \mathcal{A}_{lmn}(p,\gecc) e^{-i(m \phi+n q_r)}}\\
&= m\wphigeo(p,\gecc) + \wradgeo(p,\gecc) \Re\frac{\sum_{n} n \mathcal{A}_{lmn}(p,\gecc) e^{-i n q_r}}{\sum_{n} \mathcal{A}_{lmn}(p,\gecc) e^{-in q_r}}\label{eq:omega22smr}.
\end{align}
 From this we note that the waveform frequency is exactly $2\pi$ periodic in $q_r$, and consequently the radial period is exactly $2\pi/\wradgeo$. A less obvious observation is that the average of the second term in~\eqref{eq:omega22smr} vanishes after averaging over a radial period. A sufficient condition for this to be true is
\begin{equation}\label{eq:phasecondition}
\lvert\mathcal{A}_{lm0}(p,\gecc)\rvert > \Big\lvert\sum_{n\neq 0} \mathcal{A}_{lmn}(p,\gecc) e^{-in q_r}\Big\rvert,
\end{equation}
 since this guarantees that $\sum_{n} \mathcal{A}_{lmn}(p,\gecc) e^{-in q_r}$ is confined to a half of the complex plane  and must return to the same complex argument after one period. The condition~\eqref{eq:phasecondition} is clearly satisfied for low eccentricity orbits since $\mathcal{A}_{lmn} = \mathcal{O}(\gecc^n)$.  However, condition~\eqref{eq:phasecondition} is easily violated by high eccentricity zoom-whirl orbits. Nonetheless, we observe empirically that the average of the second term~\eqref{eq:omega22smr} vanishes in all geodesic waveforms used in this work.

\begin{figure}[tbp!]
	\includegraphics[scale=0.233,trim=0 20 0 0]{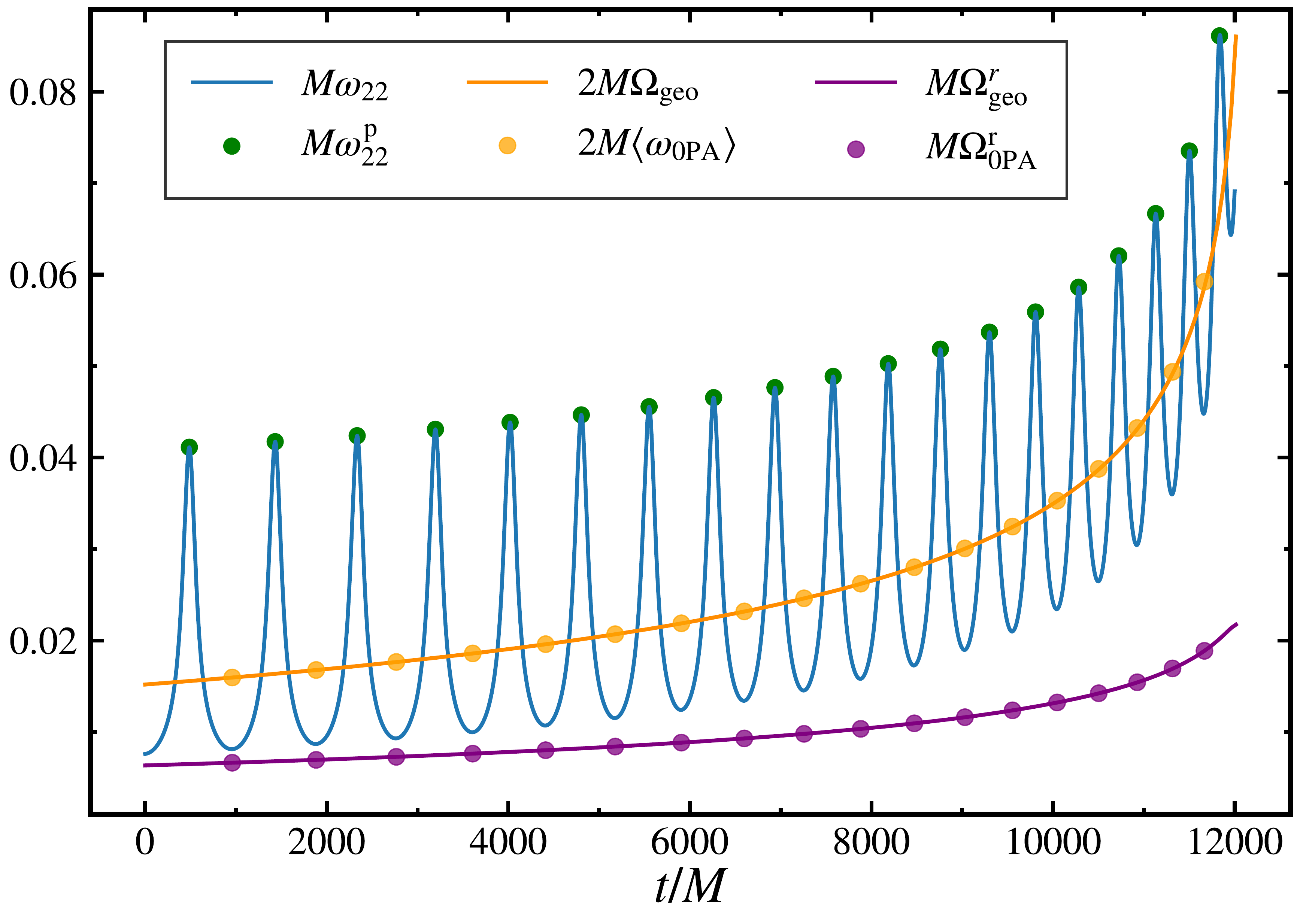}
	\caption{
Frequency extraction procedure applied to an SMR waveform at 0PA at equal mass.  The solid curves arise directly from the SMR inspiral and its dynamics.  The filled circles are the result of applying our frequency extraction procedure to the maxima of the instantaneous frequency $\omega_{22}(t)$.  Even at equal mass where the inspiral is fastest, the recovered orbital averaged azimuthal and radial frequencies agree well with the geodesic frequencies of the underlying SMR dynamics.  This figure is analogous to Fig. 4.
	}
	\label{fig:OmegaphiSMR}
\end{figure}

We thus find that in the $\smr\to 0$ limit we have exactly,
\begin{equation}
    \wradPA =\wradgeo,\quad\text{and}\quad\wphiPA = 2\wphigeo.
\end{equation}
This, of course, does not come as a surprise, since this is precisely what the frequency recovery procedure of Secs.~\ref{subsec:NR-eccentricty} and \ref{subsec:NR-Quantities} was designed to achieve. However, using the SMR 0PA inspiral waveforms we can now investigate what happens for finite values of $\smr$ when the system is evolving. Figure~\ref{fig:OmegaphiSMR} shows both the frequencies, $\wradPA$ and $\wphiPA/2$, recovered from a SMR 0PA waveform at equal mass ($\smr=1/4$) and the geodesic frequencies, $\wradgeo$ and $\wphigeo$, inferred from the underlying inspiral dynamics. Even at equal mass there is hardly any perceivable difference between the two sets of frequencies.

\begin{figure}[tbp!]
	\includegraphics[scale=0.24,trim=0 42 0 0]{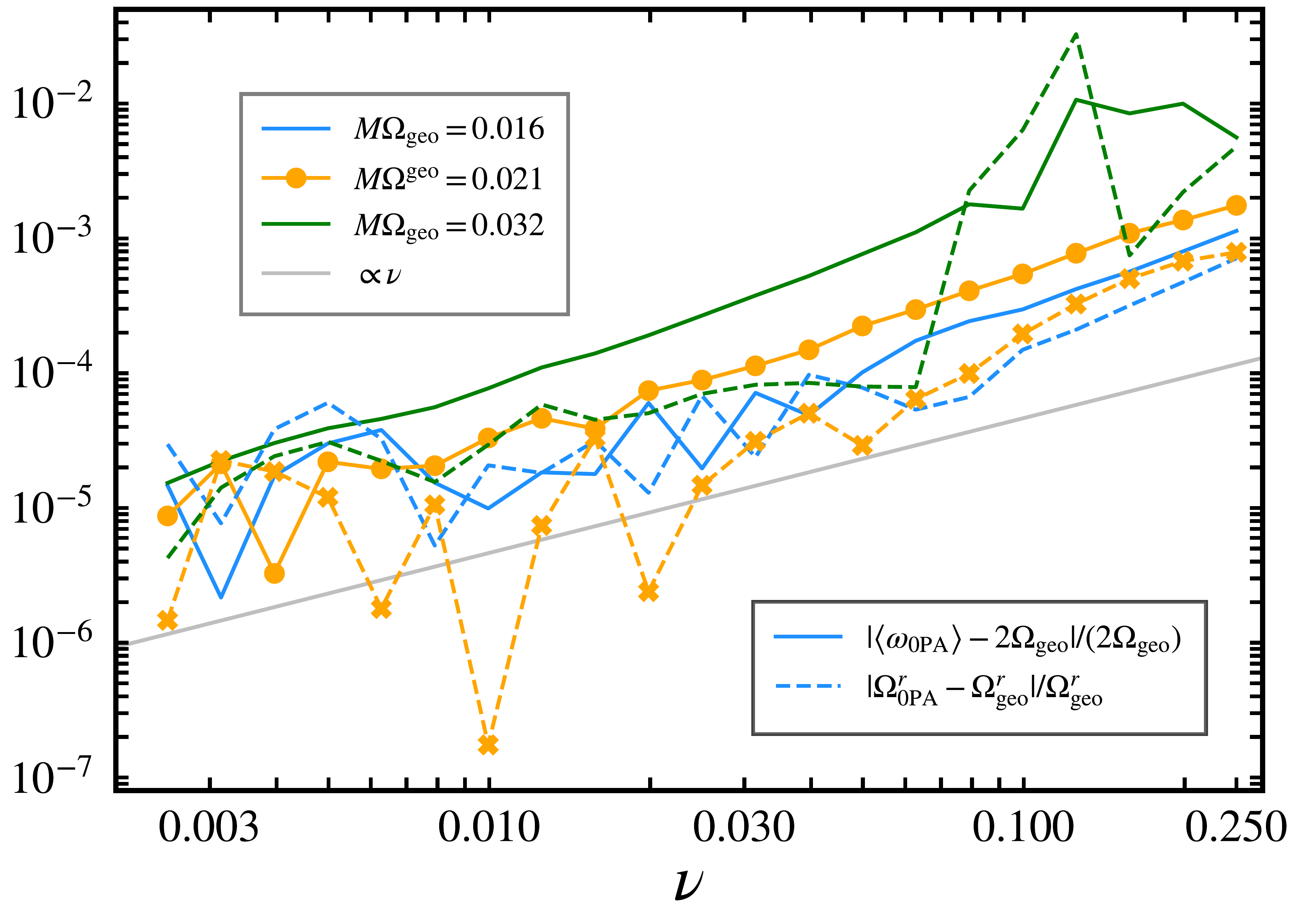}
	\caption{Absolute relative difference between the frequencies from the waveform (azimuthal $\wphiPA/2$ and radial $\wradPA$), and the frequencies from the geodesic inspiral (azimuthal $\wphigeo$ and radial $\wradgeo$), as a function of symmetric mass ratio $\smr$ at three selected points along the inspiral from Fig.~\ref{fig:OmegaphiSMR}. The frequencies extracted from the waveforms, $\wphiPA$ and $\wradPA$, have been computed using the orbit-average procedure of Secs.~\ref{subsec:NR-eccentricty} and \ref{subsec:NR-Quantities} employing the periastron passages. The gray line indicates a linearly increasing $\smr$-dependence.}
	\label{fig:OmegaphiSMR2}
\end{figure}

To compare the frequencies obtained through the two procedures more
closely we pick three frequencies along the adiabatic inspiral depicted in
 Fig.~\ref{fig:OmegaphiSMR}.
For each of these frequencies we generate a
series of adiabatic inspirals with symmetric mass-ratios varying
between $\smr=10^{-3}$ and $\smr=1/4$ going through that point (and
randomized initial values of $q_r$). For each of these inspirals we
extract the azimuthal and radial frequency from the waveform using the
procedure of Secs.~\ref{subsec:NR-eccentricty} and
\ref{subsec:NR-Quantities}. Figure~\ref{fig:OmegaphiSMR2} shows the
difference between these frequencies and the corresponding values
obtained directly from the underlying geodesic. We observe a small,
but measurable, difference between the two sets of frequencies, which
appears to grow linearly with $\smr$ and is larger for higher
frequencies. Since the SMR 0PA waveform contains no higher order
frequency corrections, this difference arises purely from unintended side
effects of the frequency recovery procedure. Some contributing factors
are the averaging over a radial period while the inspiral is evolving,
and limitations in establishing a radial period in the first place.

\subsection{Eccentricity}

To calculate the gauge invariant eccentricty $e_{\rm gw}$ for a SMR 0PA waveform we again have two options. First, we can follow the procedure of Secs.~\ref{subsec:NR-eccentricty} and \ref{subsec:NR-Quantities} to  determine the minima and maxima of $\wgw$ of the SMR 0PA waveform, and compute $e_{\rm gw}$ using Eqs.~\eqref{eq:eq7} and \eqref{eq:eqEccDef}. We will refer to this as $e_{\rm gw}^{\rm 0PA}$. 

\begin{figure}[tbp!]
	\includegraphics[width=0.95\columnwidth,trim=0 30 0 0]{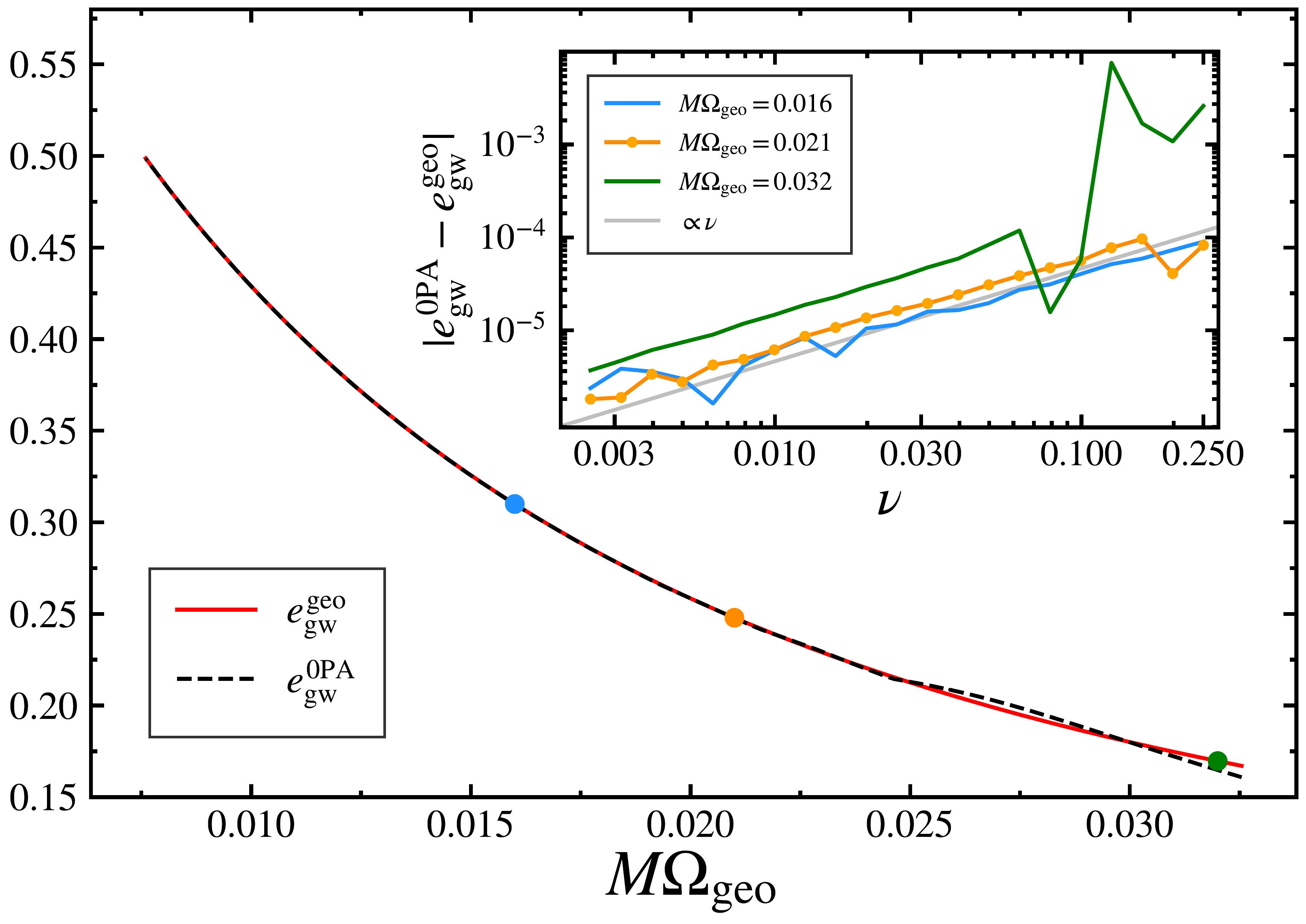}
	\caption{Comparison of $e_{\rm gw}$ obtained directly from the geodesic strain $e_{\rm gw}^{\rm geo}$ with  $e_{\rm gw}$ from an evolving adiabatic inspiral, $e_{\rm gw}^{\rm 0PA}$ as a function of the geodesic frequency $\wphigeo$. The main panel shows an equal mass inspiral ($\smr=1/4$). The inset shows the absolute difference between the two approaches at three selected points along the inspiral for varying mass-ratios. The gray lines in the inset indicate the a linearly growing symmetric mass ratio dependence.
	}\label{fig:eccenticityinspiral}
\end{figure}

Alternatively, we want to obtain  $e_{\rm gw}$ directly from the dynamical variables $p$ and $\gecc$. Unfortunately, there is no analytic closed form expression for $e_{\rm gw}$ in terms of $p$ and $\gecc$. Instead we start from the (numerical) ``snapshot'' waveform generated by a test particle going around a  geodesic with fixed $p$ and $\gecc$. The snapshot waveform $h_{lm}$ is a biperoidic function of the radial and azimuthal phases $q_r$ and $\phi$ as described by Eq.~\eqref{eq:hlmsmr}. Using the expression for $\wgw$ in Eq.~\eqref{eq:omega22smr}, we find the minima and maxima of the frequency with respect to $q_r$ and calculate the corresponding $e_{\omega_{22}}$, which can be input to \eqref{eq:eqEccDef} to provide $e_{\rm gw}$. We will refer to this quantity as  $e_{\rm gw}^{\rm geo}$.

We obtain a numerical representation of the function  $e_{\rm gw}^{\rm geo}(p,\gecc)$ by taking grid of numerical SMR solutions of the Teukolsky equation, and interpolating the result with Chebyshev polynomials to obtain $e_{\rm gw}^{\rm geo}$ with a relative accuracy of $10^{-7}$ across the relevant parameter space. Conversely, we can numerically invert this relationship to obtain a function for $p$ and $\gecc$ given $x$ and $e_{\rm gw}^{\rm geo}$.

Figure~\ref{fig:eccenticityinspiral} explores the difference between  $e_{\rm gw}^{\rm geo}$ and  $e_{\rm gw}^{\rm 0PA}$ for adiabatic inspirals. As expected, the difference between these two approaches for obtaining $e_{\rm gw}$ vanishes  in the $\smr\to 0$ limit. For $\nu\neq 0$, this difference grows again proportional to $\smr$, similarly to Fig.~\ref{fig:OmegaphiSMR2}.

\subsection{Geodesic snapshot vs. inspiral waveform quantities}
The preceding subsections have explored the difference between extracting the frequencies $\Omega_{\phi/r}$, and eccentricity $e_{\rm gw}$ from evolving adiabatic (0PA) waveforms and extracting the same information from geodesic ``snapshots'' that are not evolving at all. The relative difference between the two methods is found to be $\mathcal{O}(\smr)$. For the comparisons in the rest of this work we choose to work with the geodesic snapshot SMR quantities, since these can in general be obtained more efficiently and reliably. For the leading order comparisons this will not make a difference. However, for any higher order corrections that we infer, we must be aware that these also contain a next-to-leading order correction due to comparing NR quantities from an evolving waveform with SMR geodesic snapshot quantities.

\section{Choosing ``independent'' variables}\label{sec:IndependentQuantities}
In this section we study several options for the variables describing the state of a binary inspiral. We present some of the choices of variables made in the literature when comparing SMR and NR results, discuss their applicability in the eccentric case and, finally, describe the choice of variables which better adapt to our study.

The instantaneous state of a non-spinning eccentric binary is captured by four dynamical variables. For example, in the SMR setup these are the $(p, \gecc, q_r, \phi)$ that appear in Eq.~\eqref{eq:SMREoM}. In this work we are interested in comparing quantities that are observable for a distant observer. Since the instantaneous value of $\phi$ is completely degenerate with the position of this observer, it carries no useful information about the state of the binary. Moreover, we are presently interested in observables that are integrated over a radial cycle, eliminating $q_r$. Thus we can identify the instantaneous state of the binary with two variables, like $(p, \gecc)$. Of course,   $(p, \gecc)$ are not gauge invariant, and therefore not useful to find an NR simulation in the same instantaneous state. In order to compare the SMR and NR results, we need  a set of two variables that can fix the instantaneous state of the binary and be unambiguously computed both in NR and the SMR formalism.

One pair of variables extensively used in the literature \cite{Barack:2011ed,Lewis:2016lgx} are the azimuthal and radial frequencies $\worbavg$ and $\worbrad$ defined in analogy to Eqs.~\eqref{eq:eq102} and \eqref{eq:eq103} from the orbital frequency $\worb$.  These two frequencies  can be calculated analytically at geodesic order in the SMR formalism~\cite{Schmidt:2002qk}, and they can be extracted from NR data \cite{Lewis:2016lgx}.  However, since they are derived from the coordinate trajectories in NR,
 they are not fully gauge invariant (e.g.~Fig.~17 of~\cite{Ossokine:2015vda}).

\begin{figure}[tbp!]
	\centering
	\hspace*{0.2cm}\includegraphics[width=0.975\columnwidth]{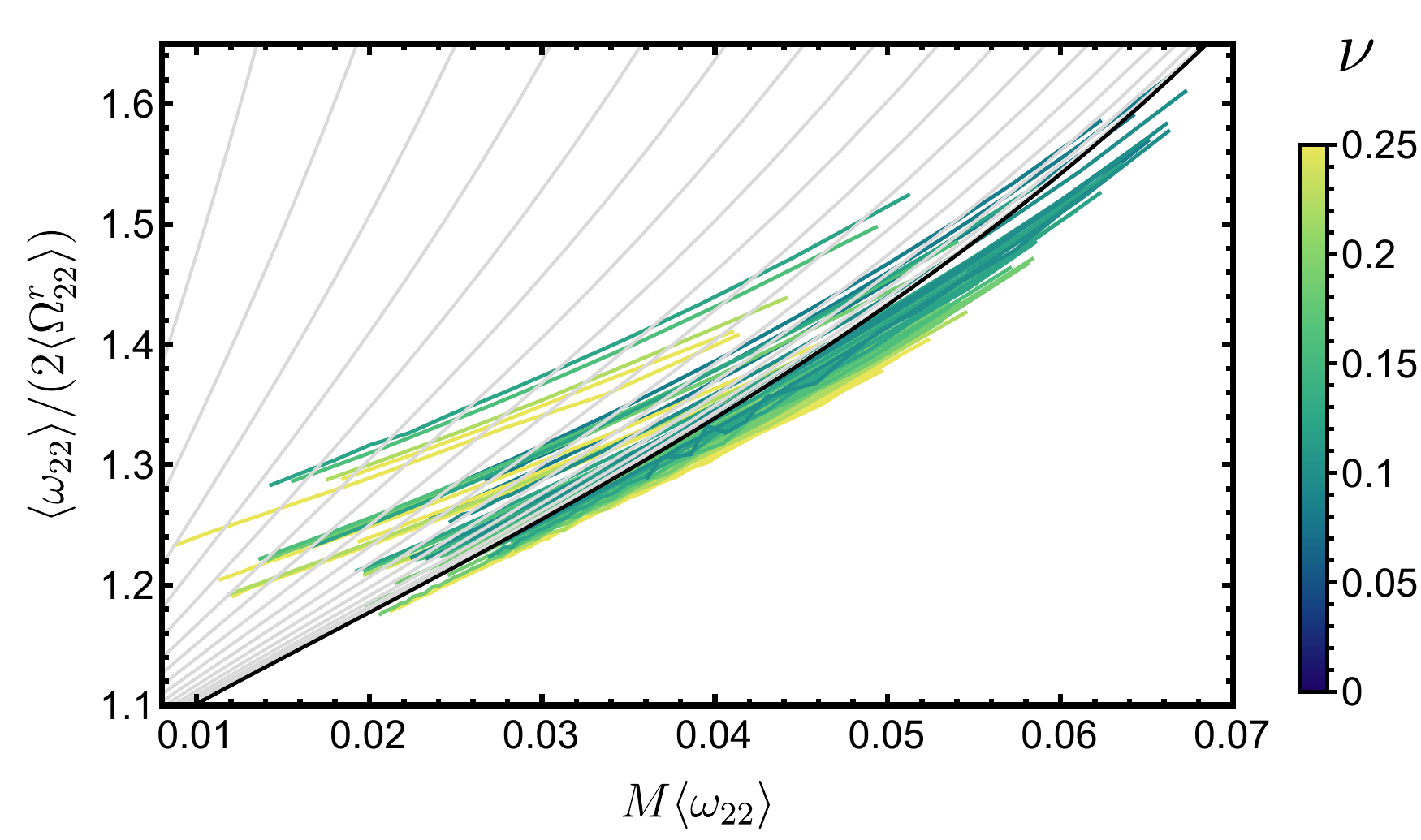}\\[0.5em]
	\includegraphics[width=0.98\columnwidth,trim=0 20 0 0]{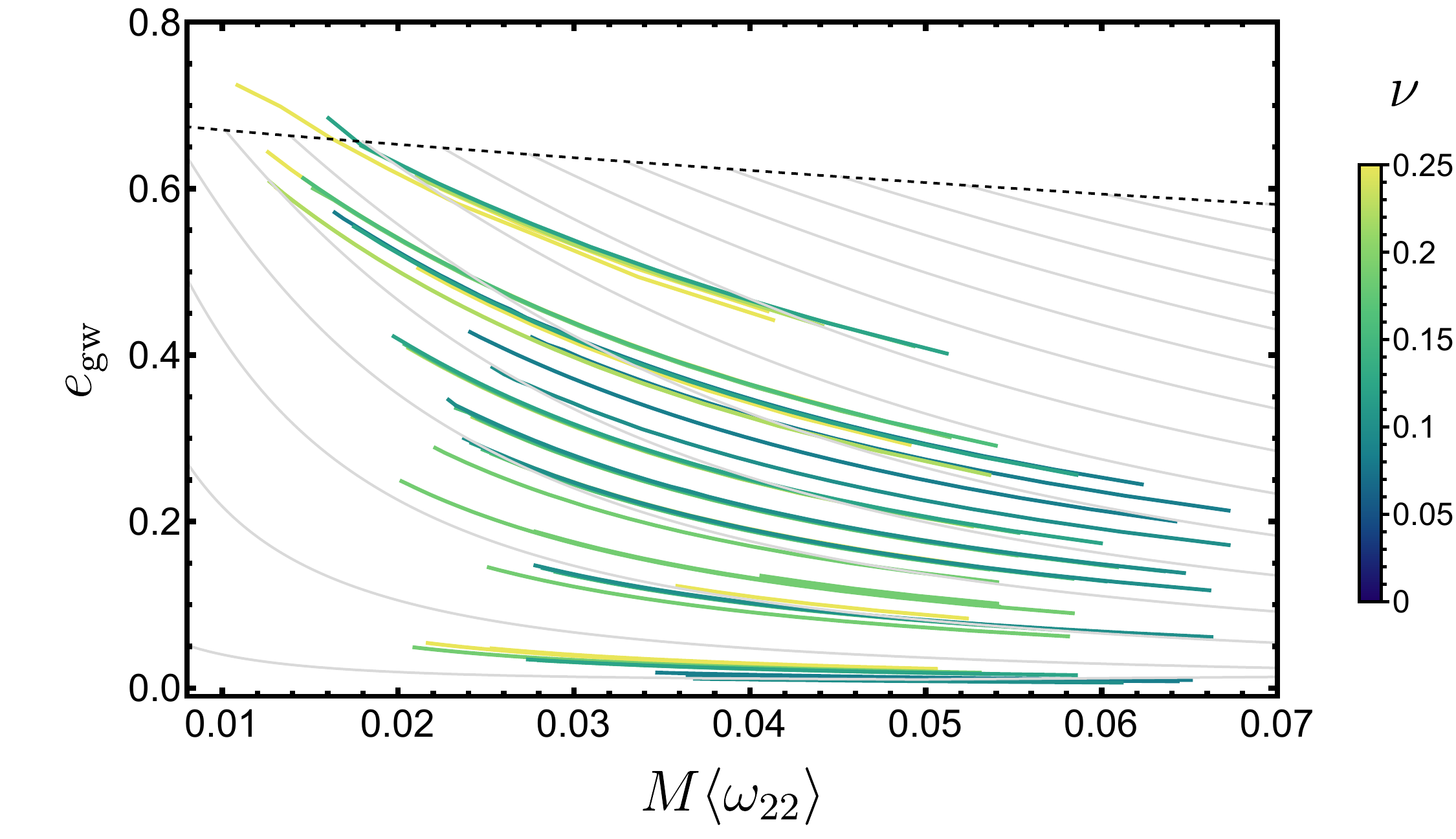}
	\caption{\label{fig:independentVariables}
    \textit{Top panel}: Ratio of the orbit-averaged azimuthal and radial frequencies computed from the (2,2)-mode , $  \wgwavg/(2 \wgwrad)$, as a function of the orbit-averaged frequency, $\wgwavg$. The colored curves represent the
NR simulations in Table~\ref{tab:tabNR}, whereas the grey shaded area indicates the region covered by Schwarzschild geodesics, bounded by the diagonal black curve representing quasi-circular geodesics.
	  \textit{Bottom panel}: Eccentricity, $e_{\rm gw}$, computed using Eqs.~\eqref{eq:eq7} and \eqref{eq:eqEccDef}, as a function of  $\wgwavg$,  for the same NR simulations as in the upper panel. While geodesics exist at all eccentricities, we have only generated SMR configurations in the grey shaded area.
In both panels each NR simulation has been color-coded according to its symmetric mass ratio $\nu$. 
	}
\end{figure}

A second possibility are  frequencies computed from the gravitational radiation instead, e.g.~$\wgwavg$ and $\wgwrad$, which are manifestly gauge invariant. Then, as shown in Secs.~\ref{sec:Eccdef} and \ref{sec:SMRevolutions}, the orbit-averaged azimuthal frequencies from the waveform and the trajectories can be related by a factor $2$, while the radial frequency stays the same.
These frequencies are plotted in the top panel of Fig.~\ref{fig:independentVariables}.
For some portions of NR simulations the ratio $\wgwavg/(2\wgwrad)$ lies below the value of the corresponding circular orbit at the same $\wgwavg$, i.e. the NR frequencies fall outside the range spanned by geodesics.
It might be possible to rectify this situation by applying a linear mass-ratio gravitational self-force correction to the NR frequencies. This would, however, result in a very convoluted analysis requiring SMR inputs on the NR side of the comparison. Thus, to avoid such a complication we discard the radial and azimuthal frequencies as independent variables to describe both NR and SMR eccentric inspirals.

A third possibility as a pair of independent variables are the binding energy, $E_b$, and the dimensionless angular momentum, $j$. This pair of variables has been extensively used for comparisons between NR simulations and effective-one-body (EOB) evolutions~\cite{Nagar:2015xqa,Ossokine:2017dge,Antonelli:2019ytb}. Both quantities can be analytically calculated at geodesic order in the SMR formalism~\cite{Schmidt:2002qk}, and they can also be extracted from the NR simulations. Nonetheless, the computation of the reduced angular momentum and the binding energy from NR simulations requires the application of some unknown offsets to both quantities.
This is due to the fact that  $E_b$ and $j$ are reconstructed by integrating the fluxes to infinity and using the initial (ADM) or final mass and angular momentum. However, the fluxes at the start of the simulations, due to junk radiation, and at the end, due to the exponential power decay during ringdown, are not very well resolved. Consequently, the obtained $E_b$--$j$ curves are generally off by a shift in the $E_b$--$j$ plane  \cite{Ossokine:2017dge}. Even after that shift is applied, it is possible that the NR data exist in the region of the $E_b$--$j$ plane that is inaccessible by geodesics. Thus, in order to avoid the introduction of systematics from the determination of the offset in   $E_b$ and $j$, here we do not consider them as independent variables for the mapping between SMR and NR configurations.

Finally, we present a combination of variables such that Schwarzschild geodesics and NR simulations lie in the same region of parameter space. These variables are the eccentricity measure $e_{\rm gw}$, defined in Eqs.~\eqref{eq:eq7} and \eqref{eq:eqEccDef}, and the orbit-averaged azimuthal frequency computed from the (2,2)-mode, $\wgwavg$, defined in Eq.~\eqref{eq:eq102}. The lower panel of Fig.~\ref{fig:independentVariables} displays $e_{\rm gw}$ as a function of $\wgwavg$, for the NR simulations in Table \ref{tab:tabNR} and for Schwarzschild geodesics. The $e_{\rm gw}$--$\wgwavg$ plane is naturally overlapping for both NR and Schwarzschild geodesics, without the need of any shifts or rescalings. This eccentricity definition by construction spans the range from 0 (circular) to 1 (parabolic orbit). This remains true at any level of the SMR approximation. Additionally, we note that given a (2,2)-mode waveform  $e_{\rm gw}$ is uniquely determined, and thus, it is a gauge invariant observable.

 However, we note that the leading order contribution to $e_{\rm gw}$ cannot be computed analytically in the SMR formalism, but requires solving the first order field equations numerically. Similarly, the next-to-leading order in mass-ratio correction to  $e_{\rm gw}$ requires the second order metric perturbation, which has not yet been calculated for eccentric orbits. Consequently, calculating the next-to-leading order contribution to the expansion of any observable at fixed $e_{\rm gw}$ and $\wgwavg$ requires obtaining the second order metric perturbation. (The only exception to this are quantities at fixed $e_{\rm gw}=0$ or $e_{\rm gw}=1$, since the higher order corrections to these values are zero by construction.)

In the limit $e_{\rm gw}\to 0$, fixing $e_{\rm gw}$ and $\wgwavg$ reduces to the usual comparisons done for quasi-circular inspirals. Hence, we consider these two variables, $e_{\rm gw}$ and $\wgwavg$, as our independent variables for the comparison of NR and SMR inspirals.

\section{Results}
\label{sec:SMRNRcomparison}
In this section we compare the energy and angular momentum fluxes, as well as the periastron advance, obtained from NR and SMR adiabatic evolutions, and provide constraints on the magnitude of the next order term in the SMR expansion for the mass ratios considered here.
We consider the orbit-averaged   energy and angular momentum fluxes from the NR simulations in Table \ref{tab:tabNR}, computed using Eqs.~\eqref{eq:eq107}---(\ref{eq:eq108}), as well as the periastron advance $K$, computed using Eq.~(\ref{eq:eq105}). 

\begin{figure*}[!]
\includegraphics[scale=0.36, trim=0 8 0 70]{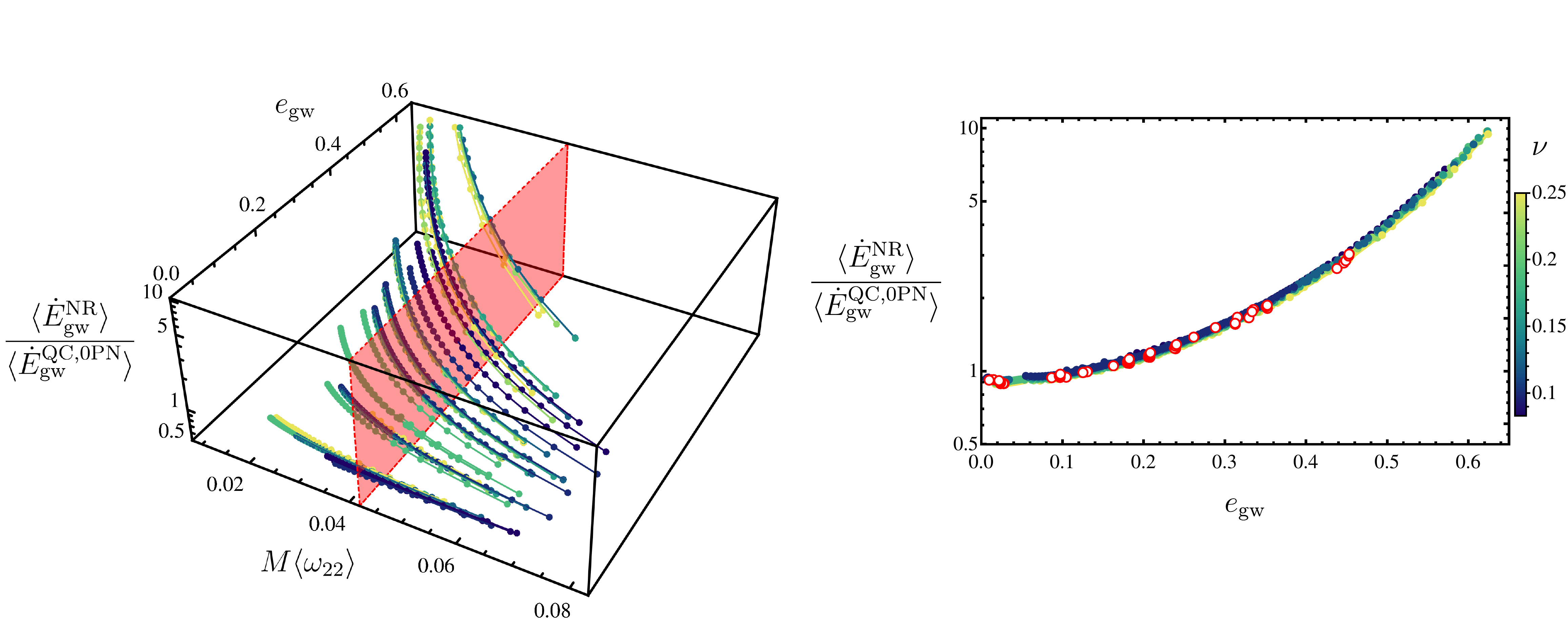}
\includegraphics[scale=0.36,trim=0 18 0 80]{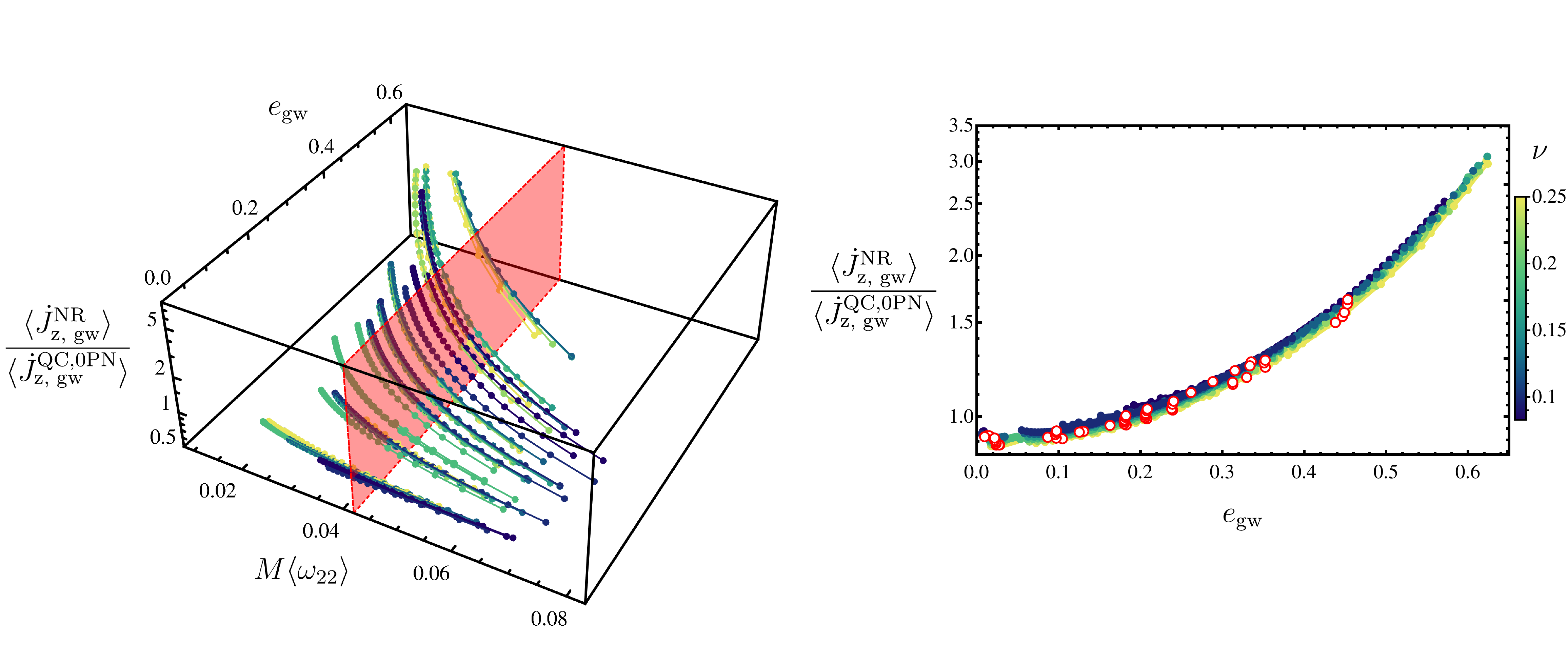}
\hspace*{1.1cm}\includegraphics[scale=0.34,trim=0 30 0 80]{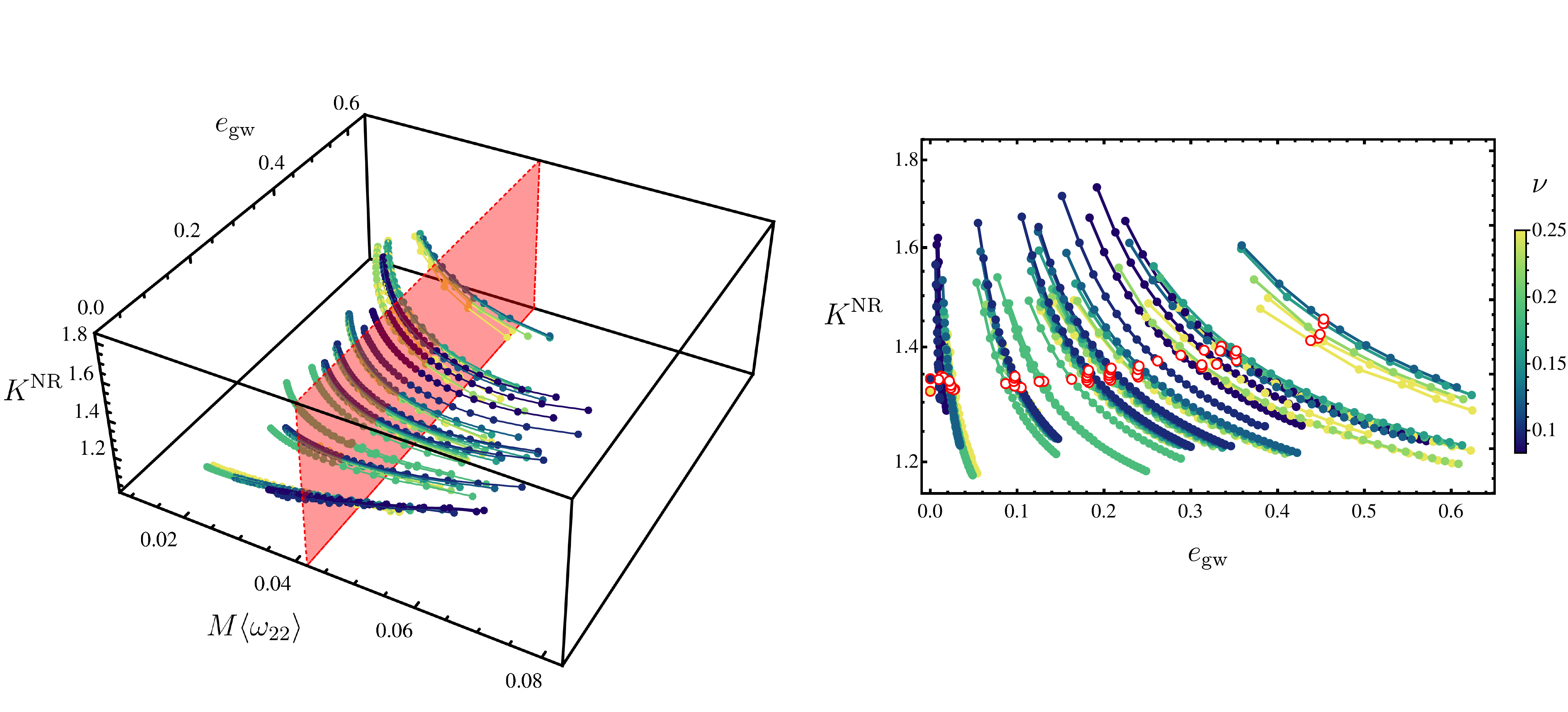}
	\caption{
	\textit{Left column:} From top to bottom energy flux, angular momentum flux and periastron advance extracted from the NR simulations in Table \ref{tab:tabNR} as a function of eccentricity, $e_{\rm gw}$, and orbit-averaged azimuthal frequency, $\wgwavg$. Each curve corresponds to a NR simulation in Table \ref{tab:tabNR}, and is color-coded by symmetric mass ratio $\nu$. The energy and angular momentum fluxes are rescaled by the quasi-circular Newtonian expressions in Eqs.~\eqref{eq:eq204}. The red planes indicate the reference frequency $M \wref = 0.042$.
  \textit{Right column:} Projection of the left plots, in the $Z$--$e_{\wgw}$ plane, where $Z$ indicates the quantity in the z-axis (fluxes or periastron advance). The red-white circles indicate the points where each NR simulation passes the reference frequency $M \wref = 0.042$.
	}
\label{fig:EdotJdotKNR}
\end{figure*}

The orbit-averaged
fluxes extracted from the NR simulations are illustrated in the top two rows of Fig.~\ref{fig:EdotJdotKNR}. In order to reduce the dynamical range of the fluxes, we rescale them with the Newtonian (0PN) quasi-circular values for these quantities~\cite{Blanchet:2013haa}
\begin{subequations} \label{eq:eq204}
\begin{align}
  \langle \dot{E}^{\text{QC,0PN}}_{\rm gw} \rangle &= \frac{32}{5}\nu^2 \worb^{10/3},\\
  \langle \dot{J}^{\text{QC,0PN}}_{z, \rm  gw} \rangle &= \frac{32}{5}\nu^2 \worb^{7/3}.
\end{align}
\end{subequations}
Here, $\worb$ denotes the orbital frequency for which we substitute $\wgwavg /2$, see Sec.~\ref{sec:Eccdef} for details.

The rescaling by Eqs.~\eqref{eq:eq204} produces a smooth dependence of the fluxes in parameter space, with practically no curves crossing each other. This is because most of the mass ratio dependence is already accounted for by the rescaling factors.
In the right-hand panels, the data are plotted as function of $\egw$ only. This projection
highlights how well the normalization accounts for the $\nu$-- and $\wgwavg$--dependence, with only the eccentricity-dependence remaining.  The eccentricity dependence qualitatively resembles the expected analytical behaviour for the energy flux for eccentric binaries with corrections of the form  $\sim \frac{1}{(1-e^2)^x}(1+a e^2 + b e^4+...)$, where $x=7/2$~or~$2$, for the energy and angular momentum fluxes respectively, while $a$ and $b$ are coefficients which can be found in \cite{Arun:2007sg, Arun:2009mc}. We do not introduce eccentric corrections to the rescaling factors as our eccentricity definition, $e_{\rm gw}$, only reduces to the temporal eccentricity, $e_t$, at Newtonian order, while higher PN order corrections may be important to reproduce the eccentricity dependence of the NR fluxes, especially at the end of the inspiral regime. Hence, we leave the exploration of the eccentricity dependence of the NR fluxes for future work.

The periastron advance is not rescaled, since  the Newtonian value is simply 1 by Kepler's first law. As a consequence,  a larger dependence of this quantity on mass ratio is observed when projected into the $K_{\rm NR}-e_{\rm gw}$ plane. While the values corresponding to a fixed reference of $\wref = 0.042$ (red circles) as  a function of eccentricity show a similar behavior as the fluxes. Overall, the inspection of the NR curves in Fig.~\ref{fig:EdotJdotKNR} indicates that most of the mass ratio dependence may be already captured by the leading order mass ratio contribution.

\begin{figure}[tbp!]
\includegraphics[width=\columnwidth,trim=10 30 20 75]{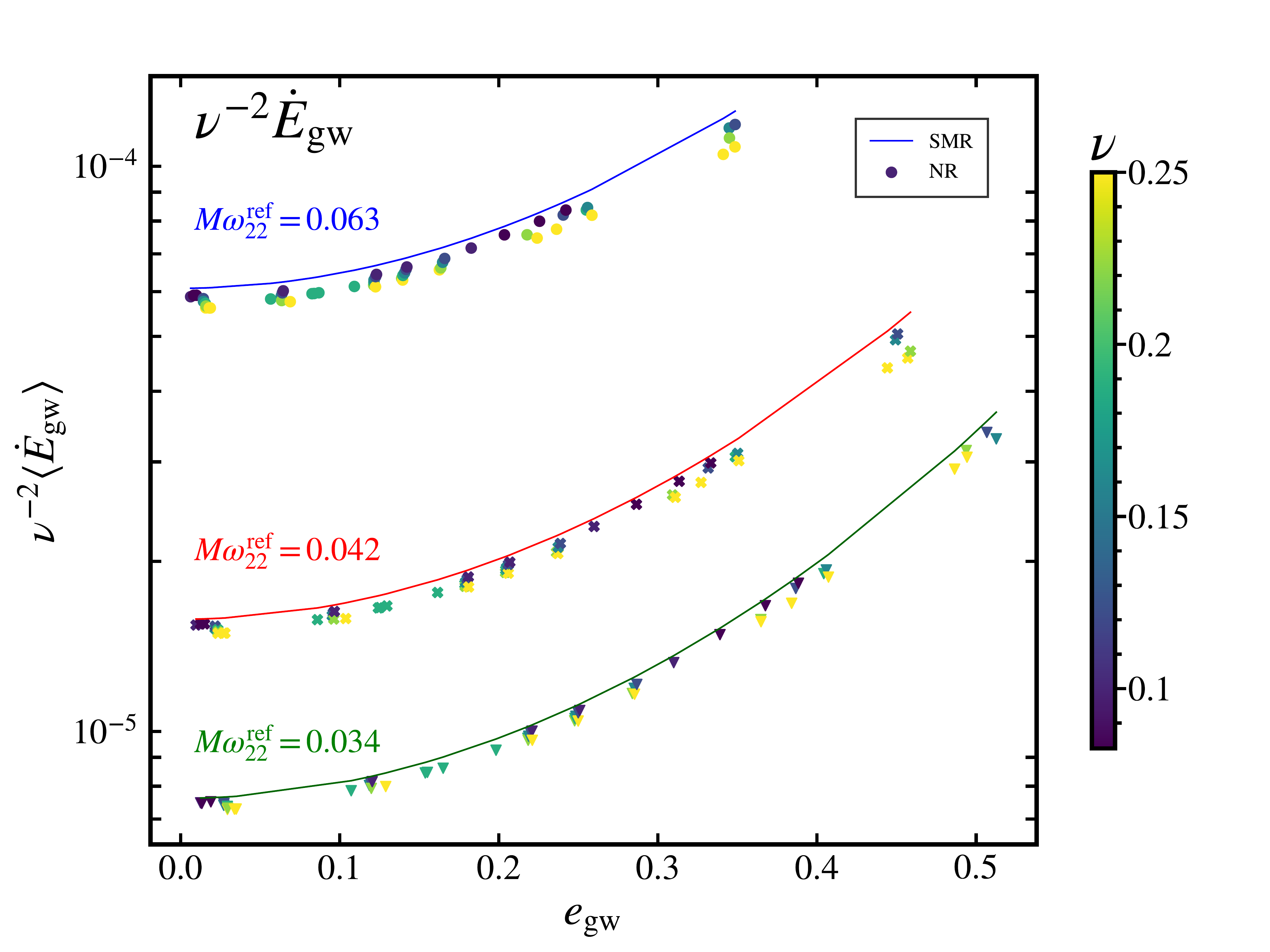}
	\caption{
	Orbit-averaged energy flux rescaled by the leading order symmetric mass ratio dependence $(\nu^{-2})$ as a function of eccentricity at three different reference frequencies, $M\wref = 0.034,0.042,0.063$. Each marker corresponds to a NR simulation at the specified reference frequency, and it is color coded by mass ratio. The solid lines are the leading order SMR energy flux.
	}
\label{fig:EdotNRSMRfixedFreq}
\end{figure}

\begin{figure}[tbp!]
\includegraphics[width=\columnwidth,trim=0 30 0 55]{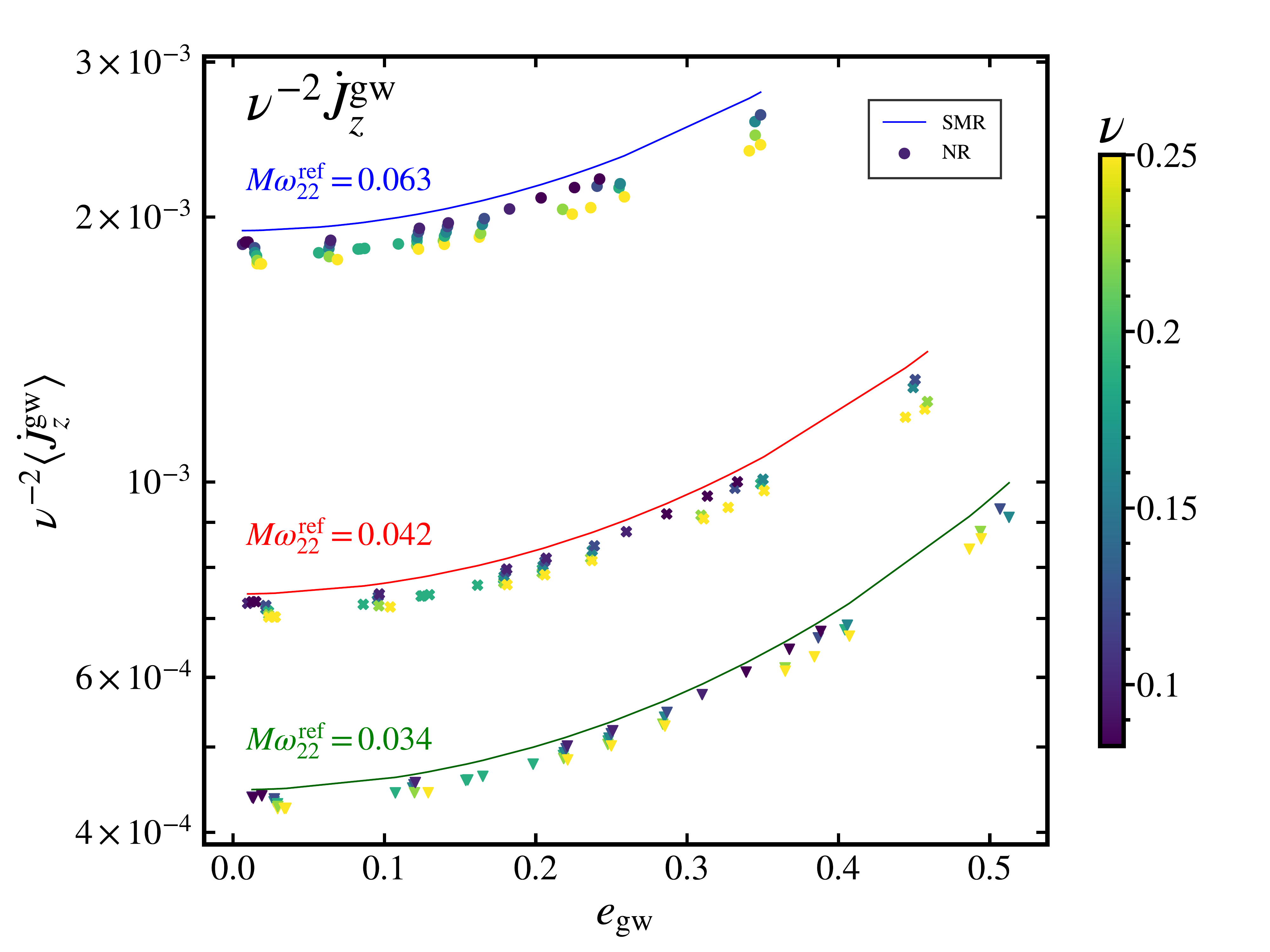}
	\caption{
	Orbit-averaged angular momentum flux rescaled by the leading order symmetric mass ratio dependence $(\nu^{-2})$ as a function of eccentricity at three different reference frequencies, $M\wref = 0.034,0.042,0.063$. Each marker corresponds to a NR simulation at the specified reference frequency, and it is color coded by mass ratio. The solid lines are the leading order SMR angular momentum flux.
	}
\label{fig:JdotNRSMRfixedFreq}
\end{figure}

\begin{figure}[tbp!]
\includegraphics[width=\columnwidth,trim=0 30 0 55]{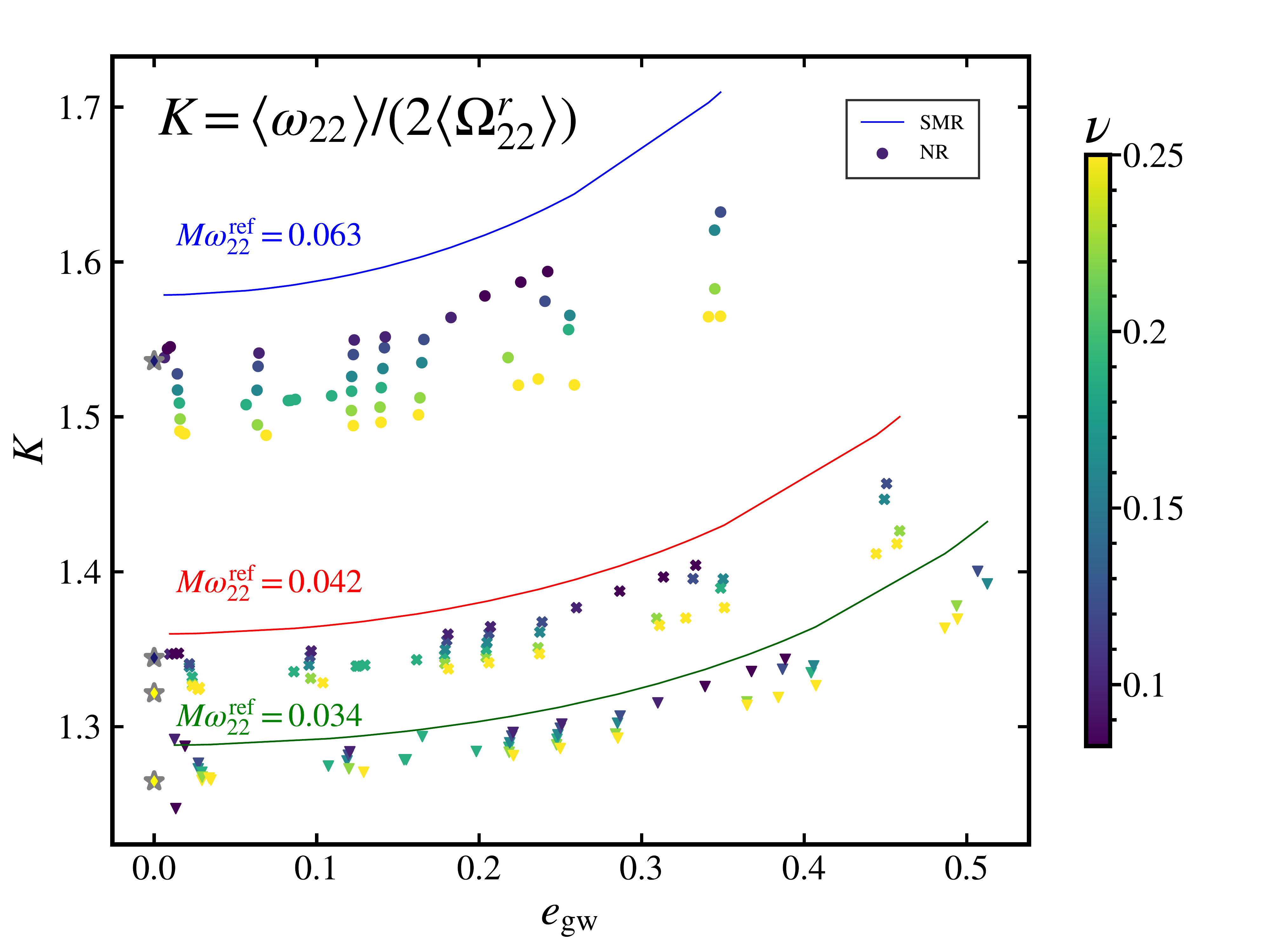}
	\caption{
		Periastron advance as a function of eccentricity at three different frequencies, $M\wref = 0.034,0.042,0.063$. Each marker corresponds to a NR simulation at the specified reference frequency, and it is color coded by mass ratio. The solid lines correspond to joining the values of the geodesic periastron precession at the same $(\nu,e_{\rm gw},\wref)$ values as the NR configurations.
	}
\label{fig:KNRSMRfixedFreq}
\end{figure}

Moving to the comparison of NR against SMR results, the NR and SMR fluxes rescaled by the leading order symmetric mass ratio squared as well as the periastron advance, are shown in Figs.~\ref{fig:EdotNRSMRfixedFreq}, \ref{fig:JdotNRSMRfixedFreq} and \ref{fig:KNRSMRfixedFreq} for three different reference frequencies representative of the full inspiral, $M\wref = 0.034,0.042,0.063$. The SMR fluxes are determined numerically from geodesic snapshots at the quantities selected in Sec.~\ref{sec:IndependentQuantities}, $(\nu,e_{\rm gw},\wref)$, as explained in Sec.~\ref{sec:SMRevolutions}. The SMR values for the periastron advance correspond to the analytic geodesic result for the periastron advance, which can be readily obtained from expressions available in the Black Hole Perturbation Toolkit~\cite{BHPToolkit},
\begin{equation}
K_{\rm SMR} = \frac{2 p \mathsf{K}\left(\tfrac{4\gecc}{p-6+2\gecc}\right) }{\pi \sqrt{p (p-6+2\gecc)}},
\end{equation}
where $\mathsf{K}$ is the complete elliptical integral of the first kind, and $p$ and $\gecc$ have been evaluated at the corresponding values of  $\wref$ and $e_{\rm gw}$.

In the case of the fluxes (Figs.~\ref{fig:EdotNRSMRfixedFreq} and \ref{fig:JdotNRSMRfixedFreq}), both NR and SMR show qualitatively good agreement, which is maintained with increasing eccentricity. This indicates that the effect of eccentricity is well captured by the SMR calculations. Additionally, the dependence on mass ratio when rescaling by the leading order symmetric mass ratio contribution is small. The qualitative agreement between the NR and SMR degrades with increasing reference frequency, as expected because higher order mass ratio corrections are larger in the strong field.
The periastron advance, shown in Fig.~\ref{fig:KNRSMRfixedFreq}, has a stronger dependence on mass ratio than the fluxes, especially at high frequencies. As in the case of the fluxes, with increasing eccentricity the agreement of the periastron advance between NR and SMR does not substantially degrade, indicating that eccentric effects are accurately described within SMR theory using adiabatic evolutions. Overall, for both fluxes and periastron advance the SMR curves overestimate the NR results for all frequencies, mass ratios and eccentricities.

\begin{figure}[tbp!]
\includegraphics[width=\columnwidth]{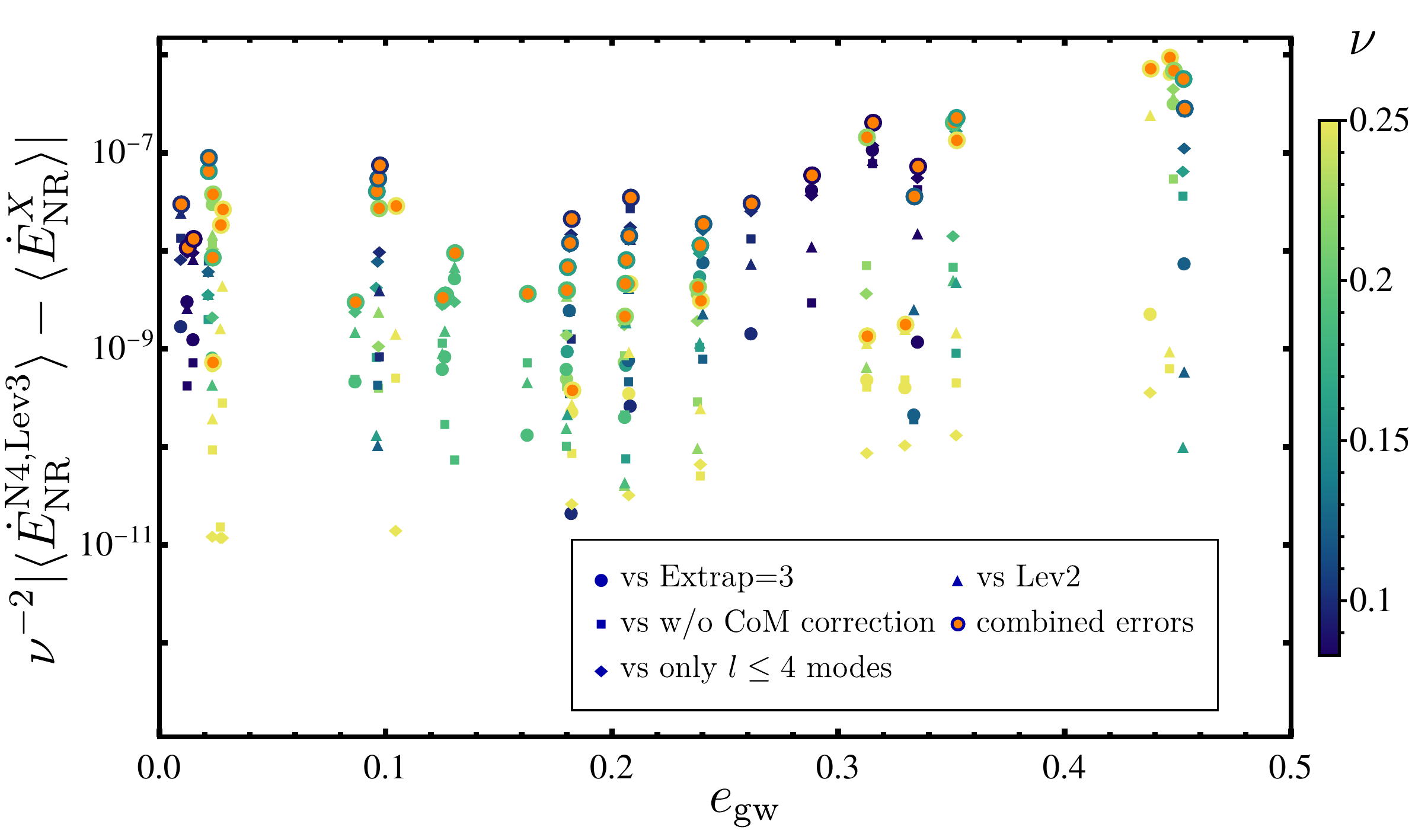}
\includegraphics[width=\columnwidth]{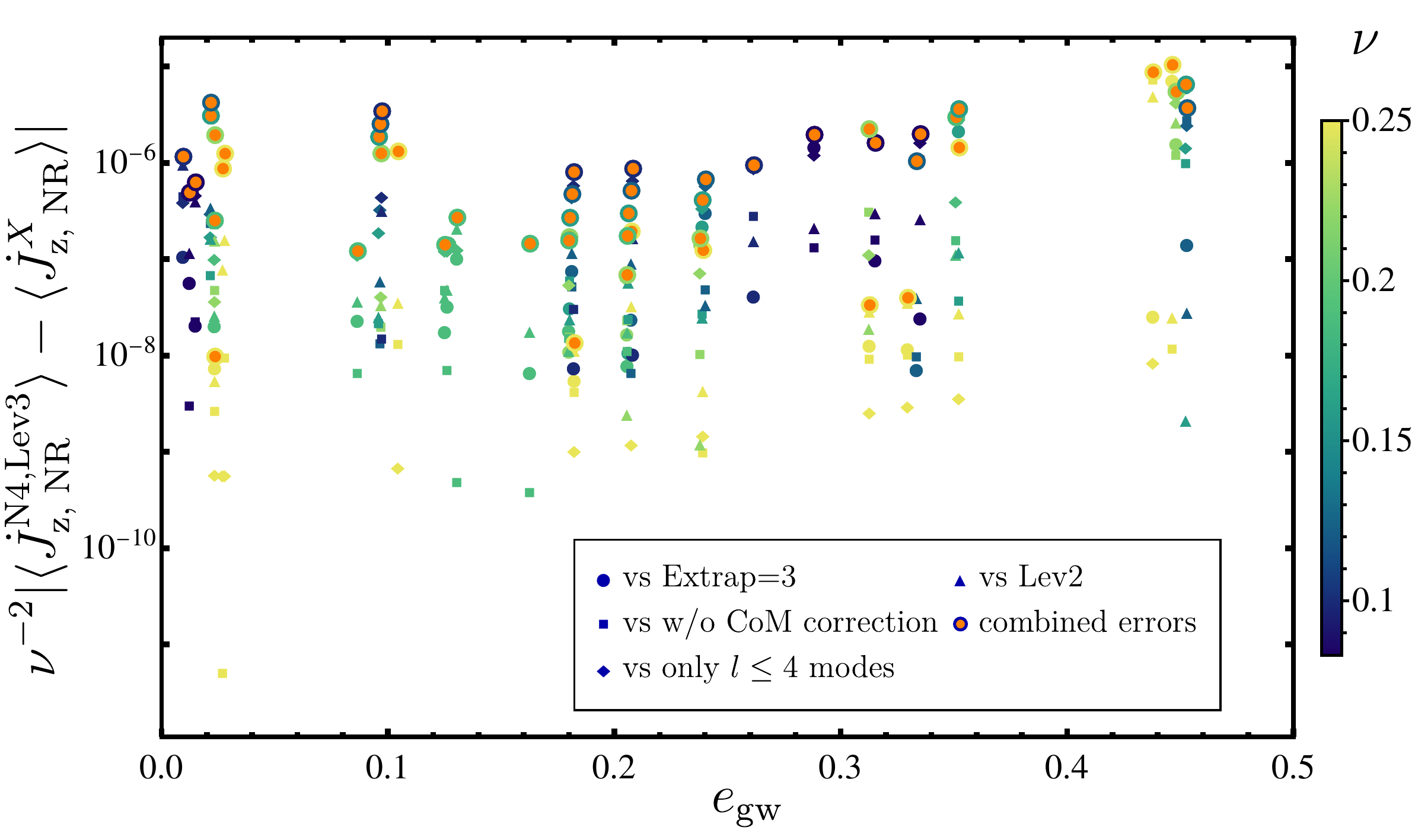}
\includegraphics[width=\columnwidth]{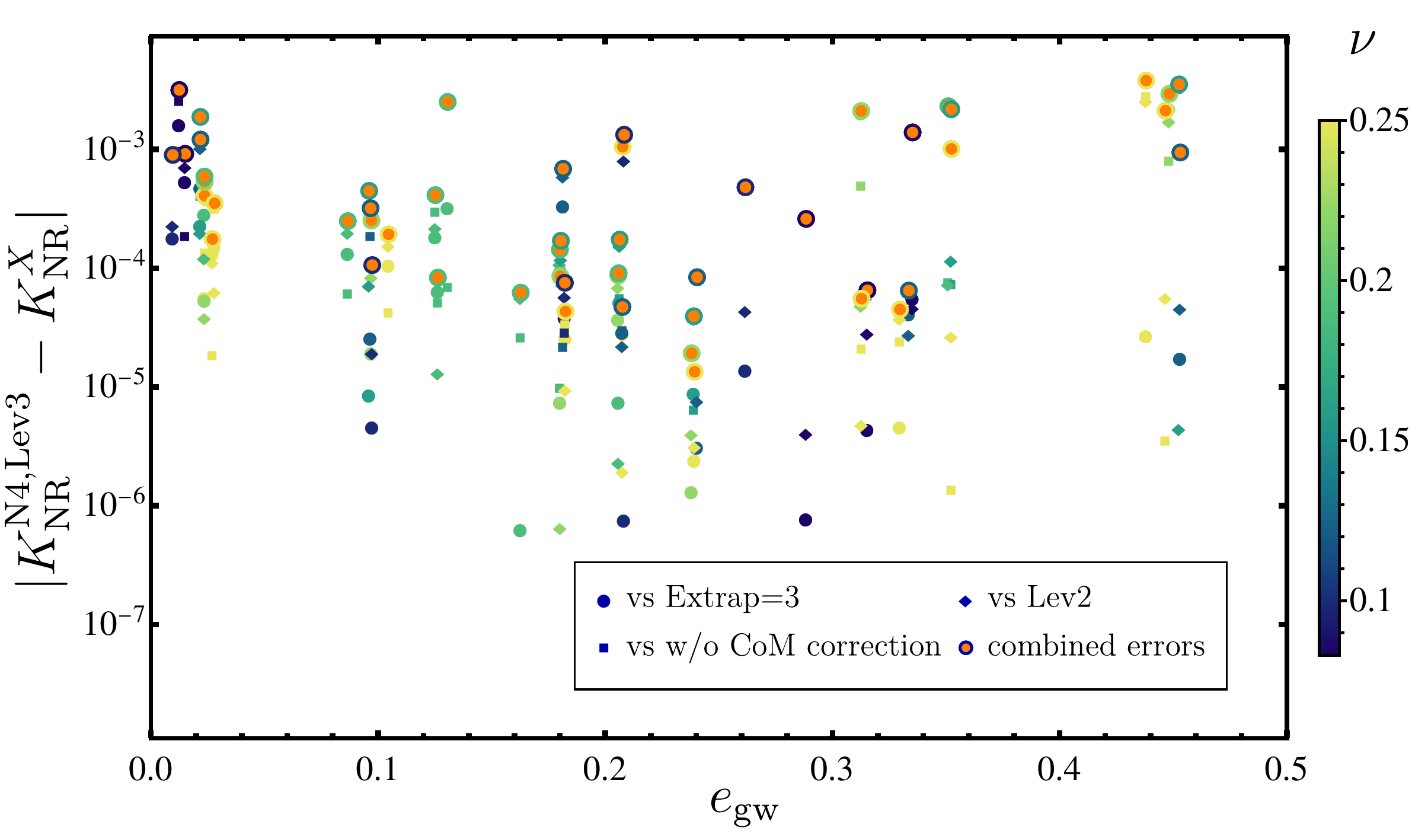}
	\caption{ Error estimates for the energy and angular momentum fluxes, and the periastron advance computed from the NR simulations in Table \ref{tab:tabNR} at a reference frequency of $\wref =0.042$.  Taking as a reference data computed from the waveform computed with highest numerical resolution (Lev3), with extrapolation order 4, CoM correction, and all the modes up to $l \leq 8$, we compute the absolute difference that arises when each one of these conditions is changed, i.e., comparing to the values computed from the waveform  with extrapolation order 3;  without CoM correction; against a lower resolution; and against the waveform with incomplete modes only up to $l \leq 4$. In the case of periastron advance the impact of higher order modes is not assessed as this quantity is computed from the (2,2)-mode. The orange circles represent the quadrature sum of the individual error contributions.}
\label{fig:NRerrorbars}
\end{figure}

Before proceeding to a more quantitative comparison of the difference
between NR and SMR, we assess the accuracy of the NR values shown in
Figs.~\ref{fig:EdotNRSMRfixedFreq}--\ref{fig:KNRSMRfixedFreq} by
comparing NR data obtained with different numerical settings.
The data in Figs.~\ref{fig:EdotNRSMRfixedFreq}--\ref{fig:KNRSMRfixedFreq} was obtained from the highest numerical resolution (Lev3) with applied center-of-mass (CoM) correction\footnote{We perform
center-of-mass correction and extrapolation of the waveforms using the
\texttt{scri} package \cite{mike_boyle_2020_4041972}, which implements
the methods developed in \cite{Boyle2013,BoyleEtAl:2014,Boyle2015a}.},
extrapolation order 4, and all
the spin-weighted spherical harmonic modes up to $l\leq 8$. For the
particular reference frequency of $\wref=0.042$, we show in
Fig.~\ref{fig:NRerrorbars} the absolute difference between the
quantities computed from this reference waveform against the same
quantities calculated from a waveform, where one of the previous
conditions is modified at a time. Precisely, the differences are
computed against a waveform without CoM correction; using
extrapolation order 3; a lower resolution (Lev2), and also in the case
in which only $l\leq 4$ modes are included. The largest differences
for the three quantities typically occur when comparing against the
lower resolution (Lev2). 
Furthermore, the individual errors are summed in quadrature for an overall error estimate for the subsequent analysis.

\begin{figure*}[tbp!]
\begin{minipage}{\textwidth}
\includegraphics[width=\columnwidth]{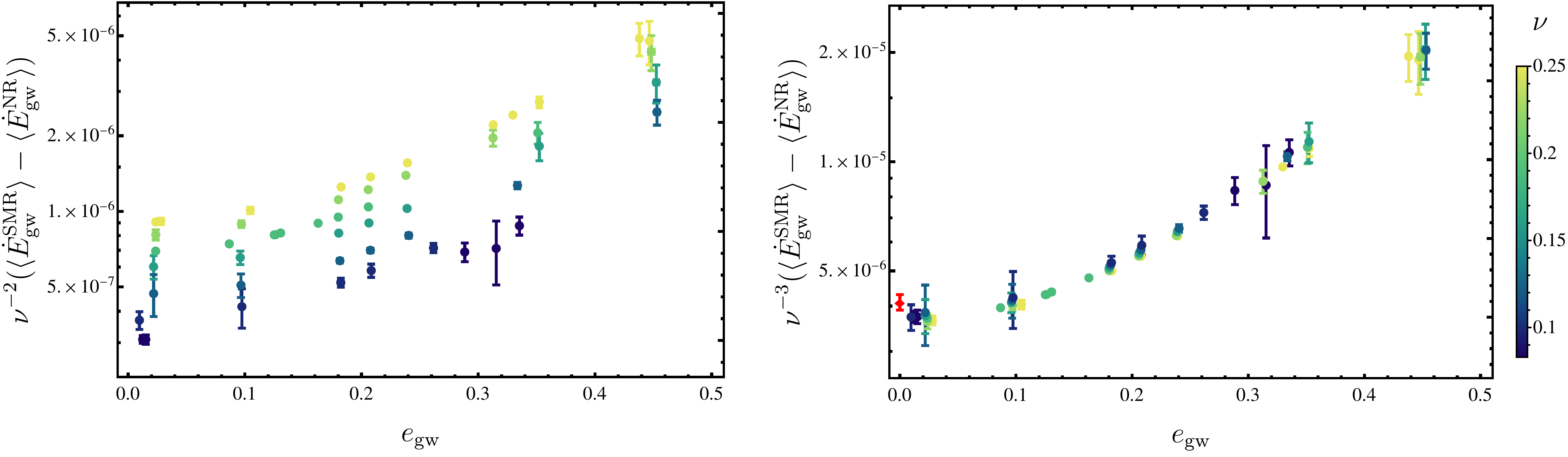}
\end{minipage}
\begin{minipage}{\textwidth}
\includegraphics[width=\columnwidth]{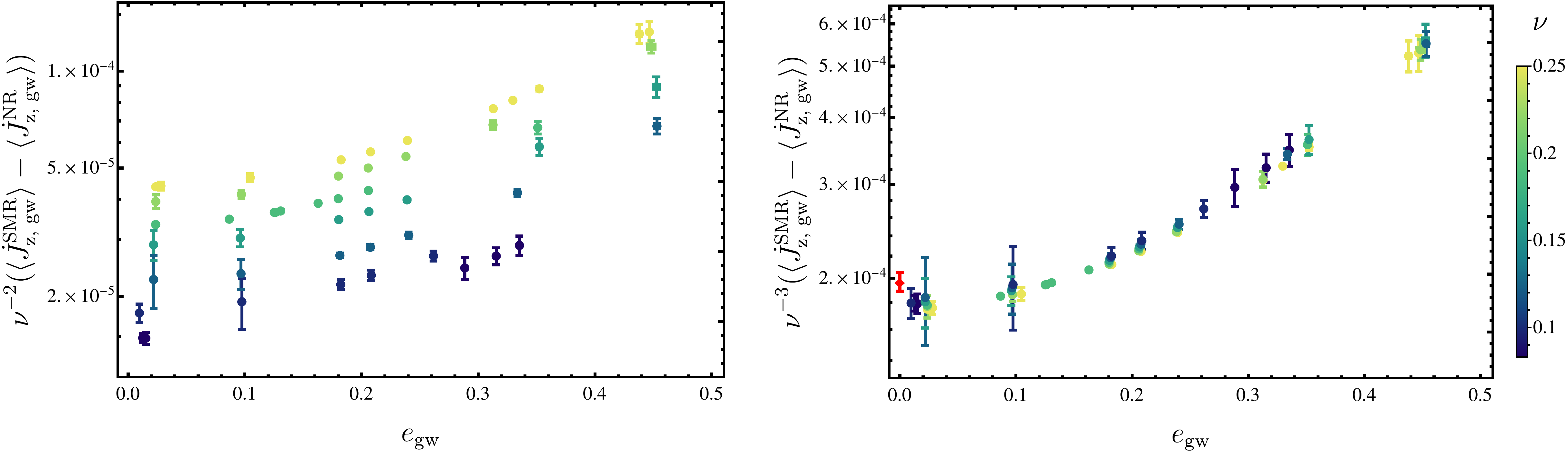}
\end{minipage}
\begin{minipage}{\textwidth}
\hspace*{0.1cm} \includegraphics[width=\columnwidth]{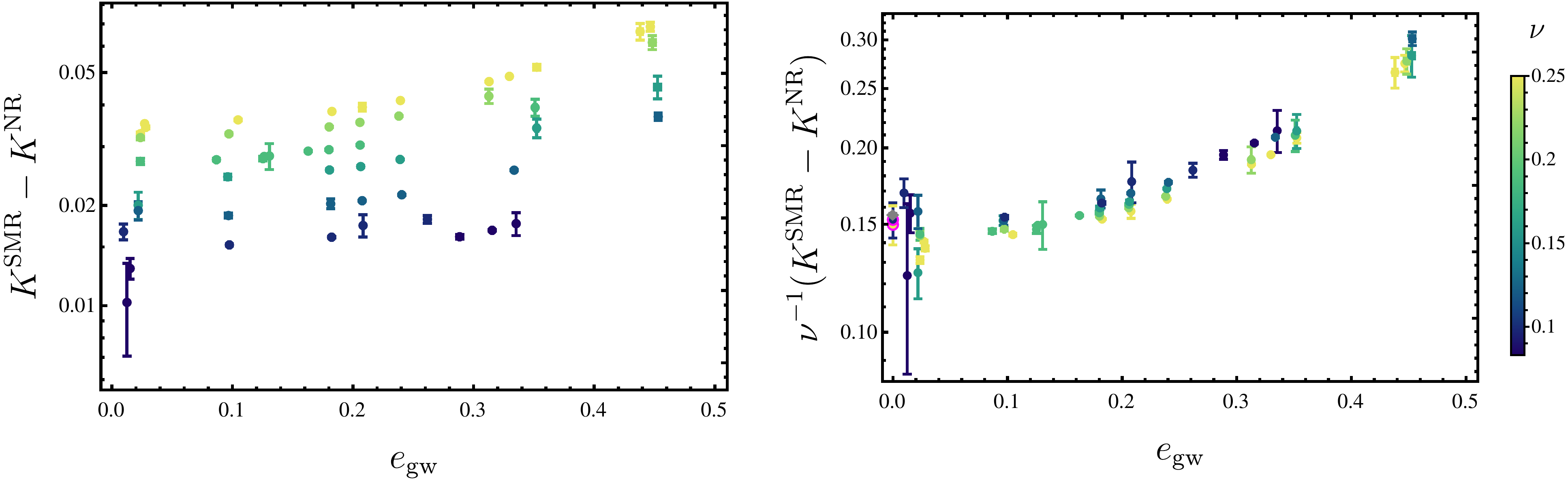}
\end{minipage}
	\caption{
	\textit{Left panels:}	Difference between the SMR and NR fluxes and periastron advance as a function of eccentricity at a reference frequency of $\wref = 0.042$. The energy and angular momentum fluxes (top and mid panels) have been rescaled by the leading order symmetric mass ratio power, $\nu^{-2}$.
	\textit{Right panels:} Same quantity as in the corresponding left plot rescaled by an additional power in symmetric mass ratio. In all panels each point is color-coded by symmetric mass ratio and carries the error bar computed in Fig.~\ref{fig:NRerrorbars}. The red dots in right top and mid panels corresponds to the quasi-circular second order self-force results from \cite{Warburton:2021kwk}. In the right bottom plot the gray dot refers to the SMR prediction for quasi-circular binaries from \cite{LeTiec:2011bk}, and the dots circled by magenta disks correspond to the periastron advance values for the $q=1,1/8$ quasi-circular NR simulations computed in  \cite{LeTiec:2011bk}.
	}
	\label{fig:NRSMRfixedFreqResults}
\end{figure*}

We now perform  a more quantitative comparison of SMR and NR results for the particular reference frequency,  $\wref = 0.042$.  The difference between the SMR and NR fluxes rescaled by the leading order power of symmetric mass ratio as a function of eccentricity are shown in the top and mid left panels of Fig.~\ref{fig:NRSMRfixedFreqResults}, while in the bottom panels the differences for periastron advance are displayed. Each data-point carries the error bar determined through the analysis in Fig.~\ref{fig:NRerrorbars}.  We see  that at 0PA order there is already  good agreement between NR and SMR results, with relative differences typically of the order $\lesssim 10\%$, with the largest discrepancies occurring at equal masses, as expected from a small mass ratio expansion.

Given the visible mass ratio trends in the left panels of Fig.~\ref{fig:NRSMRfixedFreqResults}, we rescale  by another power of symmetric mass ratio to estimate the magnitude of the unknown 1PA contributions. The right panels of Fig.~\ref{fig:NRSMRfixedFreqResults} show that this scaling collapses the three quantities into 1-dimensional curves. These 1-dimensional curves represent the next-to-leading order 1PA contribution to the respective quantity, as a function of eccentricity.  The small residual spread in mass ratio in these curves represents unknown yet higher order terms.  The fact that the right panels of Fig.~\ref{fig:NRSMRfixedFreqResults} collapse to quasi 1-dimensional curves indicates that such $\ge 2$PA contributions are small compared to the 1PA contribution.

Additionally, we have added to the right top and mid panels of Fig.~\ref{fig:NRSMRfixedFreqResults} the second order self-force results for the quasi-circular fluxes from \cite{Warburton:2021kwk}. The agreement is good, but a small shift is noticeable between the quasi-circular results with respect to our eccentric results. This feature may be a consequence due to the fact that the results from~\cite{Warburton:2021kwk} are based on a two-timescale expansion computed from the self-force dynamics, while our fluxes are averaged over a radial period of an evolving SMR waveform. However, a more detailed study is required to determine the source of this small discrepancy, which is within the error bars.

In the case of the periastron advance, when rescaling by an additional power of symmetric mass ratio in the bottom right panel of Fig.~\ref{fig:NRSMRfixedFreqResults}, we include also the quasi-circular 1PA SMR result \cite{Barack:2011ed, vandeMeent:2016hel} and the two quasi-circular non-spinning NR simulations ($q=1,1/8$)  from \cite{LeTiec:2011bk}. The eccentric results  have comparatively large error bars at small eccentricities  because the amplitude of the oscillations in $\wgw$ (from which all quantities are derived) becomes small and more difficult to resolve. A similar shift as in the case of the fluxes is present in the bottom right panel of Fig.~\ref{fig:NRSMRfixedFreqResults} between the quasi-circular results from \cite{LeTiec:2011bk} and the low-eccentricity data-points of our new analysis. We leave for future work the precise determination of such small differences between our results for the fluxes and the periastron advance, and the existing quasi-circular results from the literature.

Finally, we remark that the dependence of the fluxes and periastron advance with eccentricity resembles a functional form as expected from the PN results  \cite{Arun:2007sg, Arun:2009mc,1988NCimB.101..127D}, where for instance the eccentric corrections to the fluxes are of the form  $\sim \frac{1}{(1-e^2)^x}(1+a e^2 + ...)$, where $x$ and $a$ are coefficients to be determined. This suggests that fitting such results as a function of eccentricity and mass ratio could provide some phenomenological expressions for the unknown 1PA SMR terms as a function of eccentricity, mass ratio and frequency, as a similar eccentricity dependence is observed for other frequencies in the inspiral. We leave such a task for future work, as well as the production of new eccentric NR simulations at smaller mass ratios, which may help assess the contributions of the unknown higher order terms in the SMR perturbation theory for eccentric non-spinning binaries.

\section{Conclusions} \label{sec:Conclusions}

We have presented a new set of BBH NR simulations produced with the \texttt{SpEC} code with the objective of exploring the accuracy of the small mass-ratio expansion for eccentric non-spinning binary black holes. In particular, our study aims to extend recent work \cite{vandeMeent:2020xgc} on assessing the accuracy of the SMR theory for non-spinning quasi-circular BBH to non-spinning eccentric BBHs.

The simulations produced in this work cover mass ratios, $q \in [0.1, 1]$, initial eccentricities, $e^0_{\rm gw} \in [0.01,0.7]$, and initial mean anomalies close to apastron, $l_0 \sim \pi$. Each simulation is performed at three different resolutions, and most of them have $\gtrsim 20$ orbits, which makes our dataset the one with the longest eccentric BBH simulations to date.

These simulations are compared to waveforms produced using the gravitational self-force formalism. Using an existing frequency domain Teukolsky code~\cite{vandeMeent:2014raa,vandeMeent:2015lxa,vandeMeent:2016pee}, we have generated eccentric inspirals in a Schwarzschild background that are accurate to leading order in the SMR expansion.

As a first step   towards comparing the NR and SMR results, we adapted the orbit-average method from \cite{Lewis:2016lgx} to extract the radial and azimuthal frequencies, the energy and angular momentum fluxes, and measure the eccentricity from waveforms. We have validated this procedure to extract orbit-averaged frequencies by using the 0PA inspirals, where the geodesic azimuthal and radial frequencies are provided as an outcome of performing such evolutions. We find that the procedure of extracting the frequencies, and eccentricities produces relative differences of $10^{-5}$ in the early inspiral, while the discrepancies increase up to $~\sim 10^{-2}$ close to merger due to a combination of the boundary effects and the rapid increase of the frequencies, which is a clear limitation of the procedure. Thus, we restrict this study to the inspiral part of the waveform, and leave for future work an improvement of the extraction procedure to describe more faithfully the transition from inspiral to plunge of the signal.

We investigated eccentricity $e_{\worb}$ defined from the
  orbital frequency and eccentricity $e_{\wgw}$ defined from the
  gravitational wave (2,2) mode, and found them to systematically
  differ.  Using PN theory we have derived relations between
different definitions of eccentricity. The instantaneous orbital and
(2,2)-mode frequency are not related by the simple factor 2 for
eccentric binaries, as is the case of the orbit-averaged frequencies,
and thus, we have provided PN-accurate expressions relating both,
which produce relative differences of $\sim 10^{-2}$ when tested on NR
simulations. Furthermore, we have provided PN-accurate expressions
relating $e_{\worb}$,   $e_{\wgw}$  and the temporal eccentricity, $e_t$. We
show that in the Newtonian limit $e_{\wgw}
\sim 3 e_t /4$, so that $e_{\wgw}$ does not have the correct
  Newtonian limit.  In Eq.~(\ref{eq:eqEccDef}), we propose
a new eccentricity definition $e_{\rm gw}$ based on the (2,2)-mode
frequency, which reduces to $e_t$ in the Newtonian limit.

  Comparisons between NR and SMR require a map which associates a SMR inspiral with the instantaneous state of an eccentric NR inspiral.   
  We investigated several proposals in the literature for variables that identify the same inspiral in the NR simulations and SMR evolutions. We find that some choices used in the literature lead to the NR simulations lying outside the range spanned by the geodesic results, hampering comparisons. We propose to use as variables the orbit-averaged azimuthal frequency, $\wgwavg$, and eccentricity $e_{\rm gw}$,  measured both from the instantaneous frequency of the (2,2)-mode, which do not suffer from this limitation.

Moving to the comparison between NR and SMR results, we have focused on the energy and angular momentum fluxes, as well as the periastron advance. Overall, we find good agreement between the NR and SMR values, with relative differences typically $\lesssim 10 \%$, and no particular degradation with increasing eccentricity. 

We assess the contributions coming from the unknown higher order term in the PA expansion (1PA) by considering the difference between the NR and SMR fluxes and periastron advance. After rescaling by the symmetric mass-ratio cubed, we find that the differences collapse to one dimensional curves as a function of eccentricity with very small spread in mass ratio, see Fig.~\ref{fig:NRSMRfixedFreqResults}. This behavior indicates that the next order term in the SMR expansion (2PA) has a very small contribution compared to the 1PA term. Furthermore, we compare these differences for the fluxes and the periastron against available results in the literature for quasi-circular binaries from \cite{Warburton:2021kwk,LeTiec:2011bk}, and find that the results are consistent with our findings, except for small shifts which are within the error bars. We leave for future work the precise determination of the origin of this small feature.

The eccentricity dependence of the fluxes and periastron advance rescaled by symmetric mass ratio also suggests a functional form similar to the one predicted by the known PN results  \cite{Arun:2007sg, Arun:2009mc,1988NCimB.101..127D}. An interesting extension of the work presented here would be the modelling of these differences between the adiabatic SMR inspirals and the NR simulations, by fitting them as a function of mass ratio, eccentricity and orbit-averaged frequency $\{q, e_{\rm gw},\wgwavg\}$, and provide some phenomenological expressions which can be used to compute the unknown 1PA term for the fluxes and periastron advance. Another possible future direction is to focus on comparing the phasing between NR and SMR, and extend previous studies for quasi-circular binaries \cite{vandeMeent:2020xgc} to the eccentric case.

Future work will also include extending our set of simulations to higher mass ratios, and to gradually incorporate spins. Other applications of the simulations will include the calculation of the redshift factor \cite{Detweiler:2008ft}, extending current studies on quasi-circular binaries \cite{Albalat:2022lfz} to the eccentric case. Finally, these simulations will also be of paramount relevance to assess the accuracy of the currently existing inspiral-merger-ringdown eccentric waveform models \cite{Huerta:2017kez, Hinder:2017sxy, Liu:2021pkr,  Nagar:2021xnh, Ramos-Buades:2021adz}.

\section*{Acknowledgments}
\label{acknowledgements}

It is a pleasure to thank Arif Shaik and Vijay Varma for helpful discussions about eccentricity definitions, and Aaron Zimmerman for useful comments on the manuscript.
We also thank the authors of \cite{Warburton:2021kwk,LeTiec:2011bk} for providing the reference data points used in Fig.~\ref{fig:NRSMRfixedFreqResults}.
The NR computational work for this manuscript was carried out on the computer cluster \texttt{Minerva} at the Max Planck Institute for Gravitational Physics in Potsdam.  This work was supported by the Sherman Fairchild Foundation and NSF Grants No. PHY1912081, No. PHY-2207342, and No. OAC-1931280 at Cornell.
Hannes R. R{\"u}ter acknowledges support from the Funda\c c\~ao para a Ci\^encia e Tecnologia (FCT) within the projects UID/04564/2021, UIDB/04564/2020, UIDP/04564/2020 and EXPL/FIS-AST/0735/2021.


\appendix

\section{Numerical relativity initial conditions}\label{sec:AppendixIC}

The quasi-equilibrium, extended conformal thin sandwich initial data
used by \texttt{SpEC} requires choice of two sets of input parameters.
The first set consists of masses and spins of the two black holes.
The second set determines the orbital configuration of the two BHs.
This second set consists of an initial separation $D_0/M_0$, an
initial instantaneous orbital frequency $M_0\Omega_0$ and the initial
instantaneous radial velocity $a_0=\dot{r}/r$ (see
\cite{Pfeiffer:2007yz,Buonanno:2010yk,Ossokine:2015yla} for details).
Our task is to determine this second set of initial-data parameters
such that the subsequent evolution has an eccentricity close to a
certain desired value $e_0$ and an inspiral of reasonable length
(20-50 orbits).

Let us first point out three considerations that will influence our
procedure: First, as discussed in Sec.~\ref{subsec:NRdataset} the
present radial map used in \texttt{SpEC} cannot accommodate that the
distance between the two black holes increases by more than a factor
1.5.  We will avoid this problem by starting NR simulation near
apastron.\footnote{Very recently a new radial map has been developed
and implemented in \texttt{SpEC}, which avoids these restrictions.}
Second, there are previous results on how the tangential momentum in
an eccentric binary varies with the eccentricity.  Specifically, we
will utilize the correction factor of the tangential momentum
\cite{Ramos-Buades:2018azo}
\begin{equation}
  \lambda^0_t (r,e,\nu, \text{sign} )
  =  1+ \text{sign} \times \frac{e}{2} \times \left[ 1 - \frac{1}{r}(2+\nu)\right],
\label{eq:eq08}
\end{equation}
where $r$ is the orbital separation and $ \text{sign} = \pm 1 $ is the
sign of the correction \cite{Ramos-Buades:2018azo}. While this
correction has been derived in the low eccentricity limit, it has been
shown \cite{Ramos-Buades:2019uvh} to be useful to determine the
initial parameters in eccentric moving puncture simulations.  We
average the correction with both signs to arrive at
\begin{equation}
  \bar{\lambda}^0_t (r,e,\nu )
  =  \frac{1}{2} \times\left[   \lambda^0_t (r,e,\nu,-1)
    +  \frac{1}{ \lambda^0_t (r,e,\nu,+1)}  \right],
    \label{eq:eq9}
\end{equation}
as in Eq.~(2.3) of \cite{Ramos-Buades:2019uvh}.  The third
consideration concerns the choice of coordinates: our \texttt{SpEC}
simulations start from superposed harmonic Kerr (SHK) data
\cite{Varma:2018sqd}, whereas Eq.~(\ref{eq:eq08}) was derived in
Arnowitt-Desner-Misner transverse-traceless (ADMTT) coordinates.
Therefore, we will also employ a coordinate transformation from ADMTT
coordinates to harmonic coordinates.

Overall, we proceed as follows:
\begin{enumerate}
\item Choose mass-ratio $q\le 1$, and a desired eccentricity
  $e_0$. Set spins $\bm{\chi}_i=0$, masses $m_1 = 1/(1+q)$,
  $m_2=q/(1+q)$ (so that $M_0=m_1+m_2=1$), and
  $\nu=m_1m_2/M_0^2=q/(1+q)^2$.

\item Choose a tentative initial separation as the apastron distance
  of a Newtonian binary with periastron distance of $r_p=9M_0$,
  i.e. $\tilde D_0=(1+e_0)(1-e_0)^{-1}r_p$.  If a PN
  evolution with the same parameters indicates that the inspiral may be too short, increase $\tilde D_0$.
    \item Compute 3.5PN quasi-circular estimates for the tangential and
      radial momenta, $p^0_t,p^0_r$ in ADMTT coordinates using
      Eqs.~(A2) and (2.16) in \cite{Ramos-Buades:2019uvh}.
    \item Calculate the correction factor $\bar\lambda_t$ using Eq.~\eqref{eq:eq9}.
    \item Construct the ADMTT position and momentum vectors in Cartesian coordinates,
    \begin{align}
      \bm{x}^{\text{ADMTT}} &= (0,\tilde D_0,0), \\
      \bm{p}^{\text{ADMTT}} &= (p^0_r, \bar{\lambda}^0_t p^0_t, 0).
    \end{align}
    Here, we placed the black holes on the y-axis.  Note that
    $\bar\lambda_t\le 1$, so that the tangential momentum is reduced,
    consistent with our goal to start at apastron.
    \item Apply the transformation from ADM to harmonic coordinates
      \cite{Damour:2000ni} to obtain the position and velocity vector
      in harmonic coordinates,
      \begin{align}
      \bm{x}^{\text{H}}  &= \bm{Y}\left[ \bm{x}^{\text{ADMTT}}, \bm{p}^{\text{ADMTT}}\right], \\
      \bm{v}^{\text{H}}  &= \bm{V}\left[ \bm{x}^{\text{ADMTT}}, \bm{p}^{\text{ADMTT}}\right],
    \end{align}
      where $\bm{Y}$ and $\bm{V}$ are operators mapping the ADM
      coordinates to harmonic coordinates expanded up to 3PN order
      \cite{Damour:2000ni}. (Note that the expressions in
      \cite{Damour:2000ni} are restricted to non-spinning binaries.)
    \item Read of the initial data parameters from the position
      and velocity vectors in harmonic coordinates,
      \begin{align}
        \label{eq:eq12}
        D_0 &= |\bm{x}^\text{H}|, \\
        \initNRw  &=  \frac{| \bm{x}^{\text{H}} \times \bm{v}^{\text{H}} |}{D_0^2},\\
        a_0 &= \frac{\bm{v}^\text{H}\cdot \bm{x}^{\text{H}}}{D_0^2},
      \end{align}
      where Euclidean vector operations are used.
\end{enumerate}

\begin{figure}[tbp!]
\includegraphics[width=\columnwidth]{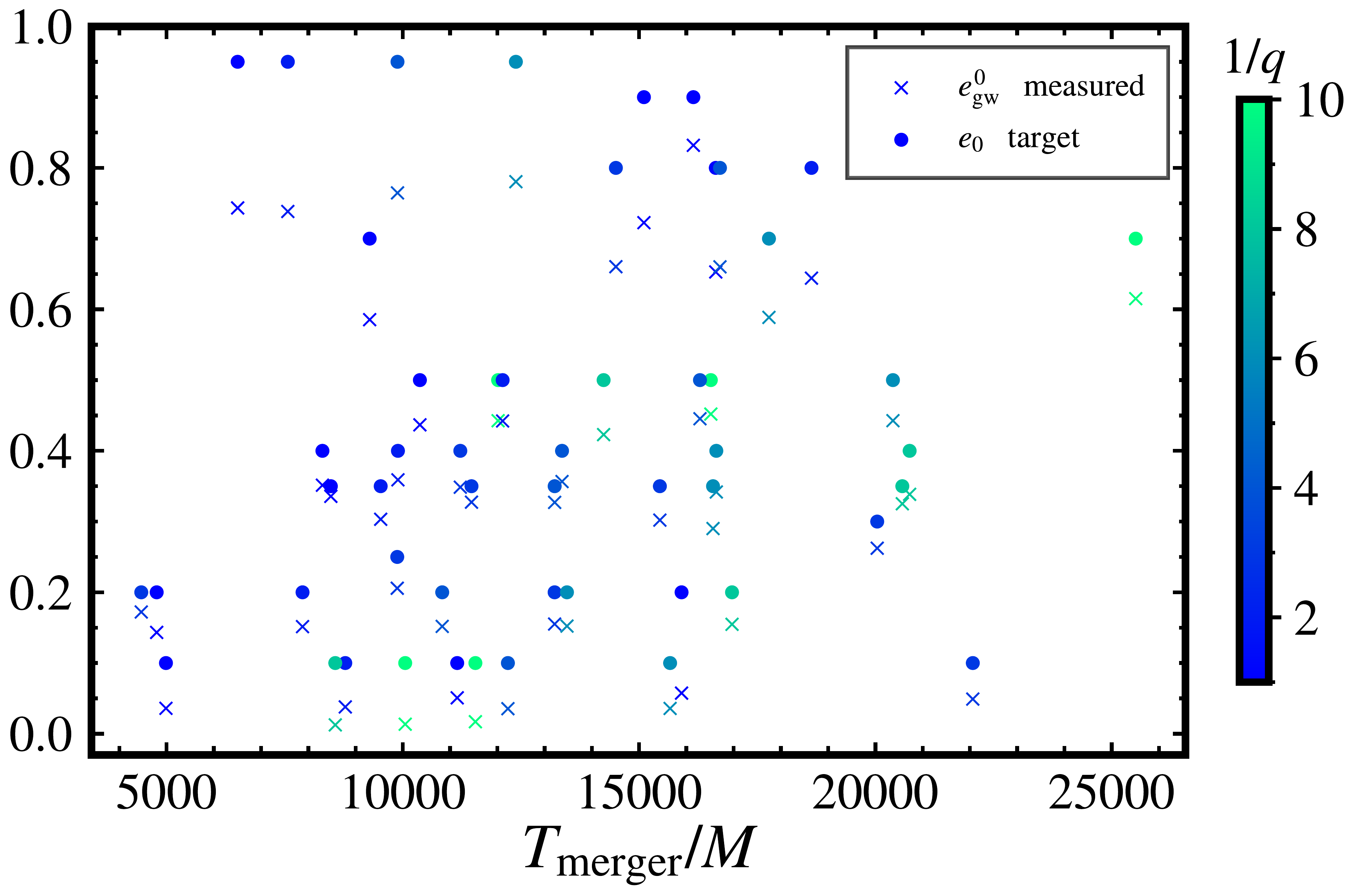}
\caption{\label{fig:EccMeasurement}
  Initial eccentricity (dots), $e^0_{\rm gw}$, as defined in
  Eq.~\eqref{eq:eqEccDef}, measured from the NR
  simulations in Table \ref{tab:tabNR}, and initial eccentricity
  (crosses), $e_0$, specified in Eq.~\eqref{eq:eq9}, as a function
  of the merger time of the simulations. Each simulation is
  color-coded according to its inverse mass ratio.}
\end{figure}

Figure~\ref{fig:EccMeasurement} compares the target eccentricity
$e_0$ with the actual eccentricity $e_{\rm gw}^0$ achieved near the
start of each simulation.  There is an offset between these
eccentricities.  We note that specially for high eccentricities the
use of the correction factor is not accurate due to the fact that it
is an expression derived in the low eccentricity limit. Furthermore,
we attribute the larger differences between our target and measured
initial eccentricities as compared to other studies like
\cite{Ramos-Buades:2019uvh} due to the assumptions on the
identification we made between harmonic and superposed harmonic
coordinates, and inaccuracies in the PN expressions for the eccentric
corrections being amplified due to the transformation from the ADM to
the harmonic gauge.

The calculation of the initial parameters presented in this section is
useful for placing points in the eccentric parameter space with a
limited accuracy.   In the future we plan to adopt an
iterative procedure to specify the desired initial eccentricity and
mean anomaly as done in \cite{Islam:2021mha, Fischer:2022a}, to
accurately and efficiently populate the eccentric parameter space.

\section{Numerical relativity waveform quality} \label{sec:AppendixNRconvergence}

In this appendix we assess the accuracy of the NR waveforms listed in Table \ref{tab:tabNR}. For each simulation \texttt{SpEC} employs multiple subdomains. The shape, size and number of subdomains is dynamically varied during the simulations according to the spectral adaptive mesh refinement (AMR) procedure  \cite{Szilagyi:2014fna,Lovelace:2010ne}. The accuracy of the simulations is controlled by a tolerance parameter which determines when AMR should add or remove grid points  within a given subdomain, and when a subdomain should be split into two, or when two neighboring subdomains should be combined into one. As a consequence, it is difficult to obtain strict convergence as a function of the AMR tolerance parameter. Convergence may fail, for instance, due to two identical simulation having different AMR tolerances in a particular subdomain modifying the number of grid points in it, or different subdomain boundaries in a particular time. Notwithstanding these issues, most simulations in the  \texttt{SXS} catalog show convergence with the AMR tolerance \cite{Boyle:2019kee}.

In this work we have run each simulation at three different AMR tolerances, henceforth called \textit{different resolutions} for brevity.
This appendix extends the error analysis of our main results with calculations of the mismatch between waveforms obtained at the highest and second highest resolutions.

\begin{figure}[tbp!]
    \centering
\includegraphics[width=\columnwidth]{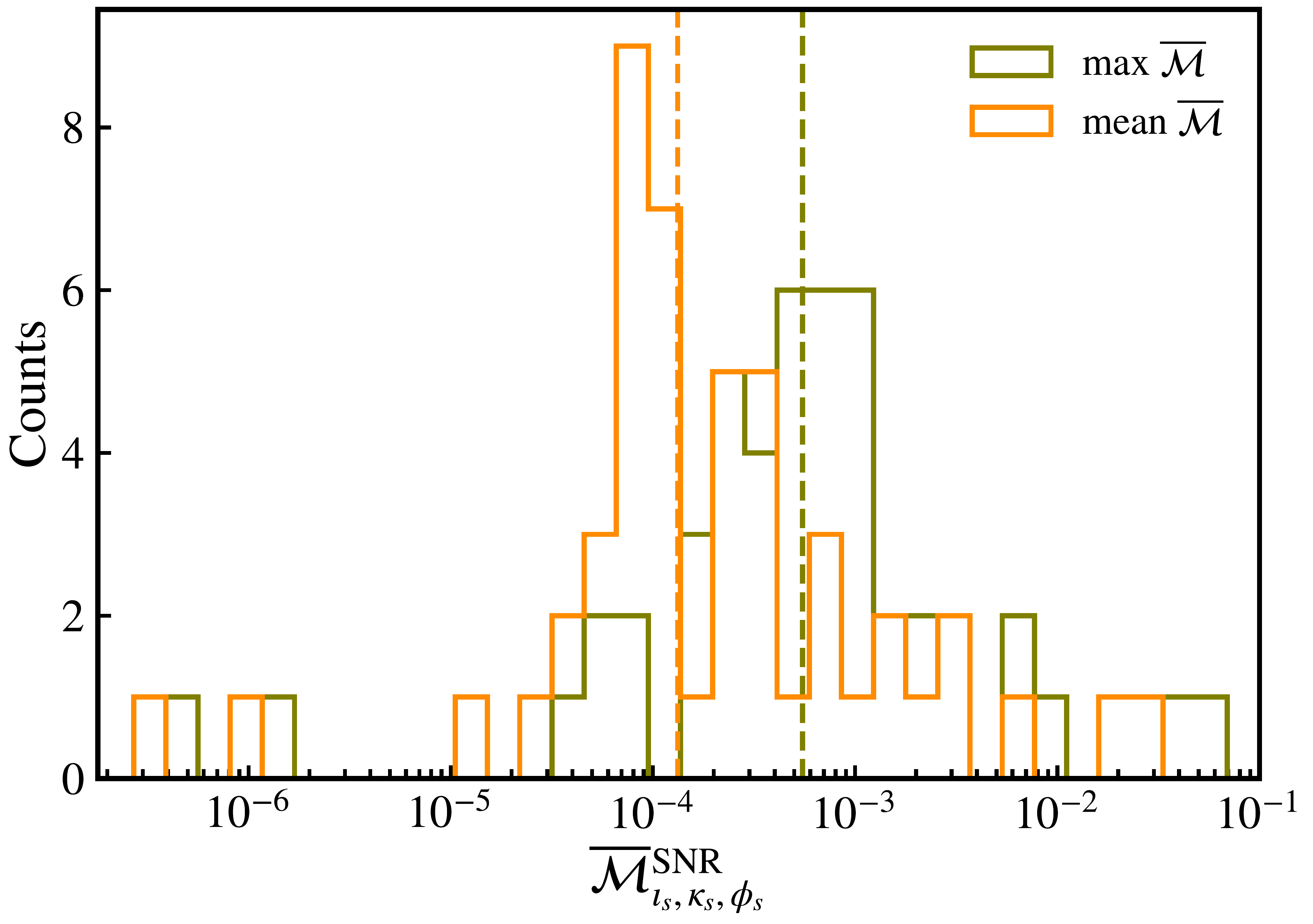}
    \caption{Histograms of the SNR-weighted mismatch between the two highest resolutions for each simulation in Table \ref{tab:tabNR}. The orange and green distributions correspond to the mean and maximum mismatch over the total mass range considered $M = [20,200]M_\odot$. The vertical dashed lines correspond to the median values of the distributions.}
    \label{fig:NRmismatch}
\end{figure}

Following Ossokine et al 2020 \cite{Ossokine:2020kjp}, we compute the SNR-weighted mismatch between waveforms computed from the highest and the next-to-highest resolutions. The mismatches are computed for binary masses $20M_\odot\le M\le 200M_\odot$, and using as a Power Spectral Density (PSD), the  Advanced LIGO's zero-detuned high-power design sensitivity curve \cite{Barsotti:2018}. When both waveforms are in band, we use $f_{\text{min}}=10$Hz and $f_{\text{max}}=2048$Hz, as the lower and upper bounds of the integral. For waveforms where this is not the case, we set $f_{\text{min}} = 1.05 f_{\text{start}}$, where $f_{\text{start}}$ is the starting frequency of the NR waveform. To represent dependence on $M$, we compute the mean and the maximum over $M$.  The results of the mismatch calculation are shown in Fig.~\ref{fig:NRmismatch}.
The vertical dashed lines denote the median values of each distribution. We note that the median value of the maximum mismatch is below $10^{-3}$, while for the mean mismatch is $\sim 10^{-4}$.  Three simulations (SXS:BBH:2517, SXS:BBH:2525, SXS:BBH:2564) have maximum mismatches above $1\%$, $\bar{\mathcal{M}}^{\text{SNR}}_{\max}={0.011,0.065,0.041}$, respectively. The highest mismatch occurs for SXS:BBH:2525
  which is both the shortest evolution in our dataset (making it more prone to systematics due to the ringdown transition in \texttt{SpEC} \cite{Scheel:2008rj,
  Hemberger:2012jz}), and which was also the first simulation produced in our dataset, so it does not take into account some improvements in \texttt{SpEC}, which have been introduced during this project (see Sec.~\ref{sec:NRsimulations} for details).
Overall, the mismatches are comparable to the ones obtained in the \texttt{SXS} catalog for quasi-circular binaries \cite{Boyle:2019kee} (see Fig.~9 there, but note that \cite{Boyle:2019kee} uses a flat PSD). This indicates that \texttt{SpEC} is capable to perform simulation of eccentric BBH with a numerical error comparable to the quasi-circular case.

\section{Calculation of $e_{\wgw}$ in PN expansions}\label{sec:AppendixeOm22}
In this Appendix we use PN theory to investigate the relation among $e_{\wgw}$, $e_{\worb}$ and the post-Newtonian $e_t$~\cite{1985AIHS...43..107D}. In the following, we set the total mass, $M=1$, to ease the notation.

\subsection{Relation $e_{\wgw}-e_{\worb}$}\label{sec:AppendixeOm22Orb}
Section~\ref{sec:Eccdef} showed that the differences between $e_{\wgw}$ and $e_{\worb}$ can be explained within PN theory.  This appendix derives the relations used there at 1PN using harmonic coordinates. As a first step, we calculate  $e_{\wgw}$ from $h_{22}$ at the 1PN order\cite{Mishra:2015bqa},
\begin{align}
\label{eq:eqD01}
h_{22} =& 4 \nu \sqrt{\frac{\pi}{5}}\left[ \hat{H}^{\text{0PN}}_{22}+\gamma \hat{H}^{\text{1PN}}_{22} \right] e^{-2 i \phi},\\
\hat{H}^{\text{0PN}}_{22} =& \frac{1}{r} - \dot{r}^2 + 2 i r \dot{r} \dot{\phi} +r^2 \dot{\phi}^2,\\ \nonumber
\hat{H}^{\text{1PN}}_{22} =& \left(\frac{9}{14}-\frac{27 \nu }{14}\right) r^4 \dot{\phi}^4+i \dot{r} \left[\left(\frac{45 \nu }{7}+\frac{25}{21}\right) \dot{\phi}+\left(\frac{9}{7}-\frac{27 \nu }{7}\right) r^3   \dot{\phi}^3\right] \\ \nonumber
& +\frac{1}{r^2}\left(\frac{\nu }{2}-5\right)+\left(\frac{26 \nu }{7}+\frac{11}{42}\right) r \dot{\phi}^2+i \left(\frac{9}{7}-\frac{27 \nu }{7}\right) r \dot{r}^3 \dot{\phi} \\
&+\frac{\dot{r}^2}{r} \left(-\frac{16 \nu }{7}-\frac{15}{14}\right)  +\left(\frac{27 \nu }{14}-\frac{9}{14}\right) \dot{r}^4,
\end{align}
where $\gamma = \frac{1}{c^2}$ is the PN order bookkeeping parameter,  $i$ is the imaginary unit, $r$ is the radial separation, $\phi$ is the orbital phase and the overdot represents a time derivative.

Taking the complex argument of Eq.~(\ref{eq:eqD01}) and expanding to 1PN order yields

 \begin{align}
  \nonumber
  \phi_{22}^{\text{1PN}} & =-2\phi + \delta \\    \label{eq:eqD03}
  &+\gamma r\dot{r}\dot{\phi}
  \frac{
    41 r^3 \dot{\phi }^2+47 r \dot{r}^2+ 235
    -2\nu\left(51  r^3 \dot{\phi}^2
     +60  r \dot{r}^2-57\right) }
  {21 (C_1^2+C^2_2)}
\end{align}
with $C_1=r^3\dot{\phi}^2-r\dot{r}^2+1$, $C_2=2r^2\dot{r}\dot{\phi}$, $\tan\delta=C_2/C_1$.

    The   frequency $\wgw$ entering the definition of  $e_{\wgw}$  can be expressed   as
\begin{equation}
\wgw \equiv \frac{d \phi_{22}}{dt}= \frac{\partial \phi_{22}}{\partial r} \dot{r} + \frac{\partial \phi_{22}}{\partial \dot{r}} \ddot{r} +\frac{\partial \phi_{22}}{\partial \phi} \dot{\phi} + \frac{\partial \phi_{22}}{\partial \dot{\phi}} \ddot{\phi}.
\label{eq:eqD04}
\end{equation}
Expanding Eq.~\eqref{eq:eqD04} at 1PN order, we obtain
\begin{equation}\label{eq:eqD05}
  \wgw^{\text{1PN}} = \mathcal{F}(\nu, r, \dot{r}, \ddot{r}, \dot{\phi}, \ddot{\phi}) = \wgw^0 + \gamma \wgw^1,
\end{equation}
where
\begin{widetext}
\begin{align}
  \wgw^0  =& \frac{-2}{C_1^2+C_2^2}\left[
    \dot{\phi}
    - \left(\ddot{r}r + (4 + \ddot{r}r^2)\dot{r}^2 - 2r\dot{r}^4\right)r \dot{\phi}
 + r^3(2 - \ddot{r}r^2 + 3r\dot{r}^2)\dot{\phi}^3 + r^6\dot{\phi}^5
 + \ddot{\phi}r^2\dot{r}(-1 + r\dot{r}^2 + r^3\dot{\phi}^2)\right],\\
\nonumber
  \wgw^1 =&\frac{-1}{21 \left(C_1^2+C_2^2\right)^2}\bigg[ (1 - r\dot{r}^2)(\dot{r}^2(-235 - 114\nu + 7r\dot{r}^2(-47 + 18\nu)) + \ddot{r}r(-235  - 114\nu + r\dot{r}^2(18(-47 + \nu) + r \dot{r}^2(-47 + 120\nu))))\dot{\phi}  \\
\nonumber
    &  + r^3(2\dot{r}^2(388 + r\dot{r}^2(875 + r\dot{r}^2(53 - 138\nu) - 84\nu) + 432\nu)
    + \ddot{r}r(-7(73 + 18\nu) + r\dot{r}^2(r\dot{r}^2(29 - 66\nu) + 2(53 + 576\nu))))\dot{\phi}^3 \\
\nonumber
    & + r^6(\dot{r}^2(1093 + 4r\dot{r}^2(47 - 120\nu) + 774\nu) + \ddot{r}r(-317
 + 90\nu + r\dot{r}^2(-59 + 156\nu)))\dot{\phi}^5 + r^9(\ddot{r}r -
 2\dot{r}^2)(-41  + 102\nu)\dot{\phi}^7 \\
 \nonumber
 & + \ddot{\phi}r\dot{r}((-1 + r\dot{r}^2)^2(-235 - 114\nu + r\dot{r}^2(-47 + 120\nu))
 + r^3(347 + 534\nu + r\dot{r}^2(810 - 624\nu + r\dot{r}^2(-29 + 66\nu)))\dot{\phi}^2\\
 & + r^6(r\dot{r}^2(59 - 156\nu) + 7(89 + 78\nu))\dot{\phi}^4  + r^9(41 - 102\nu)\dot{\phi}^6) \bigg].
\end{align}
\end{widetext}
This result is used in the main text in Eq.~\eqref{eq:eq110}. 

At the turning points apastron and periastron, $\dot{r}=0$ and $\ddot{\phi}=0$, and Eq.~(\ref{eq:eqD05}) simplifies to
\begin{align}
  \nonumber
  \wgw^{\text{1PN}\,a,p} =& \frac{2\dot{\phi}(1 -\ddot{r}r^2 + r^3\dot{\phi}^2)}{1+r^3 \dot{\phi}^2}\\
\label{eq:eqD06}
&+ \gamma  \frac{ \ddot{r} r\dot{\phi}\left[ 235 + 114\nu + (41 - 102\nu)r^3\dot{\phi}^2  \right]
  }{21(1 + r^3\dot{\phi}^2)^2}
\end{align}

  At apastron, $\ddot{r}<0$ whereas at periastron $\ddot{r}>0$. Substituting Eq.~(\ref{eq:eqD06}) into Eq.~(\ref{eq:eq7}), and replacing $\dot{\phi}=\worb$, one obtains
\begin{equation}
e_{\wgw} = \frac{\sqrt{\wgwp(r_p,\worbp, \ddot{r}_p , \nu, \gamma)} - \sqrt{\wgwa(r_a,\worba, \ddot{r}_a , \nu, \gamma)}}{\sqrt{\wgwp(r_p,\worbp, \ddot{r}_p , \nu, \gamma)} +\sqrt{\wgwa(r_a,\worba, \ddot{r}_a , \nu, \gamma)}}.
\label{eq:eqD07}
\end{equation}
where $\{r_{a,p}, \worb^{a,p}, \ddot{r}_{a,p} \}$ indicate the corresponding quantities at apastron and periastron, respectively.
Expanding Eq.~\eqref{eq:eqD07} to 1PN order yields
\begin{equation}
  e^\text{1PN}_{\wgw}   \equiv  \mathcal{G}(\nu, r_{a,p}, \worb^{a,p}, \ddot{r}_{a,p}) = e^0_{\wgw}+\gamma e^1_{\wgw},
\end{equation}
where
\begin{align}
 e^0_{\wgw} =& \frac{\alpha_-}{\alpha_+} ,\\
\nonumber
 e^1_{\wgw} =& \frac{\sqrt{\Delta(r_a,\worba,\ddot{r}_a) \worbp \left(1+r_p^2 (\ddot{r}_p +r_p (\worbp)^2)\right) }}{42 (1+r_p^3 (\worbp)^2)^{1/2}\alpha_1^2} \\
\nonumber
 & \times \left[  \frac{\ddot{r}_a r_a(-235 - 114\nu + r_a^3(-41 + 102\nu)(\worba)^2)}{(1 + r_a^3(\worba)^2)(-1 + r_a^2(\ddot{r}_a - r_a(\worba)^2))} \right. \\
& \quad+\left. \frac{\ddot{r}_pr_p(235 + 114\nu + r_p^3(41 - 102\nu)(\worbp)^2)}{(1 + r_p^3(\worbp)^2)(1 + r_p^2(\ddot{r}_p + r_p(\worbp)^2))}  \right],
\end{align}
with
\begin{align}
\label{eq:eqD08}
  &\alpha_\pm  =\Delta(r_p,\worbp,-\ddot{r}_p)^{1/2}\pm\Delta(r_a,\worba,\ddot{r}_a)^{1/2}, \\
 &\Delta (r,\worb,\ddot{r})  = 2\worb \left(1 - \frac{\ddot{r} r^2}{1+r^3 \worb^2}\right).
\end{align}
This result is used in the main text in Eq.~\eqref{eq:eq111}.

The expressions derived above can be useful to estimate $\wgw$, or the eccentricity $e_{\wgw}$, for NR simulations, where the trajectories are output in Cartesian or polar coordinates\footnote{We note that the expressions derived above correspond to harmonic coordinates \cite{Mishra:2015bqa}, while NR coordinates typically do not correspond to these ones. Thus, one should transform the harmonic coordinates to the ones used by the corresponding NR code. However, in practice we find that for our \texttt{SpEC} simulations not performing such a transformation still provides accurate results.}, and are typically cleaner quantities than the frequencies of the extracted waveform modes, especially for finite difference codes \cite{Purrer:2012wy}. Another application of Eq.~\eqref{eq:eqD07} is for eccentricity reduction/control purposes, where short evolutions are done to iteratively converge to the desired value of eccentricity. In these methods \cite{Pfeiffer:2007yz,Buonanno:2010yk,Purrer:2012wy,Ramos-Buades:2018azo} one typically chooses a trajectory-based definition of eccentricity instead of a waveform-based one due to the extra computational cost, which involves the evolution of the gravitational radiation reaching the extraction radii. However, with the expressions provided in Eq.~\eqref{eq:eqD07}, one can obtain an approximation of $e_{\wgw}$ from the coordinates.

\subsection{Relation $e_{\worb}-e_t$}\label{sec:AppendixeOmOrbet}
Most eccentric waveform models for compact binaries use PN theory to describe the inspiral regime and/or their initial    parameters \cite{Memmesheimer:2004cv,Yunes:2009yz,Huerta:2014eca,Loutrel:2017fgu,Klein:2018ybm,Tiwari:2020hsu,Moore:2018kvz,Moore:2019xkm,Tanay:2019knc,Liu:2021pkr, Placidi:2021rkh, Ramos-Buades:2021adz}. A commonly used description of eccentric orbits is the quasi-Keplerian parameterization \cite{Memmesheimer:2004cv} where three different eccentricity parameters $e_t$, $e_r$ and $e_\phi$ describe the orbit \cite{1985AIHS...43..107D}. These three eccentricities are not independent from each other, and they are all related at a given PN order.
Eccentric PN waveform models typically use the temporal eccentricity, $e_t$, as the eccentricity parameter. In the following, PN-accurate expressions between the eccentricity defined from the orbital frequency, $e_{\worb}$, and $e_t$, are computed.

In order to perform this calculation we use the 3PN expression for $\worb$, which can be found in Appendix A of \cite{Hinder:2008kv}. The calculation of $e_{\worb}$ requires the values of the orbital frequency at periastron and apastron, which correspond to values of the eccentric anomaly of $u=0$ and $u=\pi$, respectively. Thus, at the turning points, the orbital frequency can be expressed as
\begin{equation}
\worb^{a,p} = x^{3/2} \left( \worb^{\text{0PN}} + \gamma \worb^{\text{1PN}} + \gamma^2 \worb^{\text{2PN}} + \gamma^3 \worb^{\text{3PN}} \right),
\end{equation}
where $\gamma$ is the PN bookkeeping parameter.
Using the abbreviation $\varepsilon\equiv\sqrt{1-e_t^2}$, the contributions at different PN order can be written as
\begin{widetext}
\begin{align}
\worb^{\text{0PN}} =& \frac{1}{(1 \pm e_t)^2}\varepsilon, \\
\worb^{\text{1PN}} =& \pm \frac{(\nu -4)  e_tx}{\left(1 \pm e_t\right){}^2 \varepsilon},\\
\worb^{\text{2PN}} =&  \frac{\pm x^2}{12 \varepsilon^3 \left(1 \pm e_t\right){}^2} \left[\pm 9 e_t^2 + \left(-5 \nu ^2+35 \nu -48\right) e_t^3   +e_t \left(-\nu ^2 +\nu  \left(-41+72 \varepsilon\right)-180 \varepsilon+24\right) \mp 18 (-5 +2 \nu) \left(-1+\varepsilon\right) \right], \\
\worb^{\text{3PN}} =
& \frac{\mp x^3}{13440 \varepsilon^5 (1\pm e_t)^2}
\bigg[-560 \nu  (3 \nu ^2-59 \nu -36) e_t^5 \mp 70 \left(960 \nu ^2+(123 \pi ^2-10880) \nu +2880\right) (\varepsilon-1)
  \nonumber \\
  & \quad\qquad\qquad\qquad
  \mp 1680 \left(2 \nu ^3-27 \nu ^2-29 \nu +12\right) e_t^4+560
  e_t^3 \left(\nu ^2 (288 \varepsilon-334)+\nu  (389-852 \varepsilon) +960 \varepsilon+2 \nu ^3-936\right)
\nonumber \\
& \quad\qquad\qquad\qquad
  \mp 3 e_t^2 \left(1120 \nu ^2 (46 \varepsilon-85)+\nu
  \left(-239680 \varepsilon-7175 \pi ^2  +584944\right)+2240 (150 \varepsilon-227)+1120 \nu ^3\right)
  \nonumber \\
  & \quad\qquad\qquad\qquad
  +4 e_t \left(140 \nu ^2 \left(432 \varepsilon-421\right)    +\nu  \left(4305 \pi ^2
   \varepsilon-555520 \varepsilon+130796\right)+6720 \left(55 \varepsilon-23\right)+140 \nu ^3\right)\bigg],
 \label{eq:eqD091}
\end{align}
where the upper sign corresponds to apastron and the lower sign corresponds to periastron.  To derive Eq.~\eqref{eq:eqD091} we assumed that the value of $x$ is the same at apastron and periastron as it corresponds to an orbit-averaged frequency, which is evolved using the radiation reaction equations in an adiabatic evolution. This approximation may not be accurately fulfilled when post-adiabatic effects become more relevant as in the case of the binary close to merger.
Substituting  Eq.~\eqref{eq:eqD091} into Eq.~(\ref{eq:e-MoraWill})
and PN-expanding the result to 3PN order, one obtains
\begin{align}
\label{eq:eqD093}
  e_{\worb} =& e^{\text{0PN}}_{\worb} + \gamma e^{\text{1PN}}_{\worb}  + \gamma^2 e^{\text{2PN}}_{\worb} + \gamma^3 e^{\text{3PN}}_{\worb},
\intertext{where}
e^{\text{0PN}}_{\worb} =&  e_t, \\
e^{\text{1PN}}_{\worb} =&  \frac{x}{2} (4- \nu) e_t, \\
e^{\text{2PN}}_{\worb} =& \frac{x^2 e_t}{24 (1-e_t^2)}
\left[12 \left(-2+ 15 \varepsilon -4 e_t^2\right)
  +\nu  \left(13 e_t^2-72 \varepsilon+41\right) +\nu ^2 (1-e_t^2)
\right],\\
e^{\text{3PN}}_{\worb} =& \frac{x^3 e_t }{24  \left(1-e_t^2\right)^{2}} \Bigg[
24 \left(-17  +30 \varepsilon +9 \left(1-e_t^2\right) + 10 \varepsilon^3\right)
+ \nu  \left(\frac{5832}{35} +\left(\frac{123 \pi ^2}{8}-1708\right) \varepsilon
+\frac{9}{2} \left(1-e_t^2\right)-42\varepsilon^3+ 62\left(1-e_t^2\right)^2
	\right)
\nonumber\\
&\qquad \qquad\quad
+\nu ^2 \left(
	-258
	+252 \varepsilon
	+73 \left(1-e_t^2\right)
	-72 \varepsilon^3
	+\frac{21}{2} \left(1-e_t^2\right)^2
	\right)
+\frac{1}{2} \nu ^3 \left(1-e_t^2\right)^2
\Bigg].
\end{align}
\end{widetext}
We note that in the derivation of Eq.~\eqref{eq:eqD093} only the instantaneous contributions to the orbital frequency up to 3PN order have been used, while tail contributions or spin terms, which would appear beyond 1PN order, have been neglected. We leave for future work including spin effects, as well as contributions from the tail terms.

\subsection{Relation $e_{\wgw}-e_t$}\label{sec:AppendixeOm22et}
This derivation proceeds similarly to appendix~\ref{sec:AppendixeOm22Orb}, but starting from the quasi-Keplerian parametrization. We start with 1PN expressions for the (2,2)-mode waveform, $h_{22}$, in the quasi-Keplerian parameterization \cite{Mishra:2015bqa},
\begin{align}
h^{QK}_{22} & = 4 \nu x \sqrt{\frac{\pi}{5}}\left[ \hat{h}^{\text{0PN}}_{22}+\gamma \hat{h}^{\text{1PN}}_{22} \right] e^{-2 i \phi},\\
\hat{h}^{\text{0PN}}_{22} & = \frac{2}{\left[1-e_t \cos (u)\right]{}^2} \left[1 -e_t^2  + i \sqrt{1-e_t^2} e_t \sin (u)  \right. \nonumber \\
& \left. -\frac{1}{2} e_t \cos (u) \left[1-e_t \cos (u)\right] \right],\\ \nonumber \\
\hat{H}^{\text{1PN}}_{22} & = -\frac{x}{42 \left(1-e_t^2\right) \left(1-e_t \cos (u)\right){}^3} \left[(64 \nu -278) e_t^4  \right. \nonumber \\
& \left.  +(46 \nu +64) e_t^2+e_t^3 \cos ^3(u) \left((17 \nu -57) e_t^2-17 \nu -27\right) \right. \nonumber\\
& \left.  +e_t^2 \cos ^2(u) \left((114-34 \nu ) e_t^2+34 \nu +54\right) \right. \nonumber\\
& \left. +e_t \cos (u) \left((114-34 \nu) e_t^4+(207-89 \nu ) e_t^2+123 \nu  \right. \right. \nonumber\\
& \left. \left.  -405\right)+i \sqrt{1-e_t^2} e_t \sin (u) \left((272-46 \nu ) e_t^2 \right. \right. \nonumber\\
& \left. \left.  +e_t \cos (u) \left((34 \nu -114) e_t^2+50 \nu -138\right)-38 \nu -20\right) \right.\nonumber\\
& \left. -110 \nu  +214\right]
\label{eq:eqD10}
\end{align}
  The phase of Eq.~(\ref{eq:eqD10}) can be written as
\begin{align}
\phi_{22}^{\text{1PN}} & =  \tan ^{-1}\left(\frac{A_0}{A_1}\right) + \gamma \frac{ x e_t}{B_0} \left[ \sin (u) \left((103-78 \eta ) e_t^4+(917 \right. \right. \nonumber \\
& \left. \left. -294 \eta ) e_t^2 +72 (4 \eta -13)\right)+e_t \left(2 \sin (2 u) \left((117 \eta -340) e_t^2 \right. \right.\right. \nonumber \\
& \left. \left.\left. -54 \eta +277\right) +e_t \left(21 (\eta -1) e_t \sin (4 u)+\sin (3 u)
   \left(-6 \eta  \left(e_t^2 \right. \right.\right. \right.\right.\nonumber \\
   & \left. \left.\left. \left.\left.  +13\right)+79 e_t^2+5\right)\right)\right) \right] \nonumber \\
\end{align}
where
\begin{align}
  A_0 =& 4 e_t \sqrt{1-e_t^2} \sin (u) \cos (2 \phi )\nonumber \\
  &+\sin (2 \phi ) \left( -e_t^2 \cos (2 u)
 +2 e_t \cos (u) +3 e_t^2-4\right), \\
 A_1 =& 4  e_t \sqrt{1-e_t^2} \sin (u) \sin (2 \phi )\nonumber \\
 & +\cos (2 \phi ) \left(e_t^2 \cos (2 u) -2 e_t \cos (u) -3 e_t^2+4\right), \\
 B_0 =& 42 \sqrt{1-e_t^2} \left(e_t \cos (u)-1\right){}^2 \nonumber \\
 & \left(2 e_t^3 \cos ^3(u)-e_t^2 \cos (2 u) +7 e_t^2-8\right) .
\label{eq:eqD11}
\end{align}
The time derivative of $\phi^{\text{1PN}}_{22}(x,e_t,u, \phi)$ can be expressed in functional form as,
\begin{equation}
\wgw^{\text{1PN}} \equiv \frac{d \phi_{22}^{1PN, QK}}{dt}= \frac{\partial \phi_{22}}{\partial x} \dot{x} + \frac{\partial \phi_{22}}{\partial e_t} \dot{e}_t + \frac{\partial \phi_{22}}{\partial u} \dot{u}  +\frac{\partial \phi_{22}}{\partial \phi} \dot{\phi}.
\label{eq:eqD12}
\end{equation}
The time derivatives $\dot{x}$, $\dot{e}_t$ and $\dot{\phi}$ can be found in \cite{Hinder:2008kv,Konigsdorffer:2006zt}, while for the eccentric anomaly, $u$, we use the Kepler equation at Newtonian order to write\footnote{We note that there are no 1PN order corrections to the Kepler equation, and that the first higher order PN correction enters at 2PN order \cite{Memmesheimer:2004cv,Boetzel:2017zza}.}
\begin{equation}
\dot{u}  = \frac{ \dot{l} + \dot{e}_t \sin u}{1- e_t \cos u },
\label{eq:eqD13}
\end{equation}
where $l$ is the mean anomaly, and an expression for $\dot{l}$ in the quasi-Keplerian parametrization can be found in \cite{Hinder:2008kv}. We note that the 3PN Kepler equation can be found in \cite{Konigsdorffer:2006zt,Hinder:2008kv}, however, we restrict to low PN order for simplicity of the calculations, and to avoid the introduction of the true anomaly, which substantially complicates the higher order calculations \cite{Boetzel:2017zza}.

At 1PN order, one can write the following expression for the frequency of the (2,2)-mode
\begin{equation}
\label{eq:eqD14}
\wgw^{\text{QK,1PN}} = \wgw^{\text{QK}, 0}  + \gamma \wgw^{\text{QK}, 1},
\end{equation}
where
\begin{widetext}
\begin{align}
\wgw^{\text{QK}, 0} &= -\frac{2 x^{3/2} \sqrt{1-e_t^2} \left(e_t^2 \cos (2 u)-4 e_t \cos (u)-5
   e_t^2+8\right)}{\left(e_t \cos (u)-1\right){}^2 \left(2 e_t^3 \cos ^3(u)-e_t^2 \cos (2 u)+7 e_t^2-8\right)}, \\
\wgw^{\text{QK}, 1}  &  = \frac{x^{5/2} e_t }{168 \sqrt{1-e_t^2} \left(e_t \cos (u)-1\right){}^4 \left(e_t^2 \left(\cos (2 u)-2 e_t \cos
   ^3(u)\right)-7 e_t^2+8\right){}^2}\left[\cos (u) \left((5155-1605 \nu ) e_t^8+21 (86 \nu +213) e_t^6  \right.\right. \nonumber \\
   & \left. \left.  -24 (279 \nu +838) e_t^4+16 (288 \nu -1321) e_t^2+768 (115-16 \nu )\right)+e_t \left(-4 \left(7 (102 \nu -253) e_t^6+(19781-4995 \nu ) e_t^4+(6936 \nu \right.\right.\right. \nonumber \\
   & \left. \left.  \left. -43318) e_t^2-4608 \nu
   +33120\right)+e_t^5 \cos (7 u) \left((5-15 \nu ) e_t^2-6 \nu +79\right)-14 e_t^4 \cos (6 u) \left((6 \nu -32) e_t^2-21 \nu
   +92\right)\right.\right.  \nonumber \\
   & \left. \left.    +e_t^3 \cos (5 u) \left((3 \nu +461) e_t^4+9 (166 \nu -445) e_t^2-2484 \nu +7492\right)+4 e_t^2 \cos (4 u) \left((849-174
   \nu ) e_t^4+(509-1107 \nu ) e_t^2+2016 \nu \right.\right.  \right.  \nonumber \\
   & \left. \left.\left. -4298\right)+e_t \cos (3 u) \left(-495 (\nu -5) e_t^6+(6306 \nu -27043) e_t^4+20 (1535-69
   \nu ) e_t^2-336 (32 \nu -57)\right)+2 \cos (2 u) \left(6 (95 \nu \right.\right.  \right.  \nonumber \\
   & \left. \left.\left. +708) e_t^6-3 (1553 \nu +964) e_t^4+8 (642 \nu +619) e_t^2+128 (33
   \nu -214)\right)\right)\right].
\end{align}
Evaluating Eq.~\eqref{eq:eqD14} at the turning points, apastron, $u=\pi$, and periastron, $u=0$, one obtains
\begin{equation}
\wgw^{\text{a,p}}(x,e_t,\eta)  =\frac{4 x^{3/2} \sqrt{1-e_t^2}}{(e_t \pm 1 )^2(2 \mp e_t)} \pm \gamma \frac{x^{5/2} e_t \left(11 (6 \eta -23) e_t^2
\pm (607-78 \eta ) e_t+96 \eta -690\right)}{21 \sqrt{1-e_t^2} \left(e_t^2+e_t-2\right){}^2},
\label{eq:eqD15}
\end{equation}
where the upper and lower signs correspond to apastron and periastron, respectively.

Finally, substituting the result of Eq.~\eqref{eq:eqD15} into the eccentricity definition of Eq.~\eqref{eq:eq7}, and expanding to 1PN order one obtains
\begin{equation}
e^{\text{1PN,QK}}_{\wgw} =  \frac{\sqrt{2-e_t} \left(1+e_t\right) - \left(1-e_t\right)\sqrt{2+e_t} }{\sqrt{2-e_t} \left(1+e_t\right) + \left(1-e_t\right)\sqrt{2+e_t} }
 -\gamma x e_t \frac{ (54 \eta +101) e_t^2+192 \eta -1380 }{84 \left(e_t^4-5 e_t^2+2 \sqrt{4-e_t^2}+4\right)}.
\label{eq:eqD16}
\end{equation}
This is Eq.~(\ref{eq:eq112}) from the main text.
\end{widetext}

\bibliography{references}

\end{document}